\documentclass[iop, apj]{emulateapj}
\usepackage{amsmath}
\usepackage{mathrsfs}

\shortauthors{Rappazzo}

\begin{document}

\title{Equilibria, Dynamics and Current Sheets Formation \\ 
       in magnetically confined coronae}

\author{A. F. Rappazzo}
\affil{Advanced Heliophysics, Pasadena, CA 91106, USA; franco.rappazzo@gmail.com}

\begin{abstract}
The dynamics of magnetic fields in closed regions of solar
and stellar coronae are investigated with a reduced magnetohydrodynamic
(MHD) model in the framework of Parker scenario for coronal heating.
A novel analysis of reduced MHD equilibria shows that their magnetic fields 
have an asymmetric structure in the axial direction with variation length-scale
$z_\ell \sim \ell B_0/b$, where $B_0$ is the intensity of the strong axial
guide field, $b$ that of the orthogonal magnetic field component, and $\ell$
the scale of $\mathbf{b}$. Equilibria are then quasi-invariant along the axial
direction for variation scales larger than approximatively the loop length 
$z_\ell \gtrsim L_z$, and increasingly more asymmetric for smaller variation
scales $z_\ell \lesssim L_z$. The \emph{critical length} $z_\ell \sim L_z$ corresponds
to the magnetic field intensity threshold $b \sim \ell B_0/L_z$. 
Magnetic fields stressed by photospheric motions cannot develop strong axial 
asymmetries. Therefore fields with intensities below such threshold evolve quasi-statically, 
readjusting to a nearby equilibrium, without developing nonlinear dynamics nor dissipating energy.
But stronger fields cannot access their corresponding asymmetric equilibria, hence
they are out-of-equilibrium and develop nonlinear dynamics.
The subsequent formation of current sheets and energy dissipation is \emph{necessary}
for the magnetic field to relax to equilibrium, since dynamically accessible equilibria 
have variation scales larger than the loop length $z_\ell \gtrsim L_z$, with intensities
smaller than the threshold $b \lesssim \ell B_0/L_z$.
The dynamical implications for magnetic fields of interest to solar and stellar
coronae are investigated numerically and the impact on coronal physics discussed.

\end{abstract}

\keywords{magnetohydrodynamics (MHD) --- stars: activity --- stars: solar-type ---  
Sun: corona --- Sun: magnetic topology  --- turbulence}

\section{Introduction}

Solar observations show a close association between magnetic field strength
and coronal activity. In combination with the ability 
of photospheric motions to stress the field, these
are the two key elements to understand the observed coronal 
\emph{X-ray activity} of the Sun, 
of all late-type main sequence stars, and more in general of stars with
a magnetized corona and an outer convective envelope 
\citep{2004A&ARv..12...71G, 2009LNP...778..269G}.

It has long been proposed that the work done by
photospheric motions on magnetic field line footpoints
can transform mechanical energy into magnetic energy
and transfer it in the upper corona. The photospheric
(horizontal) velocity can be split into irrotational and solenoidal
components. Only its solenoidal (\emph{incompressible}) 
component has a non-vanishing vorticity and can then twist
the magnetic field lines, injecting magnetic energy into the corona.

\cite{1960MNRAS.120...89G}
conjectured that the magnetic field would proceed through a
sequence of force-free equilibria while photospheric
vortices twist the field lines, and the stored energy could 
subsequently be released when two flux tubes with similarly 
twisted field lines  come into contact with each other,
or when the configuration would become 
somehow unstable through an undetermined mechanism 
\citep{1964NASSP..50..389G}.
\cite{1981ApJ...246..331S} compute the energy flux into
the corona due to the work done by \emph{random}  photospheric 
vortical motions on the magnetic field. They find that
the \emph{correlation time} of photospheric motions must
be of the order of the observed photospheric timescales (5-8 minutes)
or longer to obtain an energy flux large enough to sustain an active corona,
otherwise for shorter correlation times the resulting twisted field is too small.
But the magnetic energy is still supposed to be stored in a force-free field in 
equilibrium, and no physical mechanism able to dissipate this energy
and heat the corona is envisioned.

Energy stored in a magnetic field in equilibrium,
that subsequently becomes unstable and releases its energy,
is the common thread of flare models \citep{2011LRSP....8....6S, 2012ASPC..463..157M},
with the processes leading to the pre-flare magnetic energy storage
and its subsequent fast release strongly debated.
On the other hand this picture does not appear apt to describe 
the dynamics of the long-lived slender X-ray bright loops,
that in comparison to a flare show little dynamics from their large-scales down to the 
smallest resolved scale ($\sim150$\,km) of current state-of-the-art X-ray 
and EUV imagers on board Hinode, SDO and Hi-C \citep{2013A&A...556A.104P}.
While the pre-flare magnetic structure is destroyed during the flare,
the \emph{large-scale magnetic topology of the loops where the basic heating occurs, and
that make the corona shine steadily in X-ray, is not strongly modified} on
comparable timescales. 
This suggests that
the energy deposition must occur at very small scales yet observationally
unresolved. Furthermore the energy reservoir that supplies dissipation
should consist of magnetic field fluctuations (with vanishing time-average,
but non vanishing time-averaged rms) that adds up to the strong axial
magnetic field that defines the loop.

\cite{1972ApJ...174..499P, 1988ApJ...330..474P,1994ISAA....1.....P, 2012PPCF...54l4028P} 
was the first to suggest that, in contrast to previous quasi-static
models, the magnetic field brought about by photospheric vortical motions 
would be in \emph{dynamical non-equilibrium} in the case of interest to coronal heating.
Furthermore the relaxation of this interlaced fields toward equilibrium
would necessarily involve the formation of \emph{current sheets}.
The energy dissipation would then occur at small scales in the fashion of 
small impulsive heating events, so-called \emph{nanoflares},
a picture broadly used for the thermodynamical modeling of the closed corona
\citep{2006SoPh..234...41K, 2014LRSP...11....4R}.

Using a simplified Cartesian model with a strong guide field
threading a coronal loop, \cite{1972ApJ...174..499P, 1979cmft.book.....P} 
argued at first that a magnetic field could be in equilibrium only
if it were invariant along $z$ (the axial direction). Due to 
the complex and disordered nature of photospheric motions
the induced interlaced magnetic field would not be invariant, and
therefore not in equilibrium.
Next, counterexamples of magnetostatic equilibria that are not invariant along $z$
were provided \citep{1982ApJ...262..349R, 2000PhRvL..84.1914B},
and analytical investigations 
\citep{1985ApJ...298..421V, 1987ApJ...312..886A, 1997PhR...283..227C} 
argued that smooth photospheric motions cannot lead to the formation of 
current sheets, whereas only a discontinuous velocity field can form discontinuities
in the coronal magnetic field.

In particular \cite{1985ApJ...298..421V} showed that the equilibria
are the solutions of the two-dimensional (2D) Euler equation
that in general are not $z$-invariant,  thus inferring that
the field would evolve quasi-statically, continuously
readjusting to a nearby force-free equilibrium without developing
nonlinear dynamics nor forming current sheets.
Reaching opposite conclusions
\cite{1988ApJ...330..474P,1994ISAA....1.....P,2000PhRvL..85.4405P,2012PPCF...54l4028P}
pointed out that almost all field line topologies relevant
to the solar corona have a different structure from
the solutions of the Euler equation, so that
the magnetic field would be still in dynamical non-equilibrium.

Reduced magnetohydrodynamics (MHD) numerical simulations with a continuous smooth
velocity forcing at the boundaries show that the dynamics
can be seen as a particular instance of magnetically 
dominated MHD turbulence
\citep{1999ApJ...527L..63D, 2003PhPl...10.3584D, 
2007ApJ...657L..47R, 2008ApJ...677.1348R}
as proposed by earlier 2D models 
\citep{1996ApJ...457L.113E, 1997ApJ...484L..83D},
suggesting that in the forced case the magnetic field is in \emph{dynamical 
non-equilibrium} rather than close to a quasi-static evolution.
Similar dynamics are also displayed by boundary
driven simulations in the cold plasma regime \citep{1996ApJ...467..887H}
and in the fully compressible MHD case \citep{1996JGR...10113445G, 2012A&A...544L..20D}.
Furthermore \cite{2011PhRvE..83f5401R}  have shown that while
velocity fluctuations are much smaller than
magnetic fluctuations,  spectral energy fluxes
toward smaller scales are akin to those of a standard
cascade with magnetic and kinetic energies in equipartition, except
for kinetic energy fluxes that are negligible.
This implies that at scales smaller than
those directly shuffled by photospheric motions,
the small \emph{velocity field is created and shaped
by the unbalanced Lorentz force of the out-of-equilibrium magnetic field},
that in turn creates small scales in the magnetic field by \emph{distorting
magnetic islands and pushing field lines together}.
Additionally \cite{1998ApJ...497..957G, 1998ApJ...505..974D} 
have established a link between boundary driven simulations and
observed statistics of coronal activity. Indeed the bursts in 
dissipation displayed by the system,
that correspond to the formation and dissipation of current sheets, 
follow a power law behavior in total energy, peak dissipation and
duration with indexes not far from those determined observationally in X-rays.

Recently \cite{2009ApJ...696.1339W} have shown that the relaxation
of a slightly braided magnetic field (``pigtail'' braid) 
appears to evolve quasi-statically, with no formation of current sheets, 
toward an equilibrium where only large-scale current layers of thickness much larger
than the resolution scale are observed. This result would seem in contrast
with Parker's hypothesis, the results of the forced numerical simulations discussed
in the previous paragraph, and the recent results supporting
the development of finite time singularities in the cold
plasma regime \citep{2013ApJ...768....7L, 2015SCPMA..58a...2L}.

To get further insight on the dynamics of coronal magnetic fields,
\cite{2013ApJ...773L...2R} analyzed reduced MHD numerical simulations
of the relaxation of initial magnetic fields \emph{invariant along} $z$  and with 
different average twists. They identified a \emph{critical intensity threshold} for 
the magnetic field.
This is explained heuristically as due to a balance between different field line
tension forces for weak fields, while such a balance cannot be attained by stronger fields.
The non-equilibrium of stronger fields stems from this force unbalance,
and drives the relaxation forming current sheets and dissipating energy.
On the contrary weaker fields show little dynamics with no energy dissipation, confirming
that they are essentially in equilibrium. 
Such threshold can explain qualitatively the result
of \cite{2009ApJ...696.1339W}, although a quantitative comparison cannot
be made because the integrated equations (magneto-frictional relaxation
vs.\ reduced MHD) and initial topologies differ.

The magnetic intensity threshold found by \cite{2013ApJ...773L...2R} 
implies that a \emph{critical twist} exists above which dynamics develop, 
and below which the system remains very close to equilibrium.
\cite{1988ApJ...330..474P} had conjectured that a critical twist is necessary
to explain the observationally inferred energy flux in active regions
\citep{1977ARA&A..15..363W}. In fact 
the energy flux injected in the corona by 
photospheric motions is the average Poynting flux 
$\langle S_z \rangle = B_0\, \langle \mathbf{u}_\textrm{ph} \cdot \mathbf{b} \rangle$ 
(see Section~\ref{sec:energy},   Equation~[\ref{eq:ipf}])
that depends not only on the photospheric velocity $\mathbf{u}_\textrm{ph}$
and the axial guide field $B_0$,
but also on the dynamic magnetic field component $\mathbf{b}$,
with stronger intensities corresponding to higher average twists.
Thus if nonlinear dynamics were to develop for weak field
intensities, energy dissipation would keep too low the value of $\mathbf{b}$, 
and consequently the flux $\langle S_z \rangle$.
This argument is further reinforced by the fact that current sheets thickness 
decreases at least exponentially in time when nonlinear dynamics develop, reaching
the Sweet-Parker thickness \citep{1958IAUS....6..123S, 1957JGR....62..509P} on ideal timescales 
\citep[about one Alfv\'en crossing time $\tau_A$,][]{2013ApJ...773L...2R},
that current sheets are unstable to tearing modes with ``ideal'' growth rates
(i.e., of order 1/$\tau_A$) already at thicknesses larger than Sweet-Parker 
\citep{2014ApJ...780L..19P, 2015ApJ...801..145T, 2015ApJ...806..131L}, 
and that magnetic reconnection rates
can be very fast in plasmas with high Reynolds numbers
\citep{1999ApJ...517..700L, 2007PhPl...14j0703L,
2008PhRvL.100w5001L, 2009MNRAS.399L.146L, 
2010PhRvL.105w5002U, 2010PhPl...17f2104H, 
2013arXiv1301.7424B} 
and in the collisionless regime \citep{1999GeoRL..26.2163S, 2001JGR...106.3715B}.

This paper is devoted to a more detailed discussion and
analysis of the numerical simulations described by \cite{2013ApJ...773L...2R},
of additional simulations that extend our previous work 
to initial conditions non-invariant along~$z$,
and to a novel analysis of the structure of the reduced MHD equilibria,
with the goal to shed light on the topics outlined in this introduction and 
advance our understanding of coronal magnetic field 
dynamics, their relationship to dynamic non-equilibrium, MHD
turbulence, quasi-static evolution, current sheets formation and
activity of solar and stellar coronae.

The loop model along with initial and boundary conditions for the simulations 
are introduced in Section~\ref{sec:pm}.
The structure of the equilibria is analyzed in Section~\ref{sec:eq}, and  
the results of the numerical simulations are described in Section~\ref{sec:res}. 
Finally results and conclusions are summarized in Section~\ref{sec:cad},
together with a discussion of their impact on coronal physics.

\section{Physical Model} \label{sec:pm}

A closed region of the solar corona is modeled,
as in previous work \citep{2007ApJ...657L..47R}, with 
a simplified cartesian geometry, uniform 
density $\rho$  and a \emph{strong  and
homogeneous axial magnetic field} 
$\mathbf{B}_0 = B_0\, \mathbf{\hat{e}}_z$ threading the system.
Plasmas in such configurations are well suited to be 
studied in the reduced MHD regime \citep{1992JPlPh..48...85Z}.
Introducing the velocity and magnetic field potentials $\varphi$ and $\psi$, 
for which
$\mathbf{u} = \nabla \varphi \times \mathbf{\hat{e}}_z$,
$\mathbf{b} = \nabla \psi    \times \mathbf{\hat{e}}_z$, 
vorticity
$\omega = - \nabla^2_{\!_{\!\perp}} \varphi$, and the current density 
$j = - \nabla^2_{\!_{\!\perp}} \psi$, the nondimensional reduced MHD equations \citep{1974JETP...38..283K, 1976PhFl...19..134S}
are:
\begin{eqnarray}
&& \partial_t \psi = \left[ \varphi, \psi \right] 
+ B_0\, \partial_z \varphi
+ \eta_{_n} \nabla_{_{\!\!\perp}}^{2n} \psi, \quad \label{eq:eq1} \\
&& \partial_t \omega = \left[ j, \psi \right] - \left[ \omega, \varphi \right]
+ B_0\, \partial_z j
+ \nu_{_{\!n}} \nabla_{_{\!\!\perp}}^{2n} \omega. \label{eq:eq2}
\end{eqnarray}
The Poisson bracket of functions $g$ and $h$ is defined as
$[g,h] = \partial_x g\, \partial_y h - \partial_y g\, \partial_x h$
(e.g., $[j,\psi]=\mathbf{b} \cdot \nabla j$), and
Laplacian operators have only orthogonal components.
To render the equations nondimensional the magnetic fields are first
expressed as Alfv\'en velocities ($b \rightarrow b/\sqrt{4\pi \rho}$),
and then all velocities are normalized with $u^{\ast} = 1$~km\,s$^{-1}$, 
a typical value for photospheric motions.
The domain spans $ 0 \le x, y \le L_\perp$ and $ 0 \le z \le L_z$, 
with $L_\perp = 1$ and $L_z=10$.
Magnetic field lines are line-tied to a motionless photosphere 
at the top and bottom plates ($z=0$ and $10$), where a vanishing
velocity $\mathbf{u} = 0$ is in place. 
In the perpendicular ($x$-$y$) directions
a  pseudo-spectral scheme with
periodic boundary conditions and isotropic truncation de-aliasing  is used
\citep[$2/3$-\emph{rule},][]{2006spme.book.....C}, while along $z$
a second-order finite difference scheme is implemented.
The CFL (Courant-Friedrichs-Levy) condition is satisfied
through an adaptive time-step.
For a more detailed description of 
the model and numerical code see 
\cite{2007ApJ...657L..47R, 2008ApJ...677.1348R}.

Dissipative simulations use hyper-diffusion 
\citep{2003matu.book.....B}, that effectively
limits diffusion to the small scales, with $n=4$ and 
$\nu_n = \eta_n = \left(-1\right)^{n+1} / R_n$, 
with $R_n$ corresponding to the Reynolds number for $n=1$ 
\cite[see][]{2008ApJ...677.1348R}.

\subsection{Initial and boundary conditions} \label{sec:initial}

Simulations are started at time $t=0$ with a vanishing
velocity  $\mathbf{u} = 0$ everywhere, and a uniform
and homogeneous guide field $B_0$.
The orthogonal field $\mathbf{b}$ consists of 
\emph{large-scale} Fourier modes, set expanding
the magnetic potential  in the following way:
\begin{eqnarray}
&& \psi_0  = b_0
\sum_{rsm} ( 2 \mathscr{E}_{_m})^{\frac{1}{2}}\ \frac{\alpha_{rsm} \ \sin \left( \mathbf{k}_{rsm}\! \cdot \mathbf{x} + 2\pi \xi_{rsm} \right) }
{ k_{rs}\ \sqrt{ \sum_{ij} \alpha^2_{ijm} } }  
\label{eq:pot4}\\
&&\textrm{with} \quad \mathbf{k}_{rsm} = \frac{2\pi}{L_\perp}
\left( r\, \mathbf{\hat{e}}_x + s\, \mathbf{\hat{e}}_y \right)
+\frac{2\pi}{L_z} m\, \mathbf{\hat{e}}_z, \nonumber \\
&& \textrm{and} \quad k_{rs} = \frac{2\pi}{L_\perp} \sqrt{ r^2 + s^2 }, \nonumber
\end{eqnarray}
where the coefficients $\alpha_{rsm}$ and $\xi_{rsm}$ are two independent 
sets of random numbers uniformly distributed between 0 and 1.
The orthogonal wave-numbers $(r,s)$  are always
in the range $3 \le (r^2+s^2)^{1/2} \le 4$, 
while the parallel amplitudes $\mathscr{E}_m$  (with $\sum_m \mathscr{E}_m =1$)
are set to distribute the energy in different ways in the axial direction.
Given the orthogonality of the base used in Eq.~(\ref{eq:pot4})
the normalization factors guarantee that
for any choice of the amplitudes the rms of the
magnetic field is set to 
$ b = \langle  b_x^2 + b_y^2  \rangle^{1/2} = b_0$,
while for total magnetic energy 
$E_{_{\!M}} = b_0^2 V/2 \sum_m \mathscr{E}_m$,
i.e., $\mathscr{E}_m$ is the fraction of magnetic energy in the 
\emph{parallel} mode~$m$.
Two-dimensional (2D) configurations invariants along $z$  are obtained 
considering the single mode $m=0$ with $\mathscr{E}_0=1$.

\subsection{Energetics} \label{sec:energy}

From equations (\ref{eq:eq1})-(\ref{eq:eq2}), with $n=1$ and considering the 
kinetic and magnetic Reynolds numbers equal, 
the following energy equation can be obtained:
\begin{equation}  \label{eq:en}
 \frac{\partial}{\partial t}\ \frac{1}{2} \left( \mathbf{u}^2 + 
 \mathbf{b}^2  \right) = - \nabla \cdot \left( \mathbf{S} + \mathbf{F} \right)
  - \frac{1}{R} \left( j^2 + \omega^2 \right), 
\end{equation}
where $\mathbf{S}= \mathbf{B} \times ( \mathbf{u} \times \mathbf{B})$ 
is the Poynting vector, and 
$\mathbf{F}=(p+\mathbf{u}^2/2)\mathbf{u}
-(\omega\mathbf{u}+j\mathbf{b})\times \mathbf{\hat{e}}_z/R$
is an orthogonal transport-related flux. 
Integrating equation (\ref{eq:en}) over the whole box the
energy ($E$) equation is 
\begin{equation}
 \frac{\partial E}{\partial t} = S - \frac{1}{R} \int_V \mathrm{d}V 
            \left( j^2 + \omega^2 \right), \label{eq:en2}
\end{equation}
i.e., as expected, the global energy balance depends
on the competition between the energy flowing into 
the computational box from the photospheric boundaries $S$
and the ohmic and viscous dissipation. 
Because in the x--y planes periodic boundary conditions are 
implemented, and $F_z=0$, the only relevant component of the flux vectors 
is that of the Poynting vector along the axial direction $S_z$ that,
as $\mathbf{B} = B_0 \mathbf{\hat{e}}_z + \mathbf{b}$, is given by
\begin{equation}
  S_z = \mathbf{S} \cdot \mathbf{\hat{e}}_z = -B_0
        \left(  \mathbf{u} \cdot \mathbf{b} \right). \label{eq:pf}
\end{equation}
Indicating the photospheric velocity fields at the top and bottom 
boundaries  $z=0$ and $L$ with $\mathbf{u}^0$ and $\mathbf{u}^L$
for the integrated energy flux (i.e., the power) $S$ we obtain
\begin{equation} \label{eq:ipf}
  S = B_0 \int\limits_{z=L}\mathrm{d}a 
          \left( \mathbf{u}^L \cdot \mathbf{b} \right) 
     -B_0 \int\limits_{z=0}\mathrm{d}a 
          \left( \mathbf{u}^0 \cdot \mathbf{b} \right).
\end{equation}
The injected energy power is proportional to $B_0$ and depends on 
the dot product of the photospheric
velocities $\mathbf{u}^{0,L}$ and the magnetic field  $\mathbf{b}$
at the boundaries. But while $\mathbf{u}^{0,L}$ and $B_0$ have
fixed values (in our simplified model), the magnetic field $\mathbf{b}$
is determined by the linear or nonlinear dynamics developing in the 
computational box. 

Since the magnetic field component $\mathbf{b}$
can often be considered quasi-invariant along~$z$ (as described
in the following sections), as a shorthand we will indicate the difference
between the boundary velocities with $\mathbf{u}_{ph} =  \mathbf{u}^L - \mathbf{u}^0$,
so that for quasi-invariant fields the Poynting flux can be approximated as
$\langle S_z \rangle \sim S/\ell^2 \sim B_0 u_{ph} b$.

\section{Equilibria and their dynamic accessibility: Analysis} \label{sec:eq}

As discussed in the introduction, the properties of the equilibria 
of this system are pivotal to understand its dynamics 
\citep{1972ApJ...174..499P, 1988ApJ...330..474P,1994ISAA....1.....P, 
1985ApJ...298..421V, 1986ApJ...311.1001V},
therefore their structure is analyzed here in detail.
It is shown that, depending on the ratio
$b_0/B_0$ of the rms of the orthogonal component to
the guide magnetic field intensity, \emph{the equilibria can be
approximately invariant along $z$ or strongly asymmetric}.
As shown in the following this can explain why fields
with a twist below a \emph{critical} value do not form strong
current sheets, while they do at higher twists as conjectured
by \cite{1988ApJ...330..474P}. Nevertheless unlike commonly
thought, such equilibria are generally not linearly unstable
for most conditions relevant to coronal loops, since
they arise from a balance of forces in an \emph{asymmetric and
irregular topology}.

Neglecting velocity and diffusion terms, equilibria of 
Eqs.~(\ref{eq:eq1})-(\ref{eq:eq2}) are given by $\mathbf{B} \cdot \nabla j = 0$.
Since the total magnetic field $\mathbf{B}$ is given by
$\mathbf{B} = B_0 \mathbf{\hat{e}}_z + \mathbf{b}(x,y,z)$ with
$\mathbf{b} \cdot \mathbf{\hat{e}}_z = 0$, the equilibrium condition can
be written as:
\begin{equation} \label{eq:eqb}
\frac{\partial  j}{\partial z} = - \frac{\mathbf{b}}{B_0} \cdot \nabla j,
\end{equation}
where the right-hand side term corresponds to the ``2D perpendicular'' Lorentz force
component $\mathbf{b} \cdot \nabla \mathbf{b}$,
and the left hand side to the ``parallel'' $B_0 \partial_z \mathbf{b}$ field line tension
(the labels refer to their derivative, but both components are orthogonal to $B_0$, 
a more detailed discussion is in Section~\ref{sec:run0}  prior to Equation~[\ref{eq:eqrp}]).

Assigned $\mathbf{b}$ in an x-y plane, e.g., at the boundary z=0 
$\mathbf{b}(x,y,z=0) = \mathbf{b}_{bd}(x,y)$, the integration
of this equation for $z>0$ allows to compute the corresponding equilibrium in 
the whole computational box $0 \le z \le L_z$.

Now consider the 2D Euler equation \citep{euler1761principia}
\begin{equation} \label{eq:eu} 
\frac{\partial  \omega}{\partial t} = - \mathbf{u} \cdot \nabla \omega,
\end{equation}
with $\nabla \cdot \mathbf{u} = 0$.
Introducing the velocity potential $\phi$, then
$\mathbf{u}(x,y,t) = \nabla \phi (x,y,t) \times \mathbf{\hat{e}}_z$,
and vorticity $\omega = - \nabla^2 \phi$.
The two equations~(\ref{eq:eqb}) and (\ref{eq:eu}) are \emph{identical} under 
the mapping
\begin{eqnarray} \label{eq:etr}
   \left\{
   \begin{aligned}
   & \\[-1.em]
   & t \rightarrow z,\\[.1em]
   & \mathbf{u} \rightarrow \frac{\mathbf{b}}{B_0},\\[-1.3em]
   &
   \end{aligned} \right.
\end{eqnarray}
and consequently $\omega \rightarrow j/B_0$.

The related 2D Navier-Stokes equation is obtained by  adding to the right 
hand side of Equation~(\ref{eq:eu}) the dissipative term $\nu \nabla^2 \omega$,
from which the 2D Euler equation is recovered for $\nu = 0$.
The physics and solutions of the 2D Euler 
and Navier-Stokes equations have been 
studied extensively theoretically, numerically and in the laboratory, 
in the framework of 2D hydrodynamic turbulence
\citep[see reviews by][]{1980RPPh...43..547K, 2002PhR...362....1T, 2012AnRFM..44..427B}.
Unlike the 3D case, it has been shown that given a smooth 
initial condition $\mathbf{u}_0(x,y)$ at time $t=0$ the 2D Euler
equation admits a \emph{unique and regular solution} at $t>0$,
i.e., no finite time singularity develops
\citep{rose1978fully, Chemin1993, 1993CMaPh.152...19B, 2001vif..book.....M}.

In 2D in addition to \emph{energy}, also
mean-square vorticity (\emph{enstrophy}) is conserved. 
The coupled conservation constraints have a strong impact 
on the dynamics that differs considerably from its 3D 
hydrodynamic counterpart and the corresponding magnetohydrodynamic cases.
In particular, indicating the Energy with $E = (1/2) \langle \mathbf{u}^2 \rangle$
(the integrated square velocity), the enstrophy with 
$\Omega = (1/2) \langle \omega^2 \rangle$, and  the palinstrophy with
$P = (1/2) \langle | \nabla \omega |^2 \rangle$, the following energy and 
enstrophy conservation equations are obtained from the 2D Navier-Stokes 
equations \citep[e.g.,][]{2012AnRFM..44..427B}:
\begin{equation}
\frac{d E}{d t} = -2 \nu \Omega = -\epsilon_\nu(t),
\qquad
\frac{d \Omega}{d t} = -2 \nu P. \label{eq:do}
\end{equation}
Since all quantities ($E$, $\Omega$, $P$, and $\nu$) are positively defined, it follows that
$\Omega$ can at most decrease. Therefore the energy dissipation 
rate $\epsilon_\nu$ vanishes as viscosity tends to zero:
\begin{equation} \label{eq:e0}
   \lim_{\nu \to 0} \epsilon_\nu = 0.
\end{equation}

This result strongly differs from the 3D case where, in the 
K41 phenomenology introduced by \cite{1941DoSSR..30..301K}, 
for a sufficiently small viscosity the energy dissipation rate
is approximately constant  $\epsilon_\nu \sim \textrm{const}$ 
and independent from viscosity. This 
\emph{dissipative anomaly} in 3D was first pointed out by 
\cite{1935RSPSA.151..421T}, and later confirmed in laboratory experiments
\citep{dryden1942review, 1984PhFl...27.1048S, 2002PhFl...14.1288P}
and by hydrodynamic numerical simulations 
\citep{1998PhFl...10..528S, 2003PhFl...15L..21K}.

Thus, in contrast to the 3D case, Equation~(\ref{eq:e0})  implies that for small
viscosities 2D turbulence is essentially unable 
to dissipate energy at small scales. The viscous sink of 
energy is missing, because at any given time 
during the decay of an initial large-scale velocity field, for a sufficiently 
small value of $\nu$, the dissipation is arbitrarily small.

Therefore in two dimensions there cannot be a direct energy cascade. 
As proposed by \cite{1967PhFl...10.1417K},
energy must flow toward the larger scales through an
\emph{inverse cascade}.
Thus during their evolution vortices (velocity eddies) acquire
increasingly larger scales, while it is enstrophy that
develops a \emph{direct} cascade \citep{1969PhFl...12..233B} with vorticity
acquiring smaller scales. As can be seen from Equation~(\ref{eq:eu}) 
the convective derivative of $\omega$ vanishes as $\nu$ tends to zero, i.e., 
vorticity is constant in time at a point that moves with the fluid.
The direct enstrophy cascade implies that an initial patch of vorticity
gets stretched in time to form filamentary structures, so that while 
$\omega$ is convected with the fluid its gradient increases.

Most numerical and laboratory experiments include a small
viscosity, but the regularity of the solutions of the 2D Euler equation and
the vanishing of the energy dissipation rate as viscosity tends to zero
for the 2D Navier-Stokes equation allow to establish a clear
connection between the solutions of the 2D Euler and Navier-Stokes
equations. 
Namely, the time evolution of the solutions 
of the same decay problem for the 2D Navier-Stokes and Euler equations 
is the same until dissipation sets in, i.e., until sufficiently small 
scales are formed and dissipation occurs in the Navier-Stokes case.

The dual cascade picture has been investigated and supported by 2D Navier-Stokes simulations 
of forced turbulence \citep[e.g., ][]{1983JPlPh..30..479H, 1984PhFl...27.1921F,
1986JFM...170..139S, 2002PhR...362....1T, 2012AnRFM..44..427B}, of
the decay of an initial condition consisting of large-scale vortices
\citep{1980NYASA.357..203M, 1991PhRvL..66.2731M}, and by
laboratory experiments of 2D decays performed with soap films, 
\citep{1989PhyD...37..406G, 1999PhFl...11.1196B, 2002PhRvL..88s4101G, 
2003PhRvL..90j4502R}.

The decaying (initial value problem, or `relaxation') case is particularly 
relevant to the analysis carried out in Section~\ref{sec:sred}. 
Recently \cite{2013PhRvE..87c3002M} have performed a number of
numerical simulations of the decay of an initial condition with energy 
in a narrow band of wavenumbers (with length-scale $\ell$, and velocity $u_{\ell}$), 
with sufficient resolution to allow 
the development of both the inverse energy and direct enstrophy cascades.
Since a decay is inherently non-steady, in order to compare with Kolmogorov 
K41 phenomenology, they perform several simulations where the initial 
condition has the same velocity rms $u_{\ell}$, but 
different random amplitudes.
In this way different realizations are obtained, allowing them to perform \emph{ensamble
averages} that smooth out the fluctuations in a single realization, and can then be compared
more straightforwardly with the original K41 phenomenology (that as a matter of fact uses 
ensamble averages). {\it Indeed a clear dual cascade is identified also in the case 
of a decay}.
In particular at wavenumbers smaller than the wavenumber of the initial condition,
the peak of energy moves toward smaller wavenumbers with time, and energy develops an
$E_k \sim k^{-5/3}$ spectrum, following K41 phenomenology
\cite[that does not depend on the direction of the cascade -- inverse or direct -- e.g., see][]{rose1978fully}.
The characteristic dynamic timescale is given by the eddy turnover time
\begin{equation}  \label{eq:tk41}
  t_\ell = \frac{\ell}{u_{\ell}},
\end{equation}
where $\ell$ and $u_{\ell}$  initially are the length-scale
and velocity of the initial condition.

\begin{figure*}
\begin{center}
\vspace{1em}
\includegraphics[height=.40\textwidth]{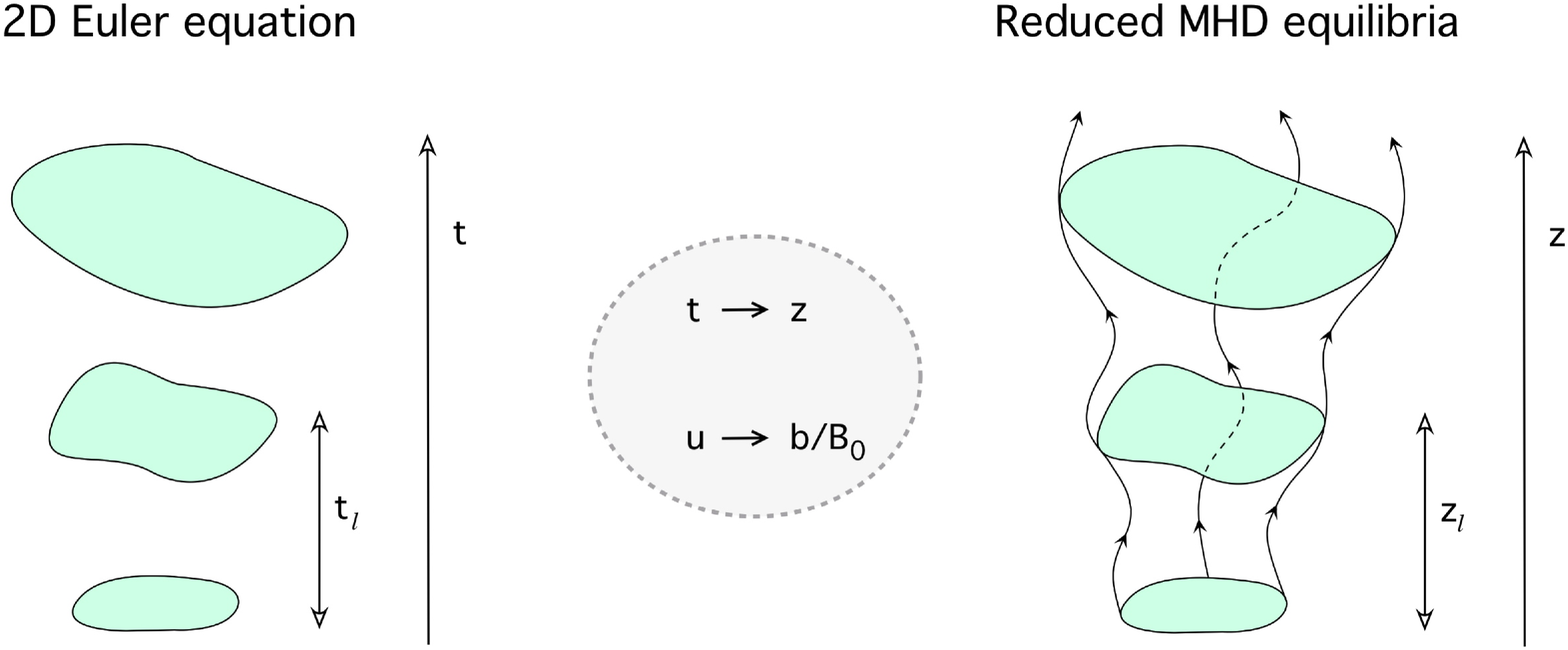}
\vspace{2em}
\caption{Schematic of the inverse energy cascade for the solutions of the 2D Euler
equation (\emph{left panel}) and for reduced MHD equilibria (\emph{right panel}). 
In the hydrodynamic case (\emph{left panel}) an eddy of size $\ell$ doubles its size in 
the eddy-turnover time $t_\ell \sim \ell/u_\ell$ (Equation~[\ref{eq:tk41}]). 
For the structure of reduced MHD equilibria (\emph{right panel}) this implies
(via the mapping $t \rightarrow z$, $\mathbf{u} \rightarrow \mathbf{b}/B_0$ 
given by Equation~(\ref{eq:etr}))
that an equilibrium solution with magnetic islands of transverse scale $\ell$
in the plane $z=0$ will acquire an axial asymmetry along the $z$-direction
characterized by the variation length-scale $z_{\ell} \sim B_0 \ell /b_{bd}$
(Equation [\ref{eq:zl}]), since magnetic islands in the plane $z=z_\ell$ 
have an approximately double transverse scale ($\sim 2\ell$).
\label{fig:fig_ic}}
\end{center}
\end{figure*}

In K41 phenomenology the eddy turnover time~(\ref{eq:tk41}) is the typical timescale for an eddy of 
size $\sim \ell$ to undergo a significant distortion due to the relative motion of its components,
thus transferring (in the 2D case) its energy at larger scales,
as schematically shown in Figure~\ref{fig:fig_ic} (left panel).
The dimensional analysis of Equation~(\ref{eq:eu})
shows that $t_\ell$ is also the timescale over which
an initial condition $\mathbf{u}_0$ with energy at scales $\sim \ell$ and  
$\langle \mathbf{u}_0^2 \rangle^{1/2}=u_\ell$
undergoes a significant distortion with
$\langle (\delta \mathbf{u})^2 \rangle^{1/2} \sim u_\ell$.
Furthermore scales are defined as logarithmic bands of
wavenumbers, e.g., $k_n \in (2^{n}, 2^{n+1}]$, with $n \in \mathbb{N}$
and scale $\ell_n \sim \ell_{\perp}/k_n = 2^{-n}\, \ell_{\perp}$
(the index $n$ will be dropped hereafter). 
In fact a single wavenumber cannot represent a scale \citep{2010PhRvL.104h1101A} 
since, from the uncertainty principle, the associated 
Fourier mode is delocalized in space and does not give rise
to a localized physical structure such as an \emph{eddy},
the building block of K41 phenomenology.
Consequently \emph{in the time interval $t_\ell$ [Equation~(\ref{eq:tk41})]
the energy of the system is transferred approximately from the scale
$\ell$ to the larger scale of double size~$2\ell$}.

Due to the structure of the absolute statistical equilibria of 
the ideal truncated system \citep{1967PhFl...10.1417K, 1980RPPh...43..547K}
the development of an inverse cascade is expected in both the 
2D dissipative (Navier-Stokes) and ideal (Euler) cases.
Generally the time evolution of dissipative and ideal
systems displays overall similar dynamics until energy reaches
the small scales, when respectively thermalization and dissipation 
sets in. This is observed even when a direct energy cascade occurs,
such as in 2D and 3D MHD, although spectral indices are observed to 
deviate from K41 in the ideal cases 
\citep[e.g.,][]{2009PhPl...16h0703W, 2013PhRvE..87a3110B}.
Since the dynamics of the 2D Euler system may depart from K41 phenomenology,
that has been developed and studied for the dissipative case,
in the following the eddy turnover time $t_\ell$
is then used as an estimate for the dynamical timescale of the 2D Euler equation solutions.
This is also justified by the dimensional analysis of an initial condition consisting 
of vortices of scale $\ell$ and velocity $u_\ell$, that indicates $t_\ell \sim \ell/u_\ell$
as the order of magnitude of the dynamically relevant timescale.

\subsection{Structure of Reduced MHD Equilibria and Dynamics} \label{sec:sred}

This phenomenology can be applied to the reduced MHD equilibria 
(Equation~[\ref{eq:eqb}]) using the mapping (\ref{eq:etr}),
as schematically shown in Figure~\ref{fig:fig_ic}.
\emph{Eddies}, that in the hydrodynamic case are velocity vortices,
correspond now to \emph{magnetic islands}.
Then, given a magnetic field $\mathbf{b}_{bd} (x,y)$  at the boundary $z=0$
with energy at scales $\sim \ell$ (i.e., a field structured in magnetic
islands of scale $\sim \ell$), the \emph{unique and regular} equilibrium solution
$\mathbf{b}_{eq}(x,y,z)$ with $\mathbf{b}_{eq}(x,y,z=0)=\mathbf{b}_{bd}(x,y)$
is characterized by an increasingly stronger \emph{inverse cascade} in the 
x-y planes for higher values of $z$ (see Figure~\ref{fig:fig_ic}, right panel), 
with the orthogonal magnetic field length-scale $\ell$ getting progressively larger 
up to doubling its value in the plane
\begin{equation} \label{eq:zl}
z_{\ell} \sim \frac{B_0}{b_{bd}}\, \ell,
\end{equation}
the analog of the eddy turnover time, derived from Equation~(\ref{eq:tk41})
using the mapping (\ref{eq:etr}).
If the magnetic field is characterized by a scale of order $\ell$ at $z=0$
it will have its energy at scale $\sim 2 \ell$ at $z = z_{\ell}$,
corresponding respectively to 
magnetic islands of scales $\ell$ and $2\ell$  
in the x-y planes $z=0$ and $z=z_{\ell}$.

\emph{Therefore the equilibrium solution $\mathbf{b}_{eq}(x,y,z)$
is generally asymmetric in the axial direction $z$, but it can 
be almost invariant or have strong variations, 
depending on the relative value of the length-scale $z_\ell$ 
compared to the loop length $L_z$}.
As long as the variation scale is larger than the loop length
($z_\ell > L_z$) the field has a weak variation along $z$, 
but for increasingly smaller values of $z_\ell$ ($z_\ell < L_z$)
the field becomes progressively more asymmetric along z 
due to the inverse cascade. Fixed the value of the guide field
$B_0$ and the scale of the magnetic field $\ell$, the 
axial variation scale $z_\ell$ is \emph{inversely proportional}
to the magnetic field intensity at the boundary $b_{bd}$:
quasi-invariant equilibria are then obtained for weak magnetic fields $b_{bd}$,
while increasingly stronger fields result in more asymmetric equilibria.

A strong asymmetry of the magnetic field $\mathbf{b}$ along $z$
is generally \emph{not compatible} with the dynamical
solutions of the reduced MHD equations~(\ref{eq:eq1})-(\ref{eq:eq2}), as
confirmed by nonlinear simulations 
\citep[for a more detailed discussion of the topics summarized in the
present and next paragraph see][]{2008ApJ...677.1348R}.
The derivatives along $z$ appear
only in linear terms in Eqs.~(\ref{eq:eq1})-(\ref{eq:eq2}). Introducing the
Els\"asser variables $\mathbf{z}^\pm = \mathbf{u} \pm \mathbf{b}$, 
and neglecting nonlinear and diffusive terms, the remaining linear terms yield the
two wave equations:
\begin{equation} \label{eq:els}
\partial_t \mathbf{z}^\pm = \pm B_0\, \partial_z \mathbf{z}^\pm,
\qquad \textrm{with} \qquad
\nabla \cdot \mathbf{z}^\pm =0.
\end{equation}
Thus fluctuations propagate along $z$ at the Alfv\'en speed $B_0$,
and a strongly asymmetric field cannot be generated, particularly
for the problem considered here with line-tying boundary conditions. 
For instance a velocity $\mathbf{u}_{ph}$ at the boundary $z=0$ implies 
the ``reflection'' condition
$\mathbf{z}^- = - \mathbf{z}^+ +  2\mathbf{u}_{ph}$
($\mathbf{z}^-$ propagates inward, $\mathbf{z}^+$ outward),
i.e., generalized Alfv\'en waves are injected and propagate
in the computational box, and \emph{only the $m=0$ mode does
not propagate} in the axial direction.

Considering an initial condition with only the guide field $\mathbf{B} = B_0 \mathbf{\hat{e}}_z$,
vanishing orthogonal component $\mathbf{b}=0$, and
a constant velocity $\mathbf{u}_{ph}$ at the photospheric boundary
that shuffles the magnetic field line footpoints,
an orthogonal component of the magnetic field invariant along $z$ is 
generated and it grows linearly in time as
\begin{equation} \label{eq:bp}
\mathbf{b}(x,y,z,t) = \mathbf{u}_{ph}(x,y) \frac{t}{\tau_A}, 
\end{equation}
where $\tau_A = L_z/B_0$ is the Alfv\'en crossing time in the axial direction.
Strictly speaking higher modes are present depending on how the
velocity $\mathbf{u}_{ph}$ is turned on, but they represent a
small contribution compared to Equation~(\ref{eq:bp}) that is the
strongly dominant term.

The solution~(\ref{eq:bp}) is obtained from Equations~(\ref{eq:els})
\citep[see][]{2008ApJ...677.1348R}, thus
\emph{neglecting nonlinear terms} in the reduced MHD equations. This is justified
as long as $b$ is small. In fact \cite{2013ApJ...773L...2R}
have shown that for initial configurations with only the
$m = 0$ mode in the axial direction and a non-vanishing
2D orthogonal Lorentz force component (with corresponding term
$\mathbf{b} \cdot \nabla j \ne 0$ in Equation~[\ref{eq:eqb}]), 
\emph{nonlinearity is strongly suppressed}, i.e., the \emph{nonlinear terms can be
neglected, for magnetic fields with intensities below the
magnetic threshold} $b \sim \ell B_0/L_z$. The magnetic field decays
only for larger magnetic fields $b > \ell B_0/L_z$, while
for smaller values the system does not form significant
current sheets and energy is not dissipated.

The intensity threshold $b \sim \ell B_0/L_z$ corresponds 
to a \emph{critical} equilibrium variation \emph{length-scale}
of the order of the loop length $z_\ell \sim L_z$ (Equation~[\ref{eq:zl}]).
Since fields with only $m = 0$ are invariant
along $z$ the correspondent equilibria length-scale $z_\ell$ can be computed
with $b_{bd} = b$. For $b < \ell B_0/L_z$ the corresponding
equilibrium, i.e., the equilibrium solution computed with $b_{bd} = b$
at the boundary, has $z_\ell > L_z$ and it is quasi-invariant along $z$.
Consequently magnetic fields with $m = 0$ and intensity below
the intensity threshold $b \sim \ell B_0/L_z$ are very close to
their corresponding equilibrium solution. \emph{It is such close
proximity to an equilibrium that suppresses nonlinearity},
that at equilibrium is indeed entirely depleted.

The emerging phenomenology for the dynamics of initially straight axial field lines shuffled by a
velocity field $\mathbf{u}_{ph}$ constant or slowly changing in time (so
that the induced magnetic field is quasi-invariant along
$z$) is therefore the following: \emph{since at first the induced magnetic field is small, the
associated variation scale is large $z_\ell \gg L_z$, 
nonlinearities are suppressed  and
the magnetic field grows as in Equation~(\ref{eq:bp}) until the
variation length-scale becomes smaller than the loop
length $z_\ell \lesssim L_z$, when nonlinearity can develop leading to the
formation of current sheets}.

The decay of initial configurations with a parallel m=0 mode are
relevant for slow photospheric motions. 
But in general the system has \emph{three characteristic timescales}:
\begin{enumerate}
\item the surface \emph{convective timescale} $\tau_{sc} \sim (\ell_{sc}/2)/u_{ph}$, 
essentially the  typical lifetime of a granule, approximately given by the ratio
of half its length-scale $\ell_{sc}$ over the photospheric velocity $u_{ph}$
(typical values for the Sun are $\ell_{sc} \sim 10^3$\,km,  $u_{ph}\sim 1$km/s,
and $\tau_{sc} \sim 5-8$\,m),
\item the \emph{Alfv\'en crossing time} $\tau_{A} = L_z/B_0$, where $L_z$ is the loop
axial length and $B_0$ the Alfv\'en length associated to the guide field, and
\item the \emph{nonlinear timescale} $\tau_{nl}$, that is investigated 
theoretically and numerically.
\end{enumerate}
For typical X-ray bright loops $L_z \sim 40\times 10^3\,$km and 
$B_0 \sim 2\times 10^3\,$km/s,
therefore the Alfv\'en crossing time $\tau_A = L_z/B_0 \sim 20$\,s
is much smaller than the photospheric timescale, with $\tau_A/\tau_{sc} \sim 0.04$.
For these loops photospheric motions are then characterized by a
\emph{low frequency}, i.e.,
\begin{equation}
\tau_{sc} \gg \tau_A,
\end{equation}
and a constant (\emph{zero frequency}, $\tau_{sc} = \infty$) 
photospheric velocity can be a good approximation, since
such slow variation of photospheric motions ($t_{sc}/\tau_A \sim 25$) introduces only 
wavelenghts much longer than the loop length itself along $z$, and
the resulting magnetic field (\ref{eq:bp}) can be considered invariant
along $z$. 

Nevertheless this condition can break down
for longer solar coronal loops and for loops on other
active stars with outer convective envelopes and magnetized coronae,
that exhibit broad variations in magnetic field intensity
and topologies \citep{2009ARA&A..47..333D,2012LRSP....9....1R},
and photospheric motion properties 
\citep{2002A&A...395...99L, 2013A&A...558A..48B},
for which $\tau_{sc} \sim \tau_{A}$, or $\tau_{sc} < \tau_A$.
In this case the resulting magnetic field will have a more
complex expression than Equation~(\ref{eq:bp}), and 
higher modes along $z$ will be present and contribute
increasingly  more, the faster the convective timescale $\tau_{sc}$ 
compared to the Alfv\'en crossing time $\tau_A$.

Therefore the structure of the magnetic 
field induced in coronal loops by photospheric granulation
will be dominated by the $m=0$ mode along $z$ (i.e., 
the field is quasi-invariant along $z$) for the typical X-ray bright 
solar loops for which $\tau_{sc} >> \tau_A$,
while for longer loops (including loops on other active stars)
higher modes ($m \ge 1$) will be increasingly more important
the smaller the timescale ratio $\tau_{sc}/ \tau_A < 1$.

\begin{table}
\begin{center}

\caption{Simulations Summary\label{tbl}}

   \begin{tabular*}{\columnwidth}{l @{\extracolsep{\fill}} lclr}
      \hline \hline\noalign{\vspace{.5em}}
      Run & z-modes & $B_0$   &   numerical grid                   &      $Re_4$  \\[.6em]
      \hline\noalign{\vspace{.5em}}
      A      &  m=0     & 10$^3$  &   2D:  512$^2$                 &   $10^{19}$  \\[.3em]
      B      &  m=0     & 10$^3$  &   3D:  512$^2$ $\times$   252  &   $10^{19}$  \\[.3em]
      C      &  m=1     & 10$^3$  &   3D:  512$^2$ $\times$   252  &   $10^{19}$  \\[.3em]
      D      &  m=0--4  & 10$^3$  &   3D:  512$^2$ $\times$   252  &   $10^{19}$  \\[.6em]
      \hline\noalign{\vspace{.6em}}
   \end{tabular*}

\tablecomments{The second column indicates the parallel Fourier modes
used in the initial magnetic field (Equation~[\ref{eq:pot4}]).
$B_0$ is the axial Alfv\'en velocity (same for all runs). The numerical
grid  $n_x \times n_y \times n_z$ is indicated in the fourth column,
it is three-dimensional for all runs except for run~A that is 
two-dimensional (with grid $n_x \times n_y$).
The last column indicates the value of the hyperdiffusion 
coefficient $Re_4=-1/\nu_4=-1/\eta_4$ used in Equations~(\ref{eq:eq1})--(\ref{eq:eq2}).
As described in the text,  each run~B--D is a collective label for  a set of 
simulations carried out with the parameters indicated in the table, 
but with initial conditions (Equation~[\ref{eq:pot4}]) differing for the values of the ratio 
$b_0/B_0$, the rms of the orthogonal magnetic field over the guide field
intensity.}
\end{center}
\end{table}

In both cases \emph{the variation scale} 
$z_{\ell} \sim \ell B_0 / b_{bd}$ (Equation~[\ref{eq:zl}])
\emph{measures the asymmetry of the equilibrium solution}. 
In both cases, even in presence of higher
modes $m \geq 1$, the dynamical solutions of the reduced
MHD equations will not be asymmetric along $z$, in contrast to
the equilibria with $z_\ell < L_z$.

In the following sections the dynamics of configurations
with different initial conditions,
with a single mode along $z$ ($m=0$ and $m=1$),
and with all modes $0 \le m \le 4$ excited, are investigated through
numerical simulations (see Table~\ref{tbl}).

The 2D photospheric velocity $\mathbf{u}_{ph}$ models photospheric motions.
These have large scales (with $\ell \sim 10^3$\,km) and are disordered, 
since they originate from turbulent convection. Thus 
generally the magnetic field $\mathbf{b}$
will have a non-vanishing 2D Lorentz force component, i.e., $\mathbf{b} \cdot \nabla j \ne 0$, since 
the opposite would imply that $j$ should be constant on the field lines of $\mathbf{b}$,
a condition too symmetric to apply to turbulent convection
(this could be realized, e.g., with an exactly 1D  magnetic shear along one direction, 
or with perfectly circular field lines).
Therefore \emph{in all  simulations presented here the 2D Lorentz force component of 
the magnetic field does not vanish, i.e., $\mathbf{b} \cdot \nabla j \ne 0$}.

Notice that when $\mathbf{b} \cdot \nabla j = 0$ 
we obtain $\partial_z j =0$ from the equilibrium Equation~(\ref{eq:eqb}), 
hence in this case the variation length-scale
is formally infinite $z_\ell = \infty$ and it is always larger than the loop length.
The inverse cascade picture of the 2D Euler equation described in section~\ref{sec:eq}
is indeed valid only for initial conditions that are out of equilibrium. 
If $\mathbf{u} \cdot \nabla \omega = 0$
in Equation~(\ref{eq:eu}) then there is no time evolution and formally $\tau_\ell = \infty$
(Equation~[\ref{eq:tk41}]).
Reduced MHD equilibria with $\partial_z j =0$ (or equivalently $\mathbf{b} \cdot \nabla j = 0$)
are those apt to describe classic linear instabilities (such as kink instabilities),
and their dynamical accessibility will be discussed in the following sections.

\section{Results} \label{sec:res}

The results of the numerical simulations are presented in
this section. All simulations, with the exception of Run~A, 
implement line-tying boundary conditions with field line
footpoints rooted in a motionless plasma in the 
photospheric-mimicking planes $z=0$ and $z=L_z$.
Run~A implements periodic boundary conditions in the axial
direction $z$, and since its initial condition is invariant
along $z$ the dynamics will not introduce any variation along
along this direction ($\partial_z = 0$) and the simulation is 
restricted to a 2D plane.
In the orthogonal directions (x--y) all runs use periodic
boundary conditions.

All simulations consider the decay (or equivalently \emph{relaxation}) 
of an initial magnetic configuration made of large-scale Fourier modes
(as described in Section~\ref{sec:initial}) with
non-vanishing 2D orthogonal Lorentz force component
($\mathbf{b} \cdot \nabla j \ne 0$).
The guide field intensity is $B_0=1000$ for all simulations,
corresponding to an Alfv\'en velocity of $1000$\,km/s, a typical
value for solar coronal loops.

\emph{Dissipative} simulations with initial conditions made of single 
parallel modes are described in Sections~\ref{sec:run0} (m=0) 
and \ref{sec:run1} (m=1), while in Section~\ref{sec:run04}
the initial condition has all modes with $0 \le m \le 4$.
A summary of the simulations is shown in Table~\ref{tbl}.

The initial magnetic fields used in these decaying simulations
can be considered as ``snapshots'' of the coronal magnetic field
at different stages of its evolution, particularly for the problem
of an initially straight field shuffled at its footpoints
by photospheric motions [see Equation~(\ref{eq:bp})]. 
The different parallel modes are related to the photospheric 
motions frequency, as discussed in the following sections.
Additionally the relaxation of magnetic fields is a topic of 
general and broad interest to solar physics 
\citep[e.g.,][]{2014masu.book.....P}.

\subsection{Runs A--B: single mode m=0} \label{sec:run0}

This section presents an extended analysis of the 
dissipative simulations carried out by \cite{2013ApJ...773L...2R}
that investigate the decay (or equivalently relaxation) of initial 
magnetic configurations with an out-of-equilibrium orthogonal magnetic field 
and parallel mode m=0 (runs A-B). 

The initial magnetic field is invariant along $z$
since only the mode m=0 is present.
It has the same structure for all runs A-B,
with same topology for the magnetic field lines of 
$\mathbf{b}$, as shown in Figure~\ref{fig:fig5}
at time $t=0$, but different values of the orthogonal
field intensity rms $b_0$ (the proportionality factor in 
equation~[\ref{eq:pot4}]) are used. Respect to the guide field $B_0$
(same for all runs) the intensity is $b_0 = 0.1\, B_0$ for run~A,
while run~B comprises a set of simulations  performed
with same parameters except $b_0$ that spans the range $0.01 \le b_0/B_0 \le 0.1$.

Runs A and B differ for the boundary conditions along z,
\emph{periodic} for run~A and \emph{line-tied} for runs~B.
In the periodic case the invariance along $z$ is preserved
during the time evolution, therefore for run~A 
Equations~(\ref{eq:eq1})-(\ref{eq:eq2}) are integrated in a
2D x--y plane with $\partial_z = 0$.

Periodicity along $z$ is appropriate as a local approximation
for the central part of very long loops, where line-tying
at the boundary has a weak influence.
In particular line-tied forced simulations, with a velocity field at the 
photospheric-mimicking boundaries, have shown that dynamics
resembles those of periodic simulations for low values
of the ratio $f_c = \ell_{sc} B_0 / L_z u_{ph}$ 
\citep{2007ApJ...657L..47R, 2008ApJ...677.1348R}.
Fixed the scale of granular cells $\ell_{sc}$ and
the photospheric velocity rms $u_{ph}$, 
line-tying has a weaker impact on longer loops
(with larger $L_z$) or loops with weaker guide fields
(smaller $B_0$).

\begin{figure}
\begin{center}
\includegraphics[width=1.\columnwidth]{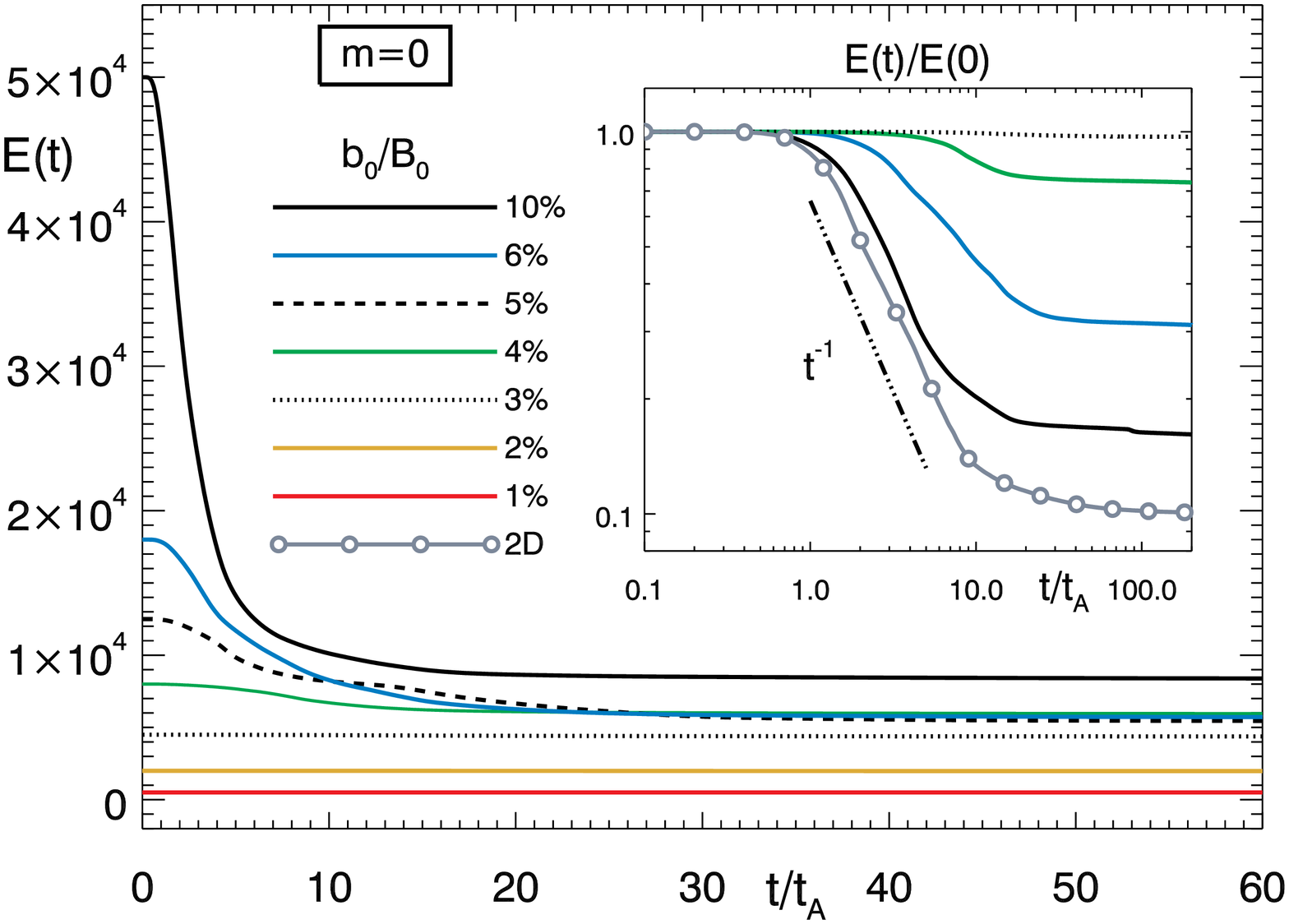}
\caption{\textsf{Runs A--B} (m=0): 
Total energy vs.\ time for line-tied
simulations with different values of $b_0/B_0$ (runs~B) and the 2D simulation 
(run~A, with $b_0/B_0 = 10\%$).
The inset shows in logarithmic scale total energy normalized with its inital
value.
\label{fig:fig1}}
\end{center}
\end{figure}

Since the 2D run~A is started with an out-of-equilibrium magnetic field
($\mathbf{b} \cdot \nabla j \ne 0$), it is therefore akin to 2D 
turbulence decay  simulations \citep{1995PhFl....7.2886H,
1997PhRvL..79.2807G, 2003matu.book.....B}
that use similar initial conditions for the magnetic field, but
additionally have an initial velocity field in equipartition. 
The dynamics are nevertheless similar and in the 2D case
total energy decays approximately as $E \propto t^{-1}$
(Figure~\ref{fig:fig1}, inset), even if the initial velocity vanishes
everywhere. Until time $t \sim 0.4\, \tau_A$ total energy
is conserved, but the non-vanishing Lorentz force transfers 
$\sim 15\%$ of magnetic energy $E_M$ into kinetic energy $E_K$,
henceforth leading to the formation of small-scales and current sheets, 
dissipating $\sim 90$\% of the initial magnetic energy, 
while kinetic energy remains much smaller
than magnetic energy throughout \citep{2013ApJ...773L...2R}. 

\begin{figure}
\begin{center}
\includegraphics[width=.85\columnwidth]{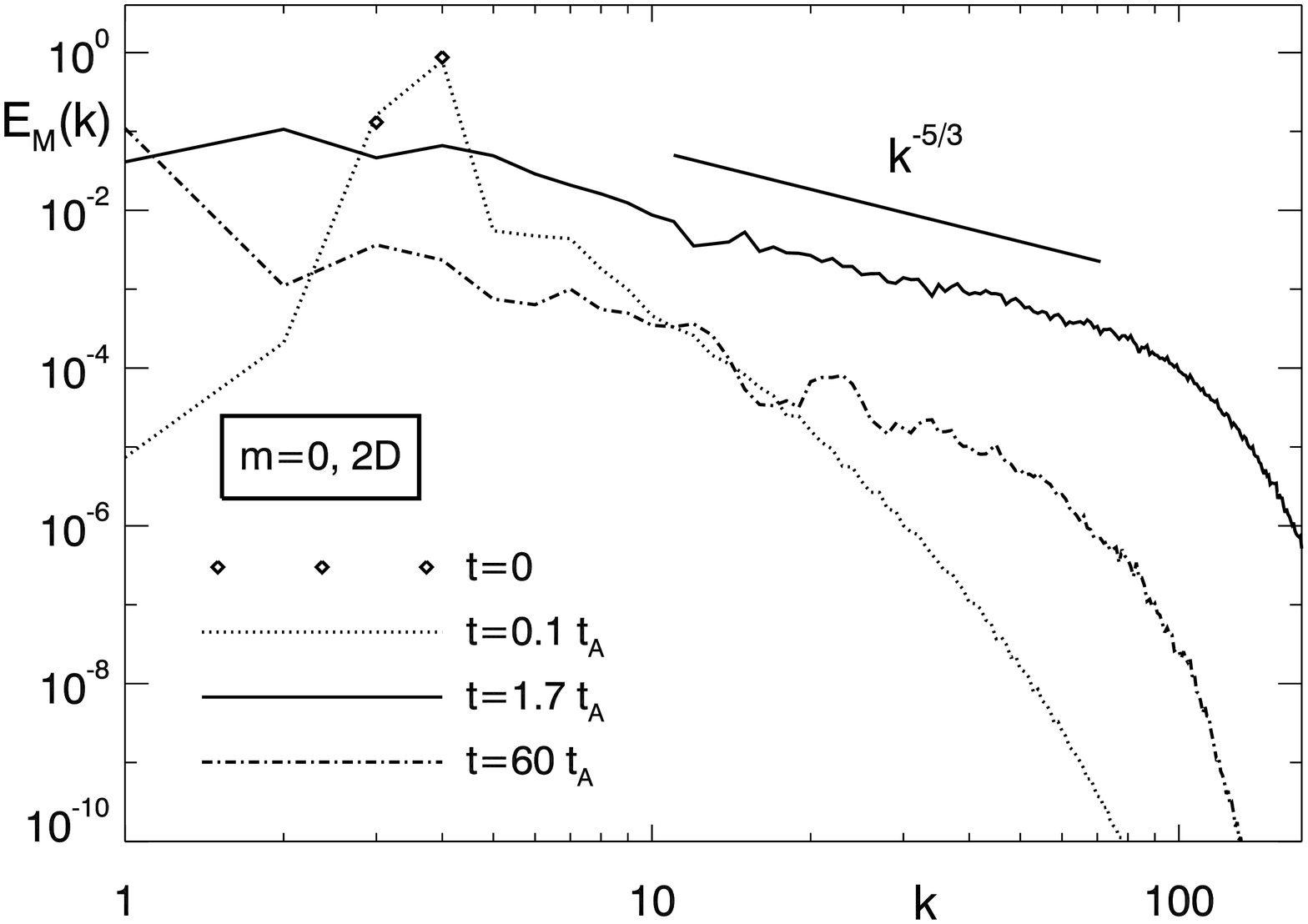} \\[1em]
\includegraphics[width=.85\columnwidth]{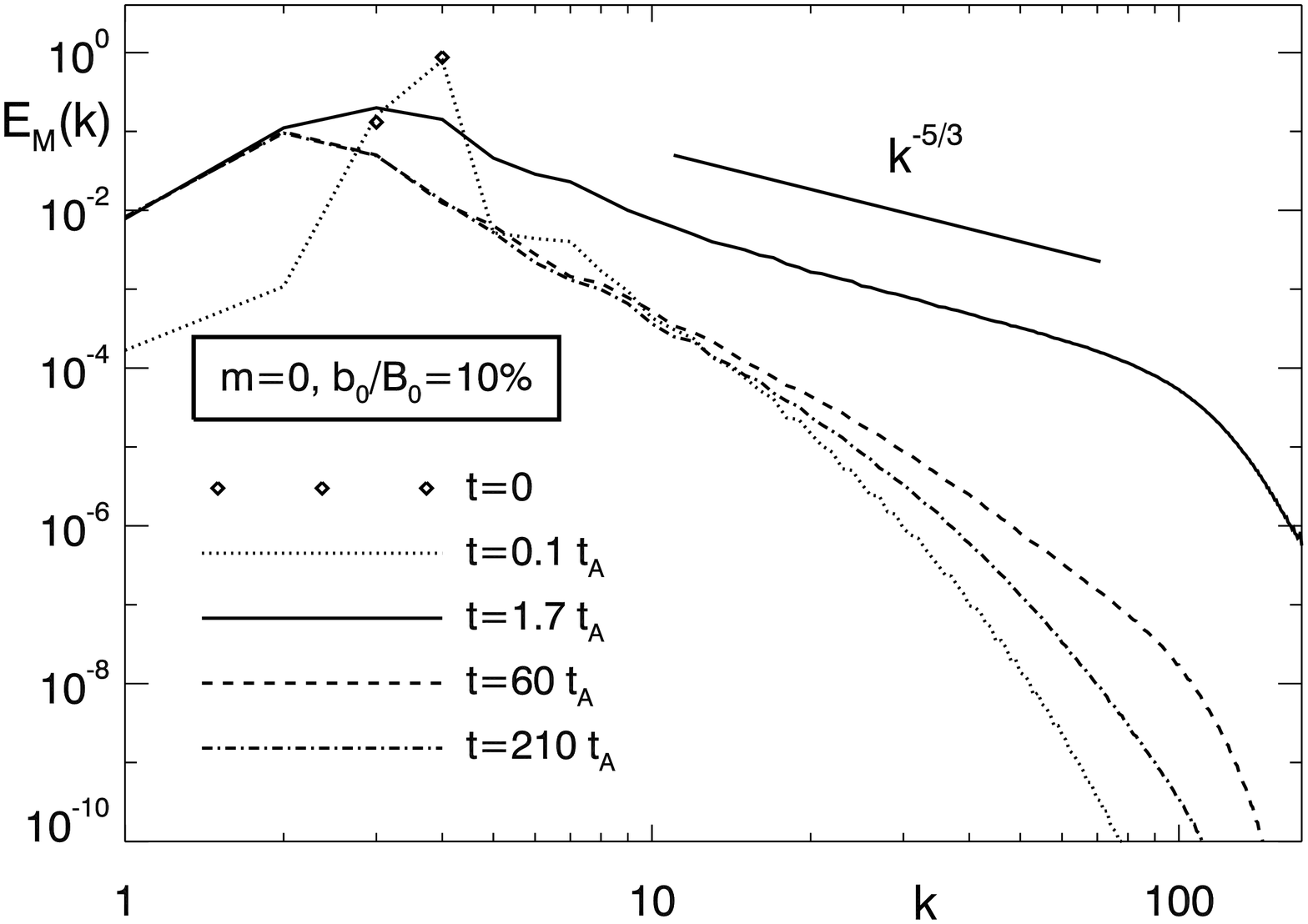} \\[1em]
\includegraphics[width=.85\columnwidth]{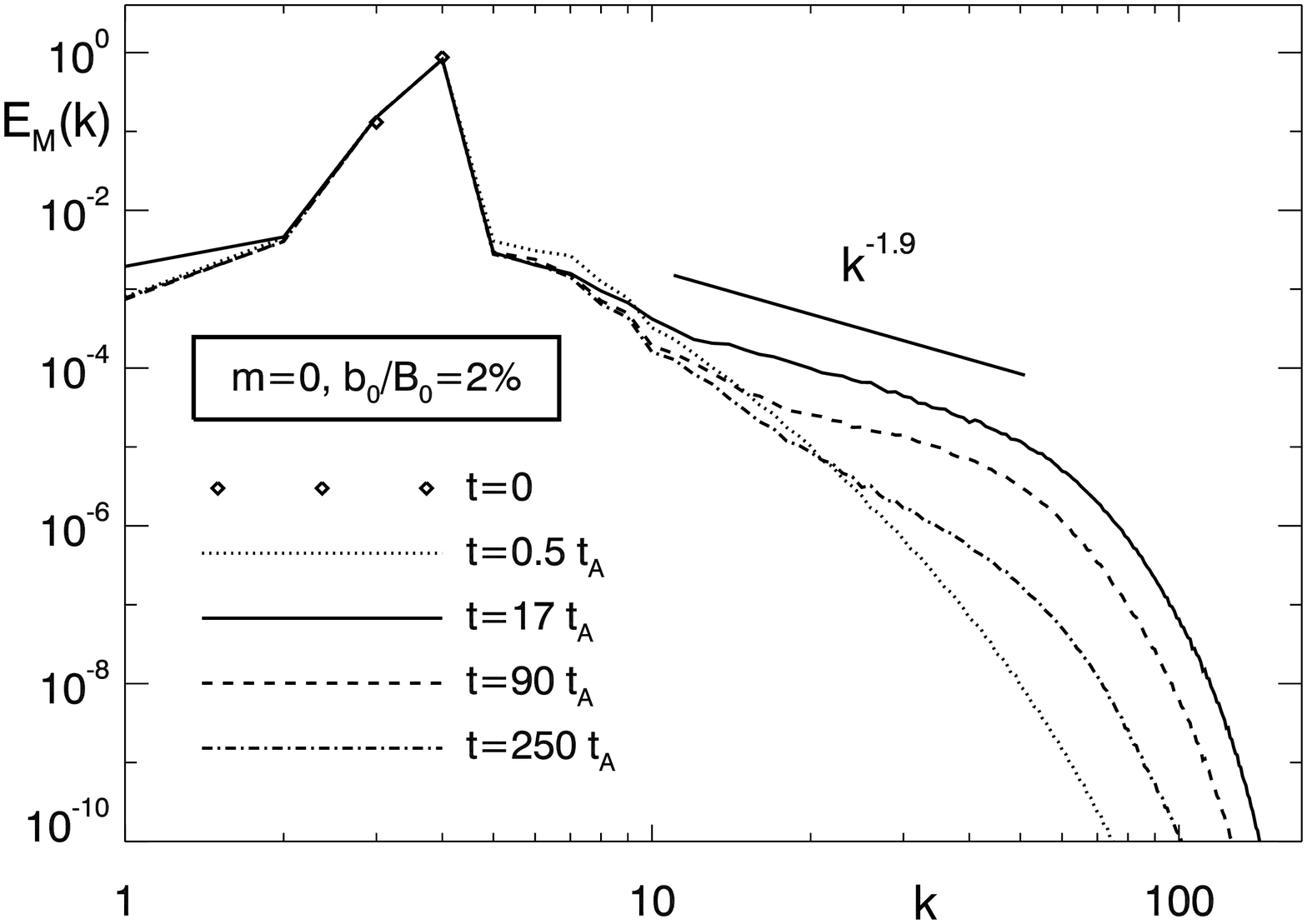}
\caption{\textsf{Runs A--B} (m=0):
Magnetic energy spectra (integrated along $z$) at selected times for the 2D run A 
with $b_0/B_0=10$\% (\emph{top panel}),
and for run B simulations with $b_0/B_0=10$\% (\emph{middle})
and $b_0/B_0=2$\% (\emph{bottom}).
Energy is normalized with its initial value at time t=0, 
when energy is present only at perpendicular modes $k=3$ and 4 (diamond symbol).
Spectra at the time of maximum dissipation for each simulation
are drawn in a continuous line.
\label{fig:fig3}
}
\end{center}
\end{figure}

\begin{figure*}
\centering
\includegraphics[width=.33\textwidth]{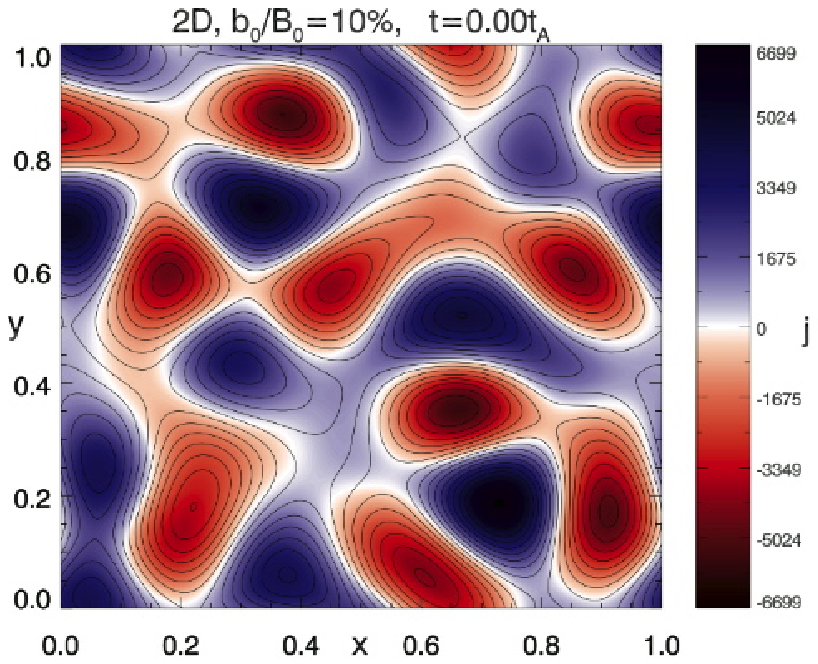}
\includegraphics[width=.33\textwidth]{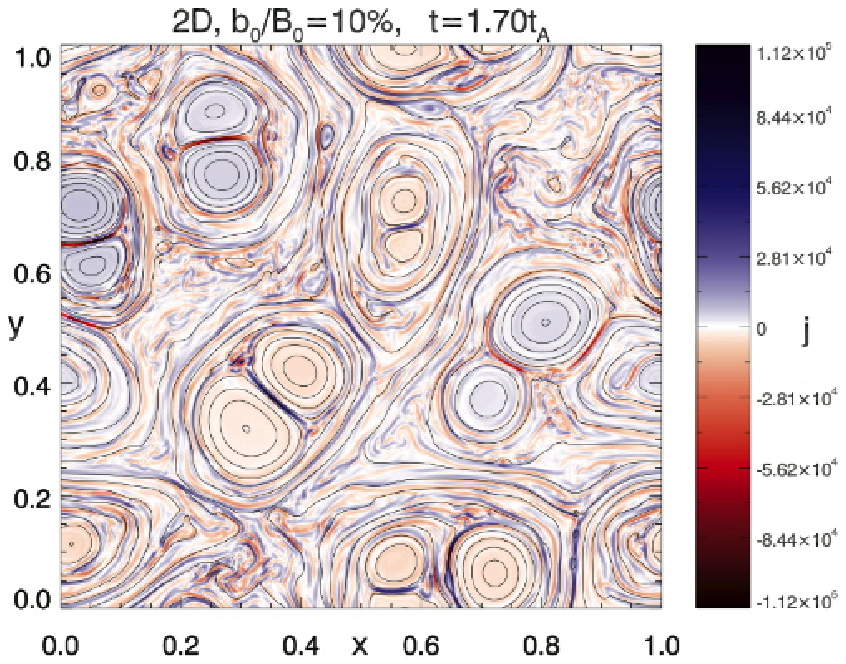}
\includegraphics[width=.33\textwidth]{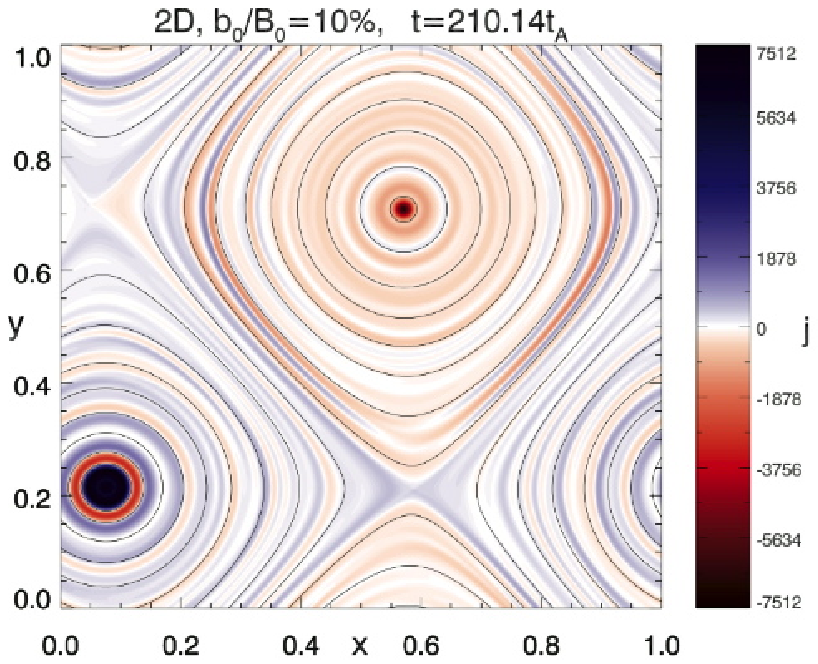}   \\[.3em]
\includegraphics[width=.33\textwidth]{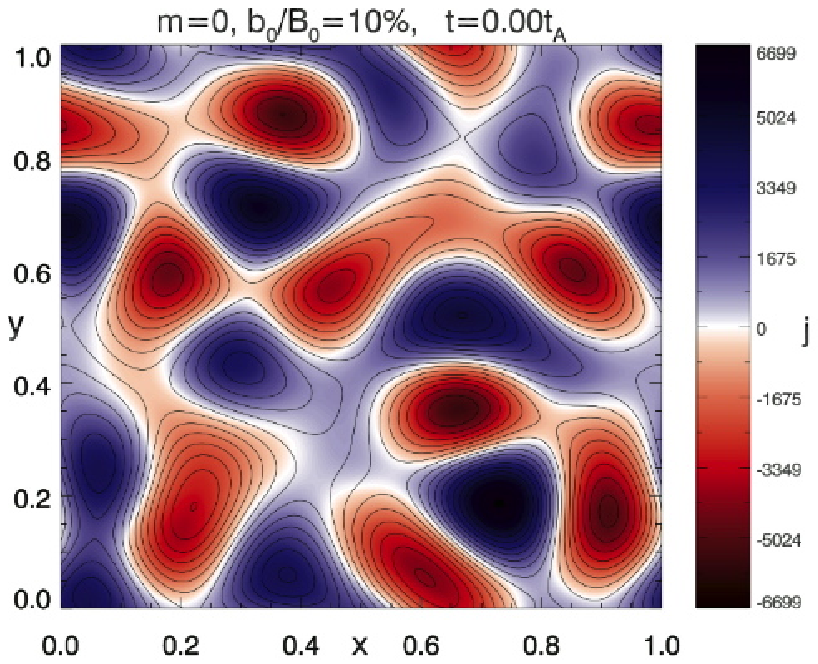}
\includegraphics[width=.33\textwidth]{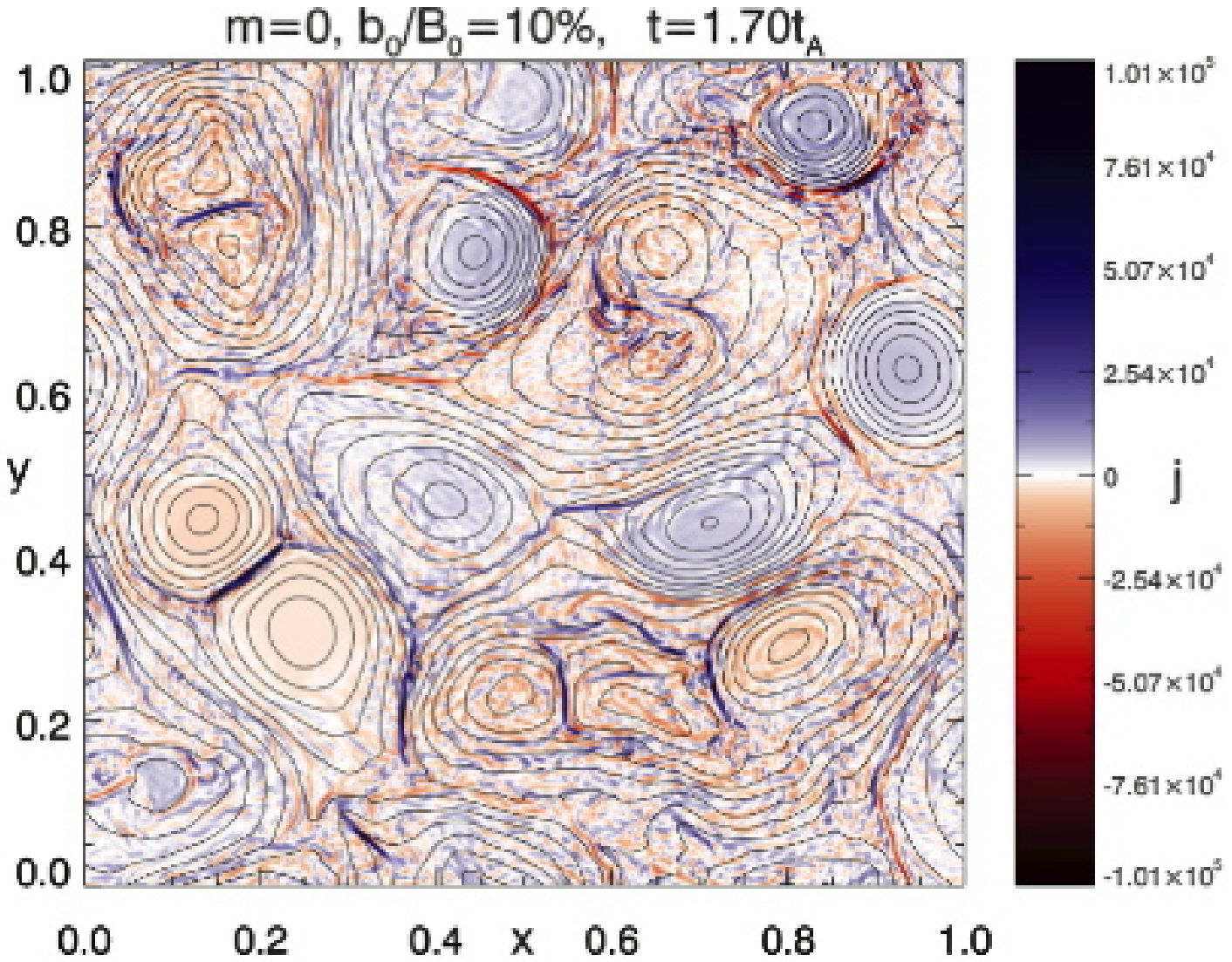}
\includegraphics[width=.33\textwidth]{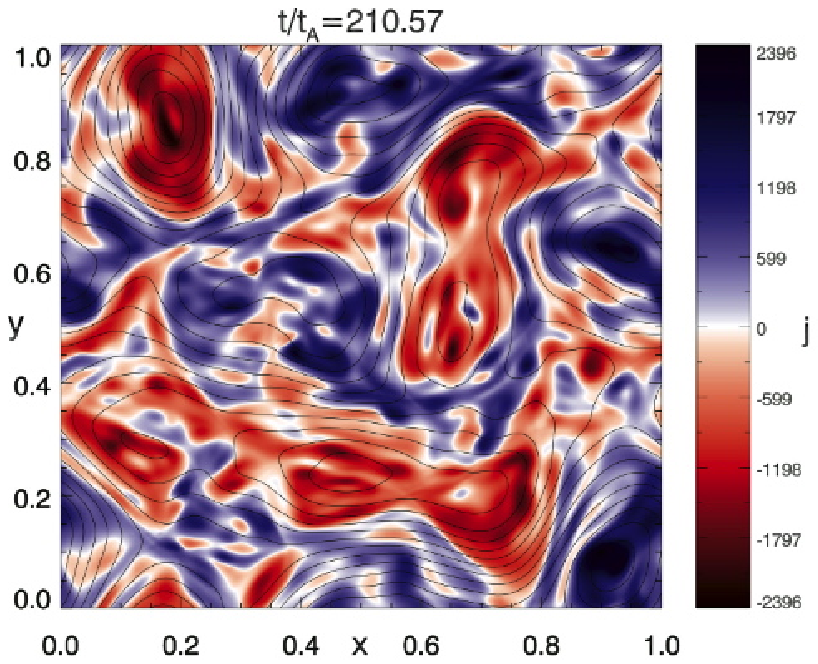}\\[.3em]
\includegraphics[width=.33\textwidth]{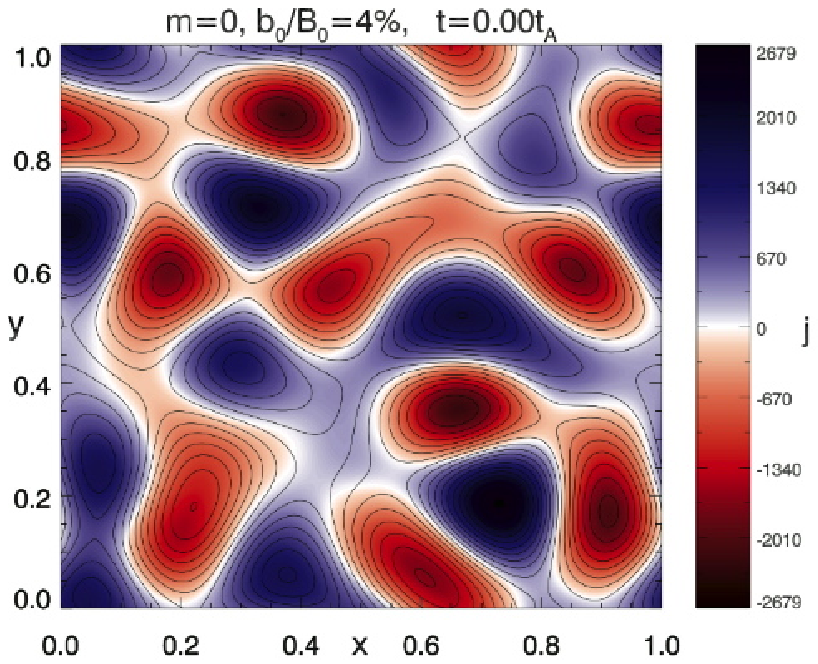}
\includegraphics[width=.33\textwidth]{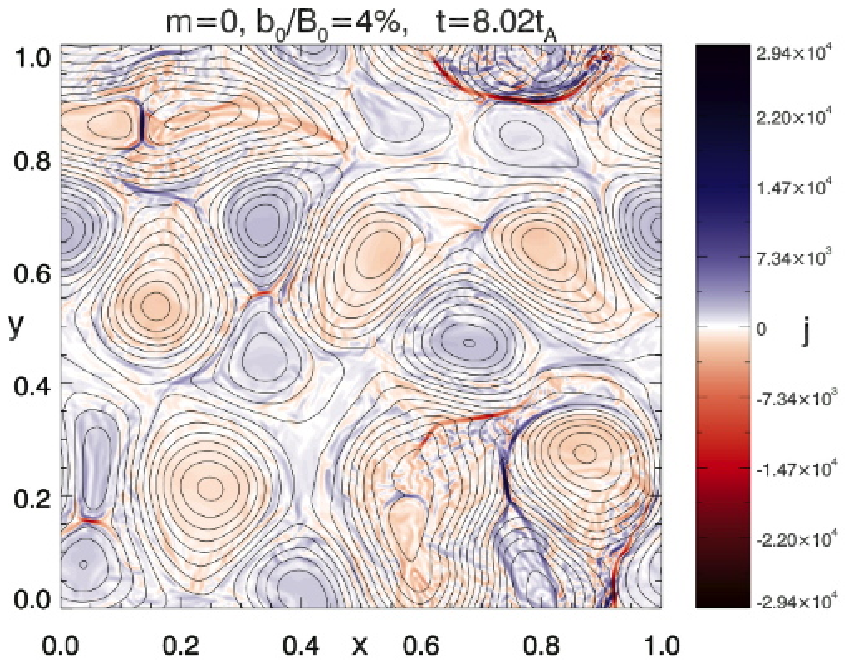}
\includegraphics[width=.33\textwidth]{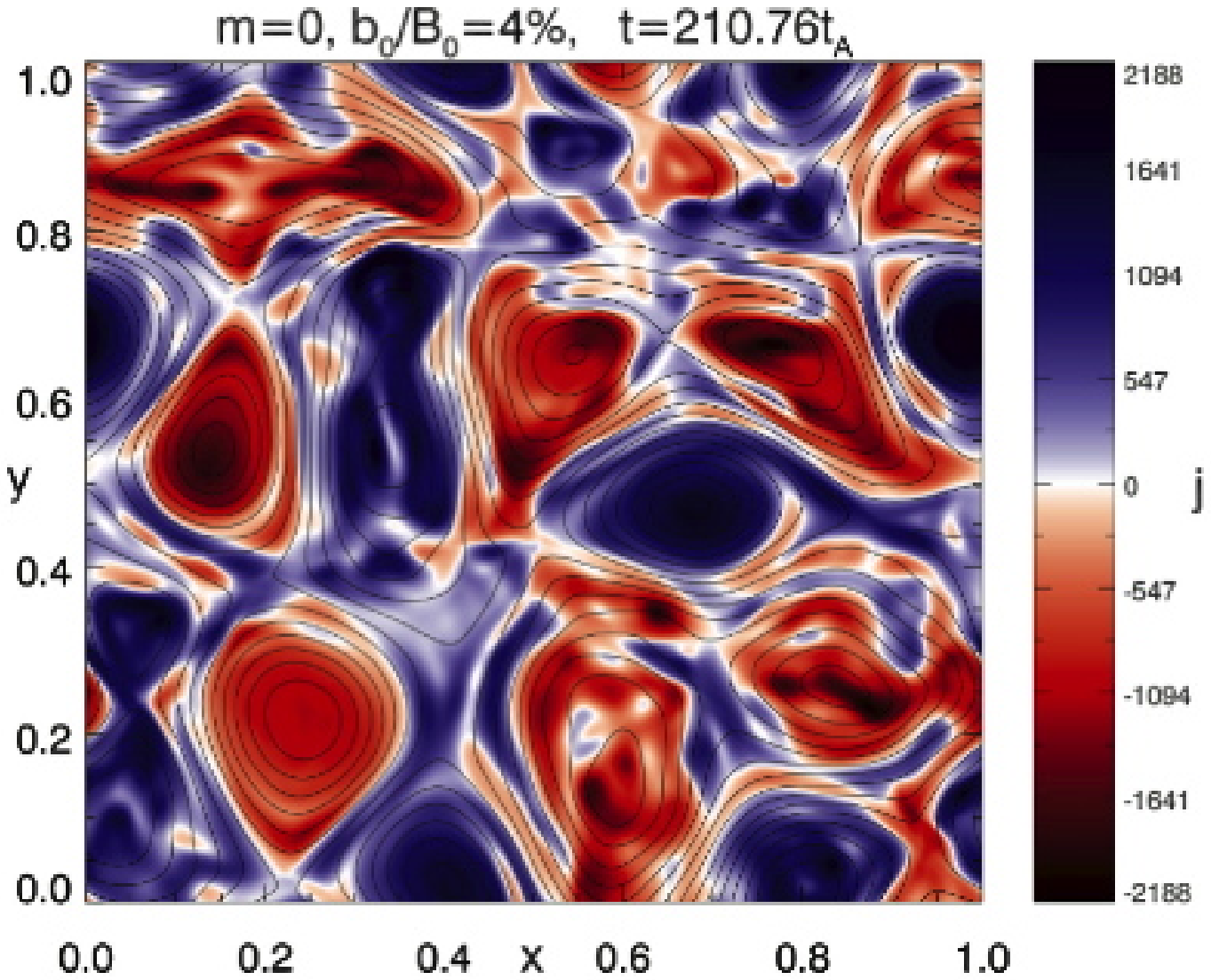} \\[.3em]
\includegraphics[width=.33\textwidth]{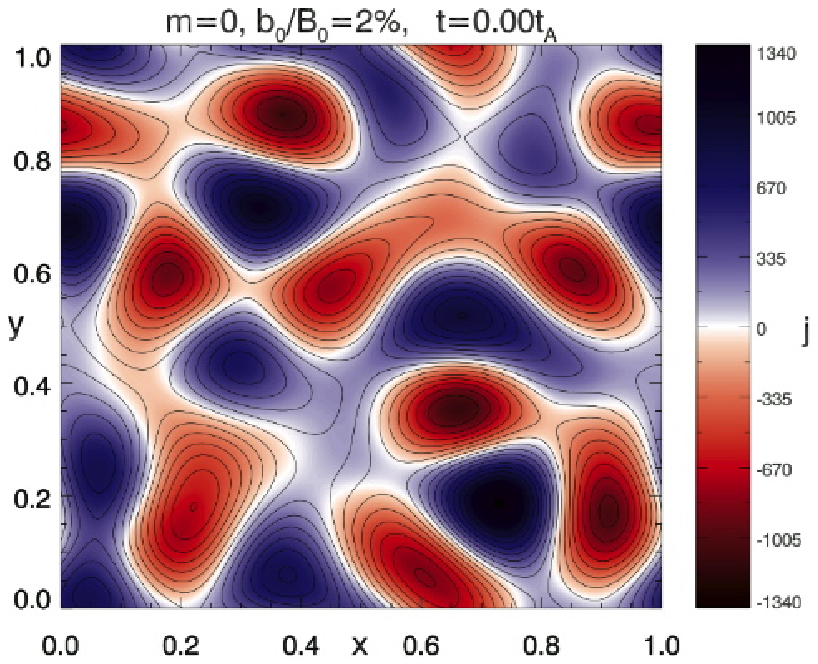}
\includegraphics[width=.33\textwidth]{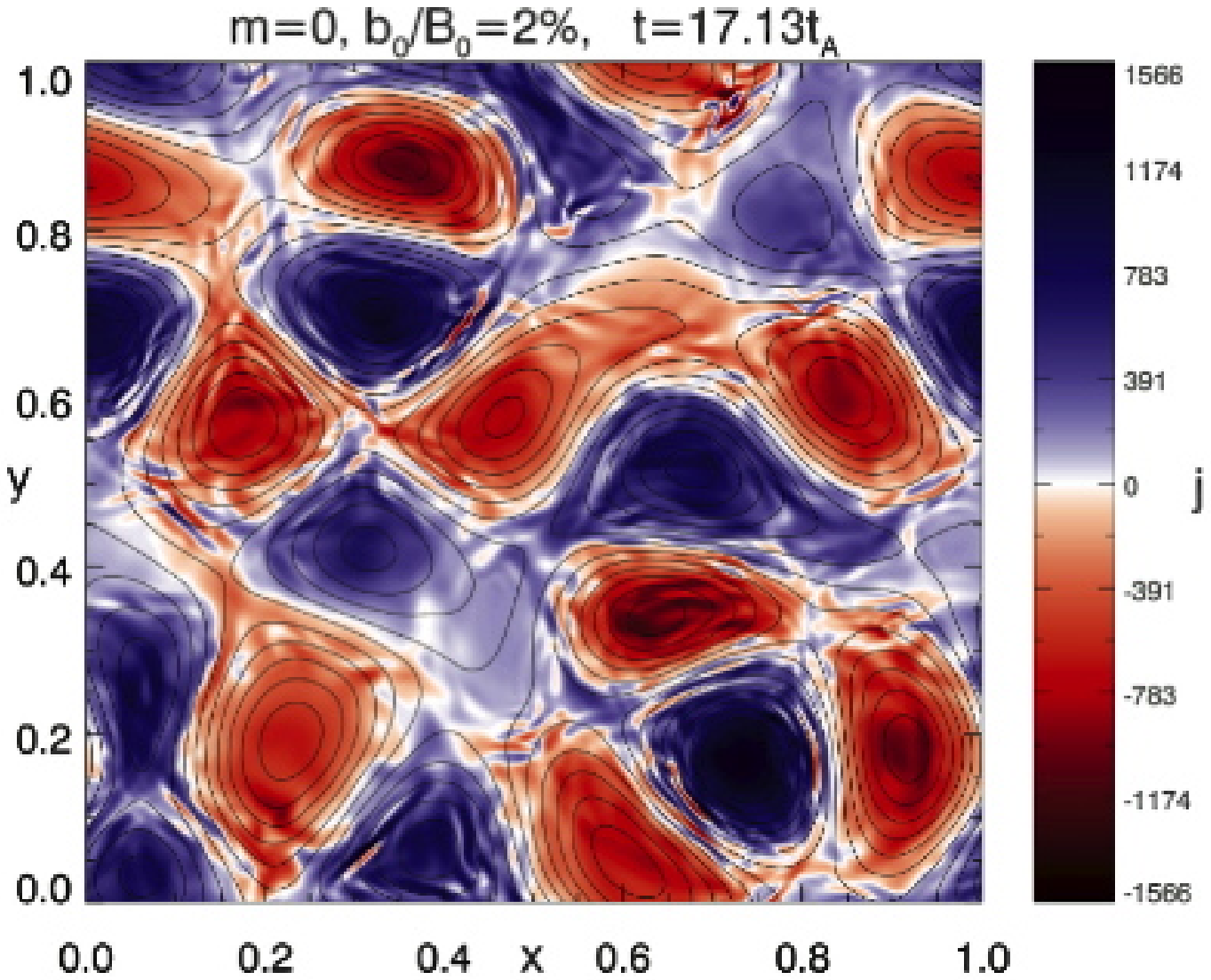}
\includegraphics[width=.33\textwidth]{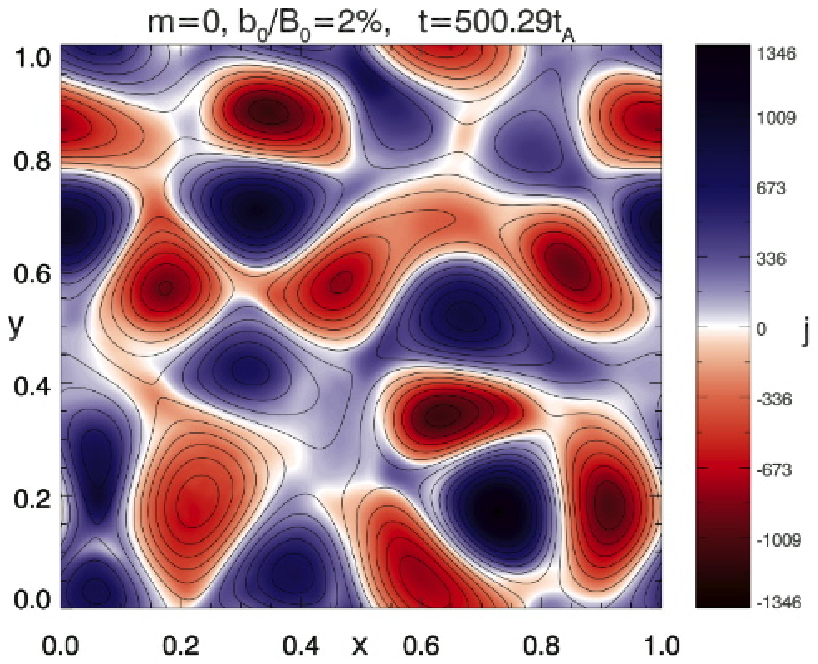}
\caption{\textsf{Runs~A--B} (m=0): Magnetic field lines of the orthogonal magnetic 
field component $\mathbf{b}$ and current density $j$ at selected times.
\emph{Top row} shows snapshots from the 2D simulation (run~A), while snapshots
from three 3D simulations (run~B) with different ratios $b_0/B_0$ are shown
in the \emph{remaining rows} (in this case the mid-plane $z=5$ is considered).
The \emph{first column}  shows the initial condition at $t=0$, same for all
except for the ratio $b_0/B_0$, the \emph{second column} shows the
fields at the time of maximum dissipation, while the \emph{third column}
shows the fields at a later time when the fields have relaxed and little if any
energy dissipation occurs.
\label{fig:fig5}}
\end{figure*}

In Fourier space magnetic energy has initially only
large-scale perpendicular modes k=3 and 4 
(Figure~\ref{fig:fig3}, top panel).
As nonlinearity develops energy is progressively transferred toward both 
small (\emph{direct}) and large scales (\emph{inverse cascade}).
In physical space the direct cascade gives rise to current sheets
(Figure~\ref{fig:fig5}, top row) that enable dissipation through magnetic
reconnection. As dissipation peaks at $t \sim 1.7\, \tau_A$, 
the spectrum extends fully toward high wave-numbers 
exhibiting an approximate $k^{-5/3}$ power-law.
At the same time a substantial fraction of total energy has already been
transferred to large-scale modes k=1 and 2 through the inverse cascade,
that in physical space corresponds to the \emph{coalescence} of magnetic
islands as magnetic reconnection occurs.
As dynamics proceeds the energy transferred at the small scales
is dissipated leading to the disappearance of current sheets and small-scales, 
so that the system finally relaxes to a state with energy mostly at large scales
(particularly in mode k=1), corresponding in physical space to 
large-scale magnetic islands and large-scale current layers 
(Figure~\ref{fig:fig5}, top row, $t\sim210\, \tau_A$).
The whole process is akin to Taylor relaxation \citep{1986RvMP...58..741T}
and \emph{self-organization} leading to the formation of large-scale structures 
\citep{1985AdPhy..34....1H}.

In the 2D case (run~A) solutions differing only for
the value of $b_0$ (the rms of the initial magnetic field in Equation~[\ref{eq:pot4}])
have a \emph{self-similar} structure. Indicating with $\psi_0(\mathbf{x},t)$ and
$\varphi_0(\mathbf{x},t)$ the solution of Equations~(\ref{eq:eq1})-(\ref{eq:eq2}) 
with initial magnetic field rms $b_0$, then the
solutions with $b'_0= \sigma b_0$, and same
random amplitudes in Equation~(\ref{eq:pot4}), 
are given by\footnote{Strictly speaking these self-similar 
solutions would require the Reynolds number to scale as $R' = \sigma R$, 
but in the high-Reynolds regime the solutions of decaying 
turbulence do not depend on the Reynolds number 
\citep{2003matu.book.....B, 1997PhRvL..79.2807G}.}:
\begin{equation} \label{eq:ss}
\psi' (\mathbf{x},t)      = \sigma\, \psi_{0} (\mathbf{x}, \sigma t), \qquad
\varphi' (\mathbf{x},t) = \sigma\, \varphi_{0} (\mathbf{x}, \sigma t),
\end{equation}
as can be verified by direct substitution.
Consequently all these solutions have a similar structure and their
time evolution differs only for the scaling factor $\sigma$.
In particular if current sheets form for a certain value of $b_0$,
they will always do for any value of $b_0$ at scaled times.
Analogously energy will exhibit a power-law decay with the \emph{same
exponent} because $E' (t) = \sigma^2 E_0 (\sigma t)$, implying that
if $E(t) \propto t^{-\alpha}$ then $E'(t) \propto t^{-\alpha}$.

When the same initial condition is used with \emph{line-tying} boundary 
conditions, the system is no longer invariant along $z$, as now the
velocity must vanish at the top and bottom plates $z=0$ and $z=L$, therefore
the velocity cannot develop uniformly along z as in the periodic case.

The time evolution of total energy for line-tied
simulations with different values of $b_0$ is shown in 
Figure~\ref{fig:fig1}. While the dynamics of the system
with $b_0/B_0 = 10\%$ is similar to the 2D 
case with energy dissipating $\sim 84\%$ of its initial value, 
the behavior is increasingly different for lower
values of $b_0$, with progressively less energy getting dissipated. 
For $b_0/B_0 \lesssim 3\%$ no significant
energy dissipation nor decay are observed.
Additionally, also for the decaying cases their dynamics are strongly
suppressed once energy crosses this threshold. As shown
in Fig.~\ref{fig:fig1} no energy decay is observed below
$E \sim 5 \times 10^3$, corresponding to a ratio 
$\langle b^2 \rangle^{1/2}/B_0 \sim 3\%$.
The inset in Fig.~\ref{fig:fig1} shows that
energy decays with different power-law indices
for lower values of $b_0/B_0$,
hence time self-similarity is lost and the impact of line-tying 
on the dynamics is more complex than a simple delay 
as in the 2D case (Equations~[\ref{eq:ss}]).

Magnetic energy spectra (integrated along $z$, Figure~\ref{fig:fig3}) show similar
dynamics for the 2D (\emph{top panel}) and the 3D case
with $b_0/B_0 =10\%$ (\emph{middle panel}). In particular 
the spectra are fully extended toward the small scales as 
dissipation occurs, corresponding to the formation and dissipation 
of current sheets in physical space (Figure~\ref{fig:fig5}, \emph{second row}). 
While similar behavior is observed for 
magnetic field intensities $b_0 > 3\%$, \emph{below this threshold}
the spectra do not extend to the high wave-numbers
where energy gets dissipated (Figure~\ref{fig:fig3}, \emph{bottom panel}, $b_0/B_0 = 2\%$),
i.e., \emph{current sheets do not thin below the critical diffusive 
thickness that allows magnetic reconnection and energy
dissipation to occur}. As shown in Figure~\ref{fig:fig5} at the
peak of dissipation (\emph{central column}) both current maxima
and the number of current sheets decrease for smaller $b_0/B_0$,
and for $b_0/B_0 = 2\%$ no current sheets are formed, but only
a few ripples are visible in the magnetic field (enhanced in the current) 
and are eventually dissipated on long timescales.


Furthermore, spectra show that the inverse energy cascade
decreases from the 2D to the 3D case with $b_0/B_0 = 10\%$,
in fact while  in the asymptotic state of the 2D run ($t=210\, \tau_A$) most of the
energy is in the $k=1$ mode and higher modes are much smaller,
in the 3D case the mode $k=2$ has the higher value.
For 3D simulations with smaller $b_0/B_0$ the inverse cascade
is progressively fainter and does not occur for $b_0/B_0 \lesssim 3\%$,
as shown in Figure~\ref{fig:fig3} (\emph{bottom panel}) for $b_0/B_0 = 2\%$,
where no significant energy is found in perpendicular modes smaller 
($k \le 2$) than those present at time $t=0$ ($k=3$, $4$).

In physical space  (Figure~\ref{fig:fig5}) the inverse cascade corresponds
to larger-scale magnetic islands in the asymptotic state.
Since a larger fraction of magnetic energy is dissipated for higher values
of  $b_0/B_0$ and in the 2D case, more magnetic flux is reconnected, 
thus leading to increased \emph{coalescence} and larger magnetic islands.
Additionally the field line topology in the relaxed state is substantially
unaltered respect to the initial condition for $b_0/B_0 \lesssim 3\%$ 
(cf.\ the first and last columns in Figure~\ref{fig:fig5}), 
with higher variations for higher values of $b_0/B_0$ and most of
all in the 2D case.

A key difference distinguishes the 2D and 3D asymptotic topologies.
Although all simulations are started with a non-vanishing 2D Lorentz 
force component ($\mathbf{b} \cdot \nabla j \ne 0$), the 2D simulation relaxes to an 
approximate equilibrium with $\mathbf{b} \cdot \nabla j = 0$,
but all 3D simulations relax to an orthogonal field with $\mathbf{b} \cdot \nabla j \ne 0$,
regardless of how much energy is dissipated during the decay 
(none for $b_0/B_0=2\%$, and an 84\% energy decay for the $b_0/B_0=10\%$ run~B).
Further analysis is presented in the following to understand the nature 
of these equilibria and how they are approached.

In the 2D case the equilibrium condition~(\ref{eq:eqb}) becomes 
simply $\mathbf{b} \cdot \nabla j =0$.
This requires the current density $j$  to be constant along the field lines
of $\mathbf{b}$ (or equivalently that the isosurfaces of $j$ are also isosurfaces
of the magnetic potential $\psi$). This condition is generally satisfied only
in highly symmetric configurations such as one-dimensional magnetic shears,
e.g., $\mathbf{b} = f(y)\, \mathbf{\hat{e}}_x$, or rotationally invariant
fields as $\mathbf{b} = f(r)\, \mathbf{\hat{e}}_\theta$ ($f$ is a generic function in cartesian
or cylindrical coordinates). 
In the 2D case the field lines are mostly circular in the asymptotic state (Figure~\ref{fig:fig5})
and the isosurfaces of the current density and of the magnetic potential (the field lines
of $\mathbf{b}$) overlap. In the 3D case the full equilibrium equation~(\ref{eq:eqb})
has to be considered, and the fact that for run~B with $b_0/B_0 < 3\%$ no significant
dynamics occurs with the initial orthogonal magnetic field essentially unaltered
even though its 2D Lorentz force does not vanish, i.e., $\mathbf{b} \cdot \nabla j \ne 0$, 
implies that $\partial_z j$ increases only slightly from its initial vanishing value. 

\begin{figure}
\begin{center}
\includegraphics[width=.79\columnwidth]{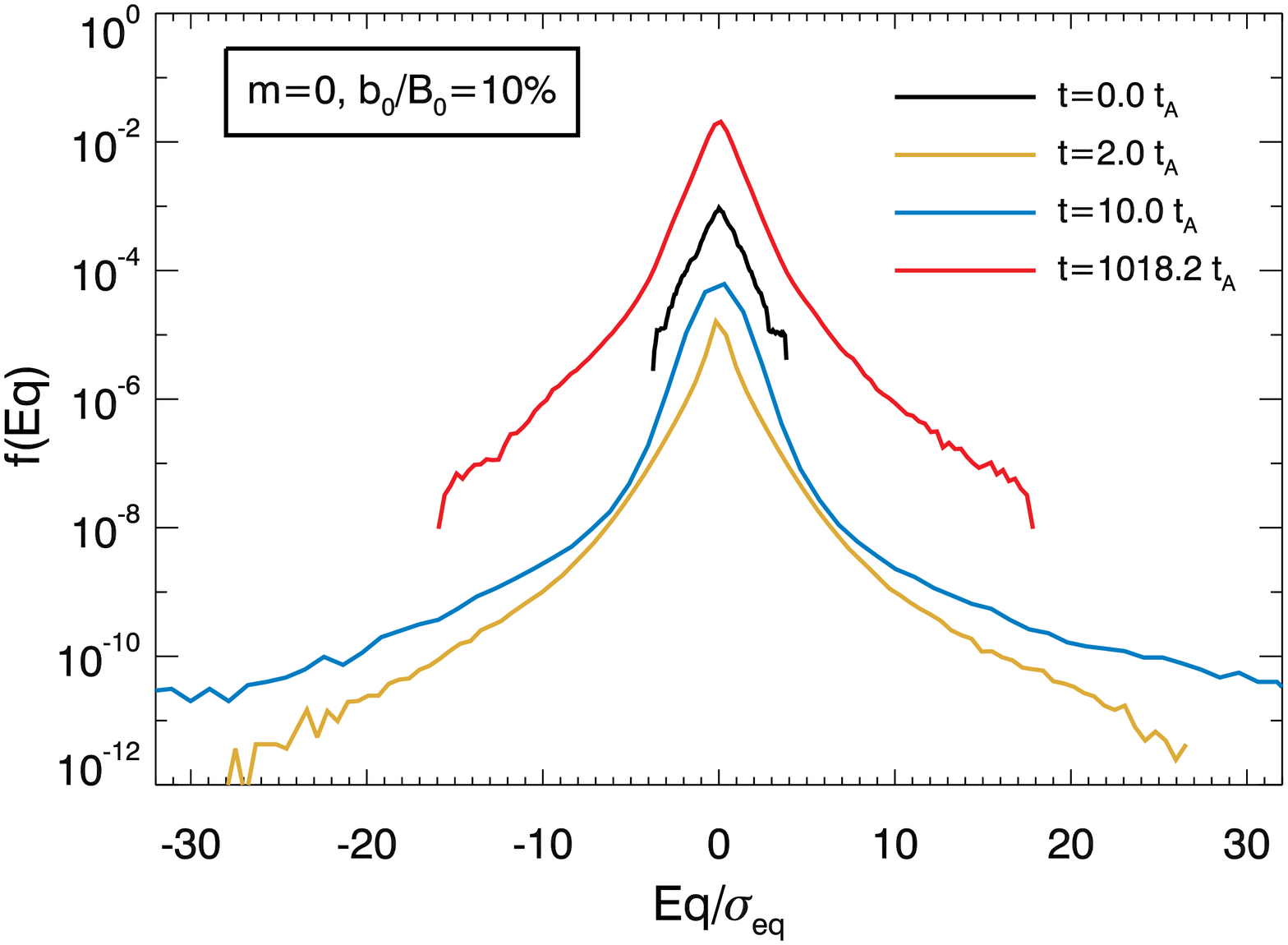} \\[.5em]
\includegraphics[width=.79\columnwidth]{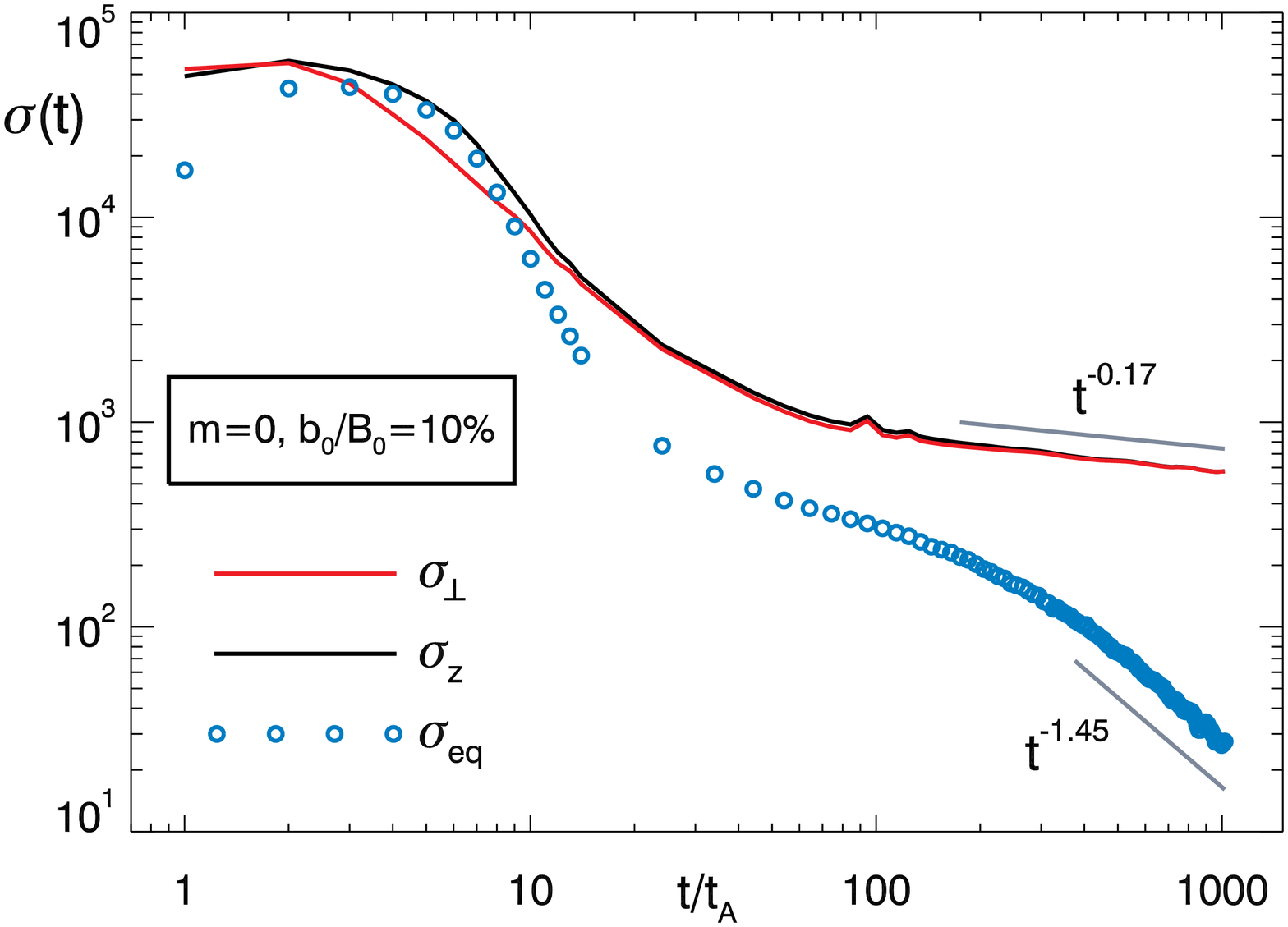} \\[.5em]
\includegraphics[width=.79\columnwidth]{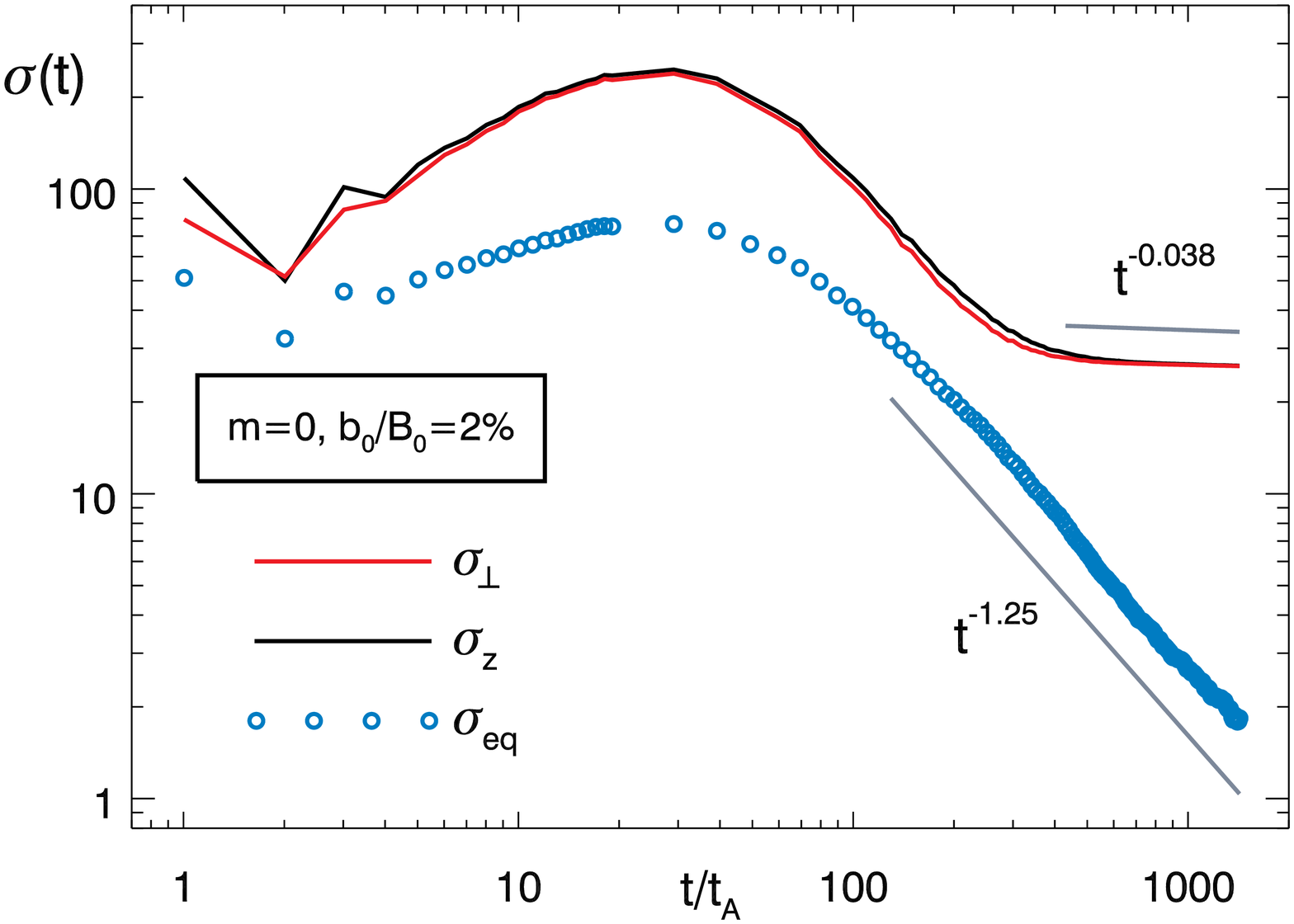} \\[.5em]
\includegraphics[width=.79\columnwidth]{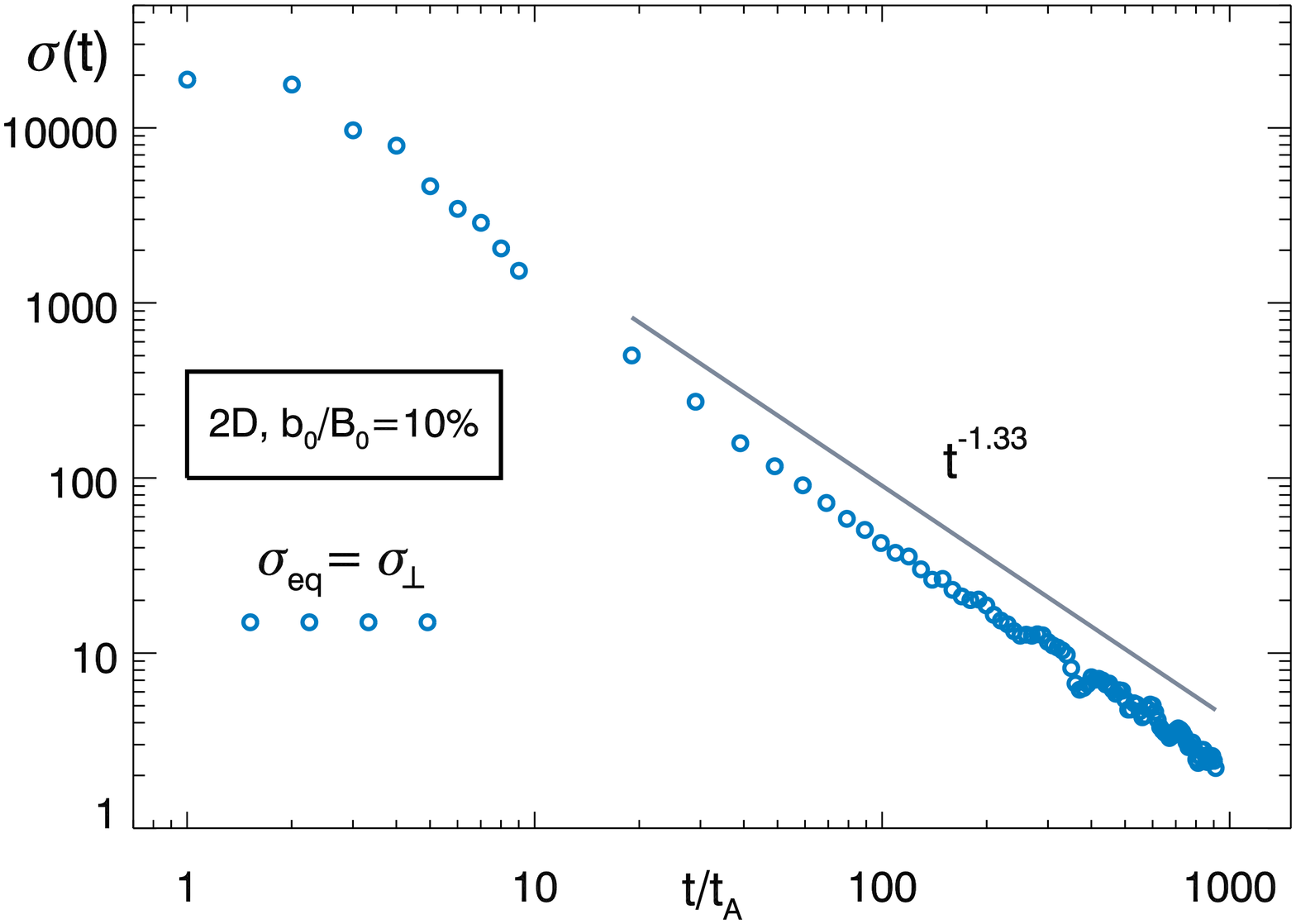}
\caption{\textsf{Runs A--B} (m=0):
Probability density functions (PDFs) of the reduced MHD equilibrium equation (\ref{eq:eqb})
at selected times for the run~B with $b_0/B_0=10\%$ (\emph{top panel}),
shown in a semi-log plot.
To accommodate the large variation of the standard deviation the abscissa displays the 
values normalized with the standard deviation at the corresponding time.
The remaining panels display in logarithmic scale the standard deviations
as a function of time for the 2D simulation (\emph{bottom panel}) 
and the 3D runs~B with $b_0/B_0=2\%$ and $10\%$ (\emph{middle panels}).
For the 3D cases the orthogonal ($\sigma_\perp$), parallel ($\sigma_z$) 
and total ($\sigma_{eq}$) standard deviations are shown.
\label{fig:fig_p1}
}
\end{center}
\end{figure}

The dynamical approach of the system to equilibrium is further investigated
with the probability density functions (PDFs) of the equilibrium equation terms.
These PDFs are histograms of the quantities of interest normalized so that
the resulting function $f(q)$ multiplied by the bin size $\Delta q$
gives the fraction of points in the computational box where the specific 
quantity $q$ has its value in the interval $[q-\Delta q/2, q+\Delta q/2]$,
and the integral $\int \mathrm{d}q f(q) =1$.
PDFs have been computed for both the left and right hand terms in the
equilibrium equation (\ref{eq:eqb}), and for their difference, indicated with
\begin{equation} \label{eq:eqeq}
Eq (\mathbf{x}) = \frac{\partial j}{\partial z} + \frac{\mathbf{b} \cdot \nabla j}{B_0}.
\end{equation}
This is a three-dimensional scalar function that measures how
far the system is from equilibrium locally at point $\mathbf{x}$,
vanishing at equilibrium and with higher absolute values the larger the departure
from equilibrium.
The PDFs have been computed for all the simulations with $m=0$ (for 
the 2D case only the term $\mathbf{b} \cdot \nabla j / B_0$ is present), and they 
all exhibit similar behavior.
All these PDFs have vanishing averages, therefore their standard deviations
can be defined as:{\setlength\arraycolsep{-15pt}
\begin{eqnarray}
&&\sigma_{z} = \left\langle \left( \frac{\partial j}{\partial z} \right)^2 \right\rangle^{1/2}, 
\qquad 
\sigma_{\perp} = \left\langle \left( \frac{\mathbf{b} \cdot \nabla j}{B_0} \right)^2 \right\rangle^{1/2}, 
\label{eq:rmsz} \\[.7em]
&&\sigma_{eq} = \left\langle \left( \frac{\partial j}{\partial z} +  \frac{\mathbf{b} \cdot \nabla j}{B_0} 
\right)^2 \right\rangle^{1/2}, \label{eq:rmseq} 
\end{eqnarray}
}respectively for the first and second terms and the whole Equation~(\ref{eq:eqeq}),
labeled as \emph{parallel} ($\sigma_z$),  \emph{orthogonal} ($\sigma_\perp$)
and \emph{total} ($\sigma_{eq}$).

In Figure~\ref{fig:fig_p1} (\emph{top panel}) the PDFs of the equilibrium 
function $Eq (\mathbf{x})$ are shown in a semi-log plot at selected 
times for run~B with $b_0/B_0 = 10\%$. 
The abscissa is rescaled with the standard deviation of the PDFs to improve
visualization, since they exhibit large variations in time.
The PDF of $Eq$ is generally super-Gaussian (with a peak around zero and ``tails''
farther out),
particularly close to maximum dissipation time ($t=2\, \tau_A$),
however the central part appears closer to a Gaussian distribution at later times ($t=10\, \tau_A$),
and particularly in the final asymptotic stage ($t=1018.2\, \tau_A$).
Although long tails are present at most times, in all cases $\sim 95\%$
of points lies within two standard deviations from zero 
(from $94\%$ at $t=0$ to $96\%$ at $t=10\, \tau_A$).

An appropriate quantitative measure 
of the \emph{distance of the system from equilibrium} is given by
the \emph{standard deviation} of $Eq$, shown in the mid panels of Figure~\ref{fig:fig_p1}
along with $\sigma_z$ and $\sigma_\perp$ for runs~B with $b_0/B_0 = 2\%$ and $10\%$.
In this case, since their averages vanish, the standard deviations are also 
the rms of the considered quantities.
At time $t=0$ the derivative along $z$ of $j$ vanishes ($\partial_z j =0$)
so that initially $\sigma_z =0$ for both runs,  while
$\sigma_\perp = 25.32$ and $633.09$ respectively.
But already after one Alfv\'en time $\tau_A$ they both increase 
(substantially only for the run with $b_0/B_0 = 10\%$) 
reaching similar values $\sigma_z \sim \sigma_\perp$, 
and subsequently continue to be very close
while their values decrease. The rms of the equilibrium function $Eq$,
the standard deviation $\sigma_{eq}$, decreases with time, and
it is asymptotically smaller than  $\sigma_z$ and $\sigma_\perp$. 
As shown in Figure~\ref{fig:fig_p1} all standard deviations
decrease asymptotically like a power-law with $\sigma_z \sim \sigma_\perp \propto t^{-\alpha}$,
where spectral indices are respectively $\alpha = 0.17$ and $0.038$ 
for the runs with $b_0/B_0 = 10\%$ and $2\%$, while the rms
of $Eq$ decays at a faster rate with $\sigma_{eq} \propto t^{-\beta}$,
where $\beta = 1.45$ and $1.25$. 

This implies that while the rms of $\partial_z j$ and 
$\mathbf{b} \cdot \nabla j/B_0$ have about the \emph{same value} 
and remain \emph{approximately constant}
in the asymptotic state (when their power-law decay occurs), 
\emph{equilibrium is approached} as the two terms in Equation~(\ref{eq:eqeq})
\emph{balance each other progressively more} 
throughout the computational box, thus leading to the rapid decrease of $\sigma_{eq}$.

The initial increase of the standard deviations is larger for $b_0/B_0 = 10\%$
than for the $2\%$ case, since for $b_0/B_0 = 10\%$  the system is farther
out of equilibrium in the beginning, with strong currents forming quickly and 
maximum dissipation rate occurring at $t\sim 1.7\, \tau_A$, when also standard deviations 
approximately peak. Their subsequent decrease up to $t \sim 20\, \tau_A$
is enhanced by the strong decay of magnetic energy and progressive
disappearance of current sheets, before approaching the asymptotic stage
around $t \sim 200\, \tau_A$.
Although the standard deviations for run~B with $b_0/B_0 = 2\%$
have similar behavior to the case with $b_0/B_0 = 10\%$, their
variations are much smaller, since the field just undergoes a slight adjustment, 
without any significant energy decay (Figure~\ref{fig:fig1}), current sheets formation or 
significant change in the magnetic field topology (see bottom row in Figure~\ref{fig:fig5}).

For the 2D run~A there is no parallel standard deviation $\sigma_z$.
The rms of $Eq$,  $\sigma_{eq}$,  is equal to $\sigma_\perp$,
and similarly to runs~B its time evolution follows
the power-law $\sigma_{eq} \propto t^{-\beta}$ with $\beta = 1.33$,
as shown in Figure~\ref{fig:fig_p1} (bottom panel).
It is worth mentioning that $\sigma_{eq}$ initially grows larger respect to its initial
value ($=633$) for run~B respect to run~A with same $b_0/B_0=10\%$.
The reaction of the line-tied field sets the system further out of equilibrium,
an enhancement of nonlinearity that might favor the development of
singularities in the line-tied system respect to the periodic case.

\begin{figure}
\begin{center}
\includegraphics[width=.9\columnwidth]{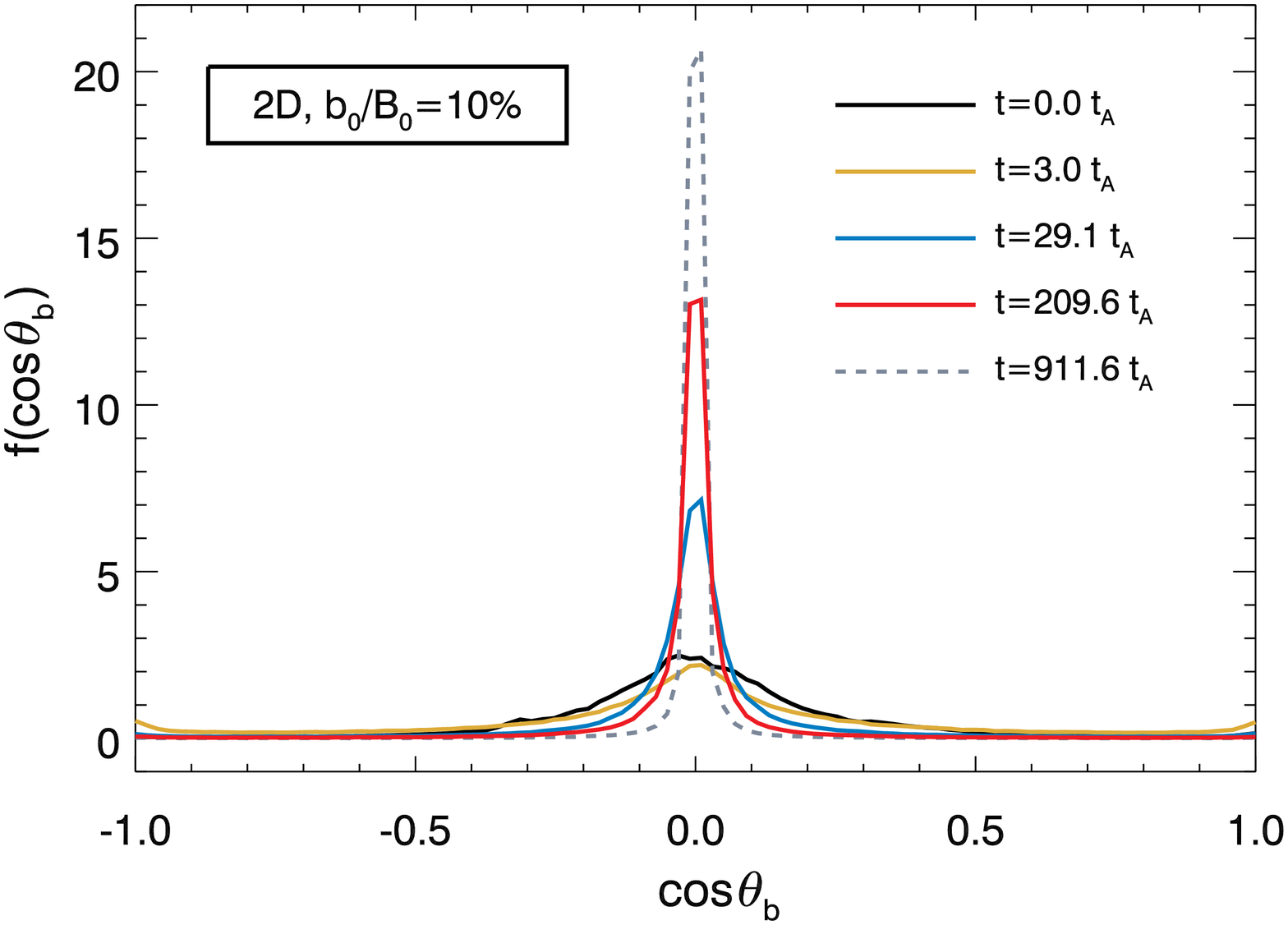}\\[1em]
\includegraphics[width=.9\columnwidth]{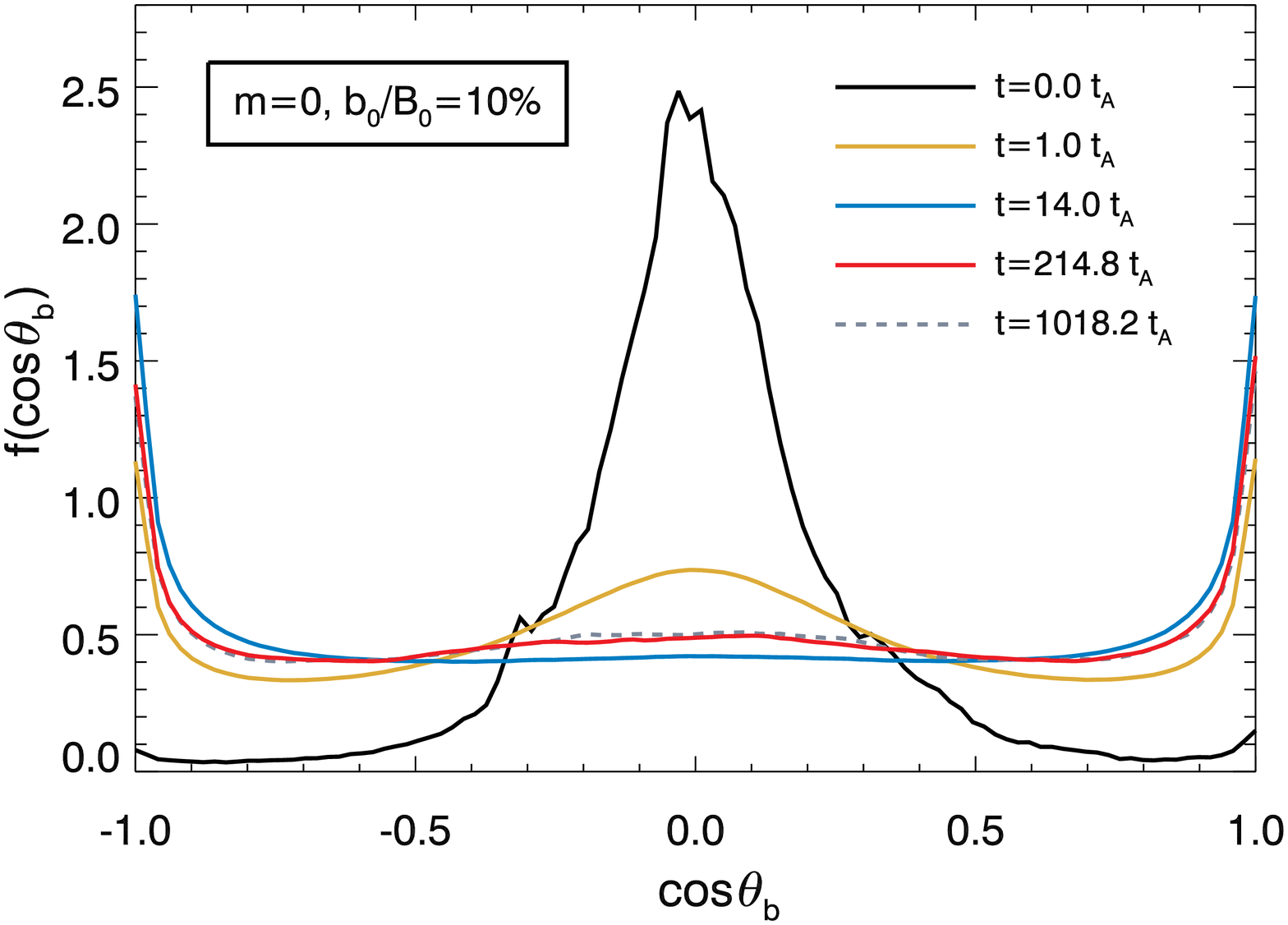}\\[1em]
\includegraphics[width=.9\columnwidth]{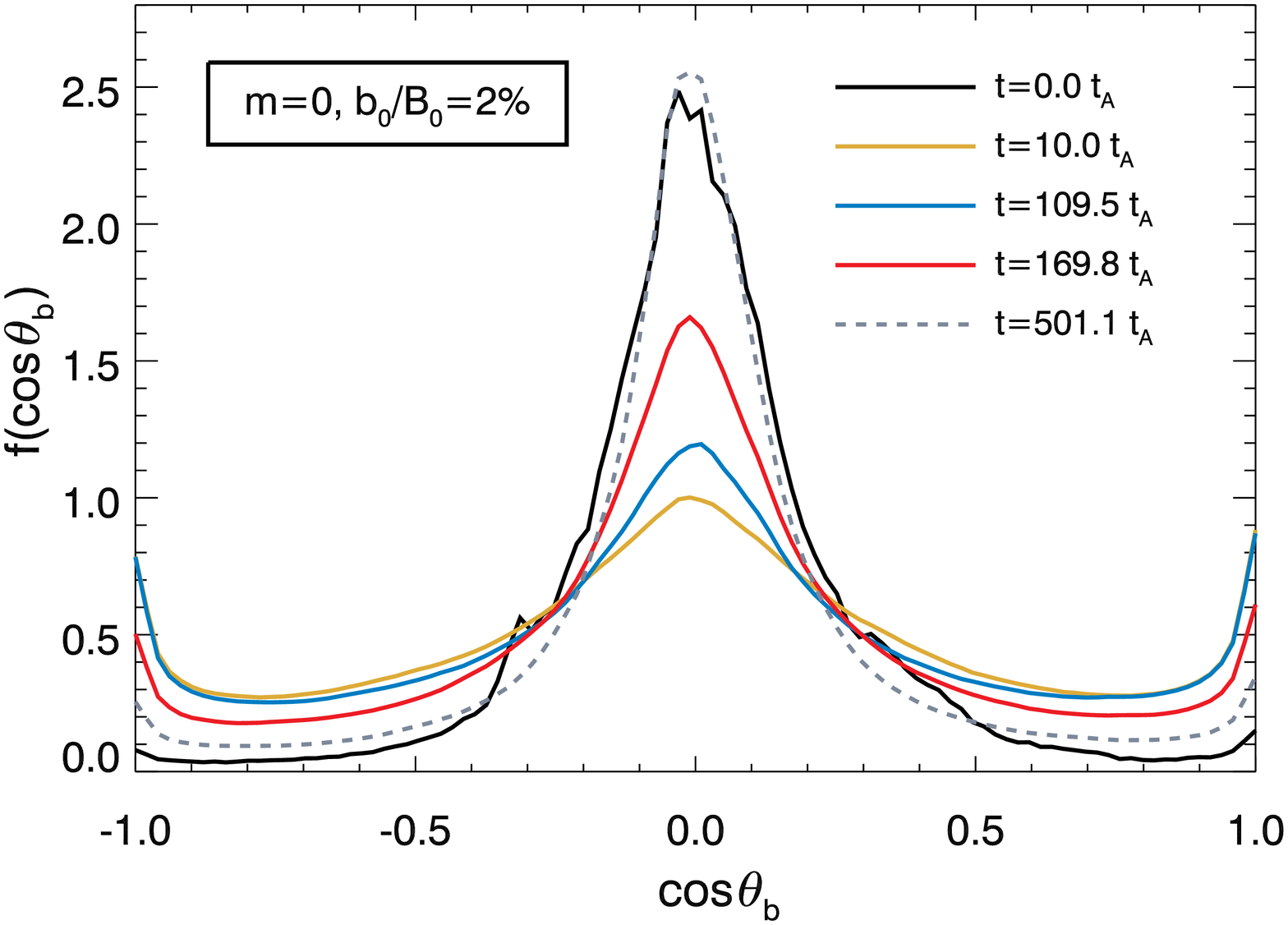}
\caption{\textsf{Runs A--B} (m=0):
Probability density functions (PDFs) of the angle $\theta_b$ between
the orthogonal magnetic field $\mathbf{b}$ and the current density gradient $\nabla j$
at selected times for the 2D run~A with $b_0/B_0=10\%$ (\emph{top panel}),
and the  3D runs~B with respectively $b_0/B_0=10\%$ (\emph{middle}) and
$b_0/B_0=2\%$ (\emph{bottom}).
\label{fig:fig_p3}}
\end{center}
\end{figure}

Further insight into the dynamics is gained analyzing the 
probability density function of the cosine of the angle~$\theta_b$ 
between $\mathbf{b}$ and $\nabla j$
\begin{equation} \label{eq:ctb}
   \cos \theta_b = \frac{\mathbf{b} \cdot \nabla j}{|\mathbf{b}|\,  |\nabla j|},
\end{equation}
shown in Figure~\ref{fig:fig_p3}.
At time t=0 the ``2D perpendicular'' Lorentz force term does not vanish for all simulations
($\mathbf{b} \cdot \nabla j \ne 0$).
Since the  orthogonal component of the magnetic field  differs for runs~A--B only for 
a proportionality factor (Equation~[\ref{eq:pot4}]),
the PDF is the same for all runs~A--B (e.g., see mid-panel in Figure~\ref{fig:fig_p3}). 
Although it is peaked around zero (corresponding to 
the 2D equilibrium condition $\mathbf{b} \cdot \nabla j = 0$) it spreads out
with significant values up to $|\cos \theta_b| \lesssim 1/2$, corresponding
to an approximate $60^\circ$ angle around $\theta_b=90^\circ$.
For the 2D run~A with $b_0/B_0=10\%$ (top panel)
the PDF initially spreads out during the strongest part of the nonlinear
stage ($ 1 \lesssim t/\tau_A \lesssim 10$) during which $\sim 90\%$ 
of the initial energy is dissipated, but afterward peaks progressively more strongly
around zero, corresponding to the equilibrium condition
for the orthogonal field $\mathbf{b} \cdot \nabla j = 0$, 
with respectively 71\% and 96\% of the grid points in the volume 
in the region $|\cos \theta_b| < 0.1$, spanning an angle of $\sim 11.5^\circ$ around
$\theta_b = 90^\circ$, at times t=29.1 and $911.6\, \tau_A$.
This confirms that the system approaches asymptotically an
equilibrium with $\mathbf{b} \cdot \nabla j = 0$ corresponding
in physical space to increasingly circular field lines progressively more 
coincident with the isosurfaces of $j$, as shown in Figure~\ref{fig:fig5} (top row).

In contrast \emph{the picture is radically different for the line-tied
simulations}. As shown in the middle panel of Figure~\ref{fig:fig_p3},
for run~B with $b_0/B_0=10\%$ 
\emph{the PDF spreads increasingly further out during the time evolution},
flattening considerably already after one Alfv\'en time and \emph{remaining
flat throughout the subsequent evolution}, when $\sim 80\%$ of magnetic energy
is dissipated, and in the following asymptotic regime, with peaks forming in 
correspondence of alignment between the two  fields ($\theta_b \sim 0^\circ$, $180^\circ$).
Therefore, in contrast with the periodic case (2D run~A) the  orthogonal
magnetic field does not approach the asymptotic equilibrium with $\mathbf{b} \cdot \nabla j =0$
(that would imply also $\partial_z j = 0$ in the 3D case, Equation~[\ref{eq:eqb}]), 
instead in the 3D line-tied case \emph{the orthogonal component of the magnetic field 
remains with a non-vanishing ``2D perpendicular'' Lorentz force term} 
($\mathbf{b} \cdot \nabla j \ne 0$).

Furthermore, if the initial magnetic field intensity is below the threshold set
out by \cite{2013ApJ...773L...2R}, as shown in the bottom panel of 
Figure~\ref{fig:fig_p3} for run~B with $b_0/B_0 = 2 \%$, the PDF
starts flattening out to some extent but then bounces back very
close to its initial profile, corresponding to a slight readjustment 
of the magnetic field as shown in Figure~\ref{fig:fig5} (bottom row),
with the orthogonal magnetic field preserving its non-vanishing 
perpendicular Lorentz force.

The results of the numerical simulations analyzed in this section
are consistent with the heuristic phenomenology laid out by
\cite{2013ApJ...773L...2R} and the more refined analysis of the 
structures of the equilibria expounded in Section~\ref{sec:eq}.  
The asymmetry along $z$ of the solutions of the reduced MHD 
equilibrium equation (\ref{eq:eqb}) can be estimated with the 
axial variation length-scale $z_\ell \sim \ell B_0/b$ 
(Equation~[\ref{eq:zl}], see also Figure~\ref{fig:fig_ic}),
where $\ell$ is the perpendicular characteristic scale (in the $x$--$y$ plane)
of the magnetic field component $\mathbf{b}$.

As discussed in Section~\ref{sec:sred}, the dynamical
solutions of the reduced MHD equations (\ref{eq:eq1})-(\ref{eq:eq2})
generally cannot exhibit strong asymmetries along $z$, in particular 
when driven from the photospheric boundaries.
On the other hand the equilibria can be strongly asymmetric or quasi-invariant
along z, depending on the relative value of the axial variation scale $z_\ell$
respect to the loop axial length $L_z$, with the \emph{critical length} given by $z_\ell \sim L_z$.
For small values of $b$ the axial variation scale $z_\ell$ is longer than the 
loop length $L_z$ and the corresponding equilibrium solution is quasi-invariant
along $z$, while for larger values of $b$ the axial scale $z_\ell$ is smaller than
the loop length and the equilibrium is more asymmetric along $z$ the larger the
magnetic field intensity $b$.

As shown in Figure~\ref{fig:fig1} no substantial energy is dissipated for 3D runs
with $b_0/B_0 \lesssim 3\%$, and just minimal dynamics occurs as the field 
slightly readjusts (Figures~\ref{fig:fig3}--\ref{fig:fig_p3}).
Since the initial magnetic field (Equation~[\ref{eq:pot4}])
has a perpendicular scale $\ell \sim L_\perp/3.87 \sim 1/3.87$ (the averaged wavenumber
of the initial condition is 3.87  and $L_\perp =1$), this threshold corresponds to a variation length-scale
for the initial magnetic field of about $z_\ell = \ell B_0/b_0 \gtrsim 100/(3 \times 3.87)$, i.e.,
$z_\ell \gtrsim L_z$ since $L_z = 10$.

Therefore for $b_0/B_0 \lesssim 3\%$ the corresponding
equilibria, computed from Equation~[\ref{eq:eqb}] with 
$\mathbf{b}_{eq}$($z$=0)~= $\mathbf{b}_0$($z$=0), as described in Section~\ref{sec:eq},
have a large variation length-scale $z_\ell \gtrsim L_z$ and are therefore quasi-invariant along $z$.
Since the initial condition has only the parallel m=0 mode it is invariant along $z$, and therefore 
very close to the corresponding equilibrium, so that nonlinearity
is strongly depleted and only a slight readjustment of the field occurs, 
with no significant energy dissipation.
On the contrary for initial conditions with $b_0/B_0 > 3\%$ the corresponding
equilibria have smaller variation length-scales $z_\ell < L_z$, hence the
equilibrium solution is strongly asymmetric along $z$, differing substantially
from the initial condition that is then necessarily out-of-equilibrium, as shown 
by the subsequent dynamics and energy dissipation.

Furthermore, as shown in Figure~\ref{fig:fig1}, also in the cases 
when decay occurs, i.e., for initial conditions with $b_0/B_0 > 3\%$, 
the magnetic field relaxes to a new equilibrium 
that approximately satisfies the condition $b/B_0 \sim 3\%$ and $z_\ell \sim L_z$. 
But while the asymptotic energy of the runs with $b_0/B_0 = 4\%$, 5\% and 6\%
are approximately the same, corresponding to a ratio $b/B_0 \sim 3.3\%$,
the run with $b_0/B_0 = 10\%$ relaxes to a slightly higher energy with
$b/B_0 \sim 4\%$. On the other hand the run with $b_0/B_0 = 10\%$
has a stronger inverse cascade (Figure~\ref{fig:fig3}), with significantly 
larger magnetic islands in the asymptotic regime (Figure~\ref{fig:fig5})
and average wavenumber $\sim 2.7$ thus obtaining again $z_\ell \sim L_z$. 
When a strong inverse cascade occurs the formation of larger perpendicular scales ($\ell$)
increases the value of the axial variation length-scale $z_\ell \sim \ell B_0/b$, thus 
attaining the equilibrium condition $z_\ell \sim L_z$ with a larger value of $b$,
and consequently a smaller dissipation of energy.

\begin{figure*}
\begin{center}
\includegraphics[width=.48\textwidth]{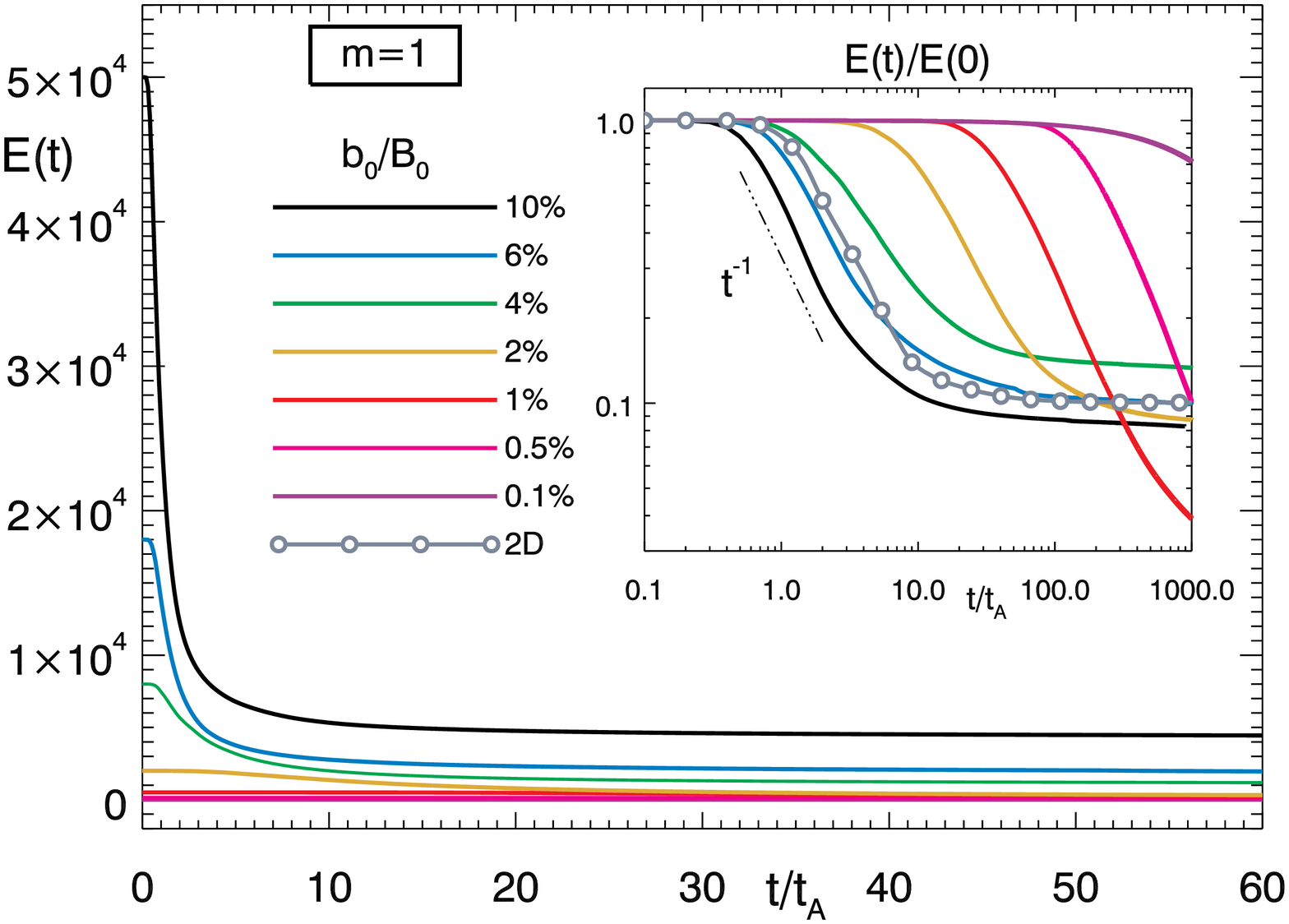} \hspace{1.8em}
\includegraphics[width=.48\textwidth]{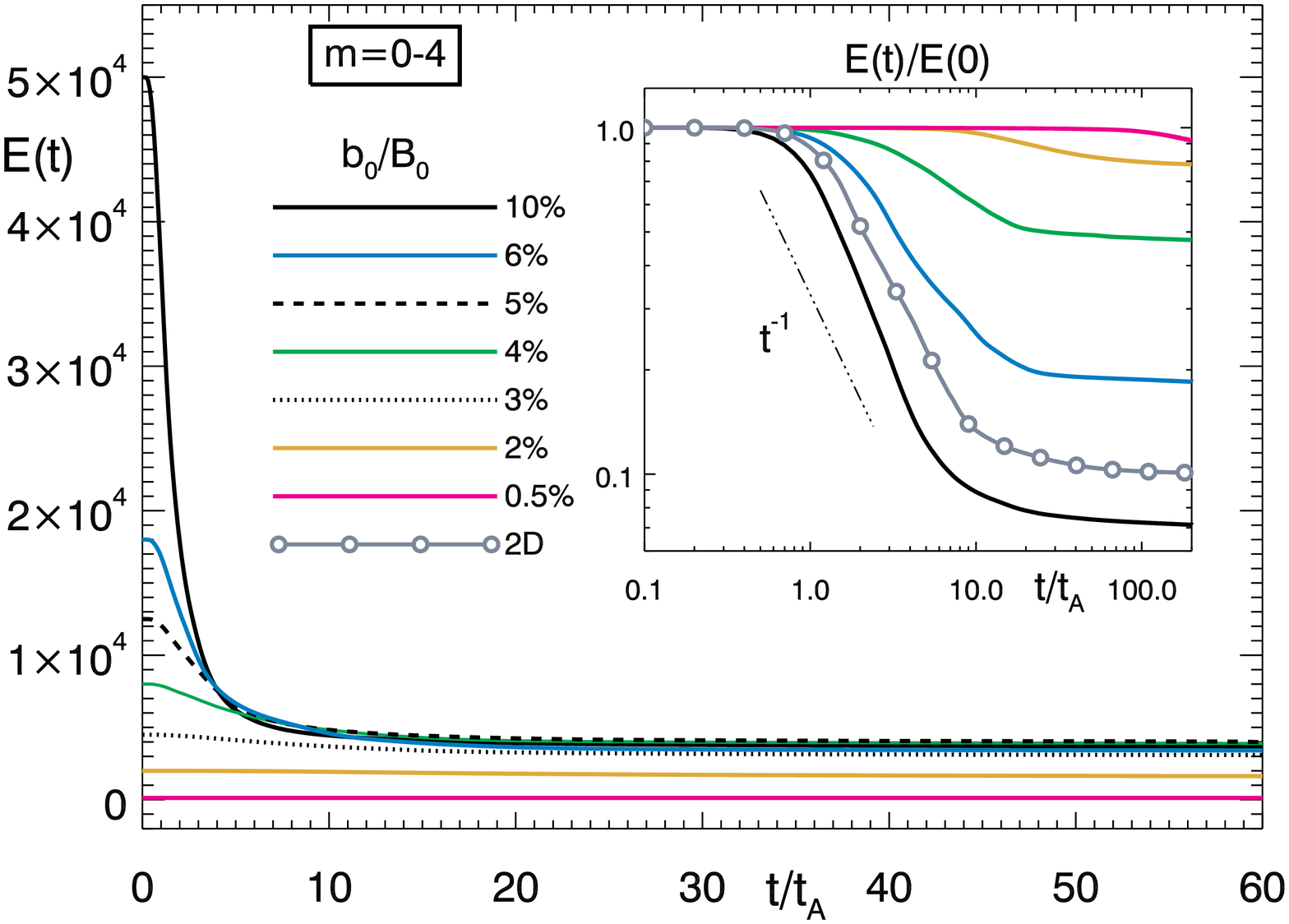}
\caption{\textsf{Runs~C} (m=1) and \textsf{D} (m=0--4): 
Total energy vs.\ time for line-tied simulations with different values of $b_0/B_0$ 
for run~C with single parallel mode m=0 (\emph{left panel}) and run~D with m=1
(\emph{right panel}). The 2D run~A with $b_0/B_0 = 10\%$ is added for comparison.
The insets show in logarithmic scale total energies normalized with their initial values.
\label{fig:figc1}}
\end{center}
\end{figure*}

The \emph{critical variation length-scale} $z_\ell$ origins from a balance
of forces because the reduced MHD equilibrium equation~(\ref{eq:eqb}) 
represents a balance between two force terms.
In the reduced MHD limit \citep[e.g.,][]{1982PhST....2...83M} the Lorentz force
splits into two terms with components only in the orthogonal $x$--$y$ planes:
the ``perpendicular'' ($\mathbf{b} \cdot \nabla \mathbf{b}$)
and ``parallel'' ($B_0 \partial_z \mathbf{b}$) field line tensions 
(plus the pressure term, determined through the incompressibility condition). 
The first term represents the field line tension 
of the orthogonal component $\mathbf{b}$,
the only one present in the 2D limit. The second is an additional
tension term due to the presence of the guide field $B_0$,
linked to the tension of the field lines of the total magnetic 
field $B_0\mathbf{\hat{e}}_z +\mathbf{b}$.
An equilibrium is attained only when these two 
\emph{counteracting} components of the Lorentz force 
balance each other satisfying Equation~(\ref{eq:eqb}).
As outlined in \cite{2013ApJ...773L...2R} these two forces are of the same 
order of magnitude for the critical intensity
\begin{equation} \label{eq:eqrp}
  b \sim \frac{\ell B_0}{L_z},
\end{equation}
corresponding to the critical axial 
variation length-scale $z_\ell \sim \ell B_0/b_{\ell} \sim L_z$
as discussed in Section~\ref{sec:sred}.

Initially also the 3D line-tied system starts to behave as in the 2D case,
with the tension of perpendicular field lines 
creating an orthogonal velocity, that coupled with all others 
\emph{nonlinear terms} are the only ones that 
can cascade energy and generate current sheets.
But this displaces the total line-tied (axially directed) field lines,
that now cannot be freely convected around as in the periodic case 
because of the line-tying constraint at the boundaries,
and  is then counteracted by the enhanced axial tension that resists bending.
Magnetic fields with intensities smaller than the threshold (\ref{eq:eqrp}) a small variation of 
$\mathbf{b}$ along $z$ (corresponding
to a variation scale larger than the loop length $L_z$) is enough
to reach an equilibrium, but for intensities larger than of $b \gtrsim \ell B_0 / L_z$
current sheets must form and energy dissipate in order to reach
the physically accessible equilibria with $b \sim \ell B_0 / L_z$, since
for larger magnetic field intensities the equilibria are strongly 
asymmetric along $z$ and therefore physically inaccessible.

The analysis in this section has considered exclusively initial conditions invariant
along the $z$-direction, with only the parallel mode m=0. In the following
sections we extend it to include field variations along $z$ with higher parallel modes..

\subsection{Runs C: single mode m=1} \label{sec:run1}

The finite length of coronal loops renders the system
akin to a \emph{resonant cavity}. A forcing velocity with frequency $\nu$ 
at the photospheric boundary, e.g., 
\begin{equation} \label{eq:vres}
\mathbf{u}(x,y,z=L,t) = \mathbf{u}_{ph} (x,y) \cos(\omega t),
\end{equation}
where $\omega = 2\pi \nu$ is the angular frequency,
will inject Alfv\'en waves at that frequency propagating in the axial direction of the loop.
In general these waves, that are continuously injected and reflected at the boundaries
(see Section~\ref{sec:sred}, Equation~[\ref{eq:els}]), will be out of phase and decorrelated along
the loop so that their sum will remain limited the whole time to values of the order
of the forcing velocity at the boundary, with no  growth in time for the amplitude
of the resulting velocity and magnetic fields.
But for the \emph{resonant frequencies} 
\begin{equation}
\nu_n =n\, \nu_A / 2, \quad \textrm{with} \quad n \in \mathbb{N}
\end{equation} 
and $\nu_A = 1/\tau_A$, the waves are in phase and they sum coherently 
\citep{1985A&A...146..199I}. 
Thus the magnetic and velocity fields in the loop
grow linearly in time similarly to the case with constant velocity 
(Equation~[\ref{eq:bp}]). For instance, considering the boundary velocity  (\ref{eq:vres})
at the resonant frequency $\nu_n$ with $n \ge 1$, the resulting fields grow approximately as
\citep{1999PhPl....6.4146E, 2010PhDT.......193R, 2015bps..book.....C}:
\begin{align}
&\mathbf{b} \sim \mathbf{u}_{ph} \cos \left(\omega_n \frac{z}{v_A} \right) \cos( \omega_n t)\,  
\frac{t}{\tau_A}, \label{eq:linres1} \\
&\mathbf{u} \sim \mathbf{u}_{ph} \sin \left( \omega_n \frac{z}{v_A} \right) \sin( \omega_n t)\, 
\frac{t}{\tau_A}. \label{eq:linres2}
\end{align}
Indeed a constant velocity can be regarded as the 
\emph{zero frequency resonance} $n=0$ of the system, that differs from resonances
with $n \ge 1$ because while the magnetic field grows linearly in time, for $n=0$
the velocity field does not grow and its value remains of the same order of magnitude
of the photospheric velocity ($u \sim u_{ph}$).

\begin{figure*}
\centering
\includegraphics[width=.33\textwidth]{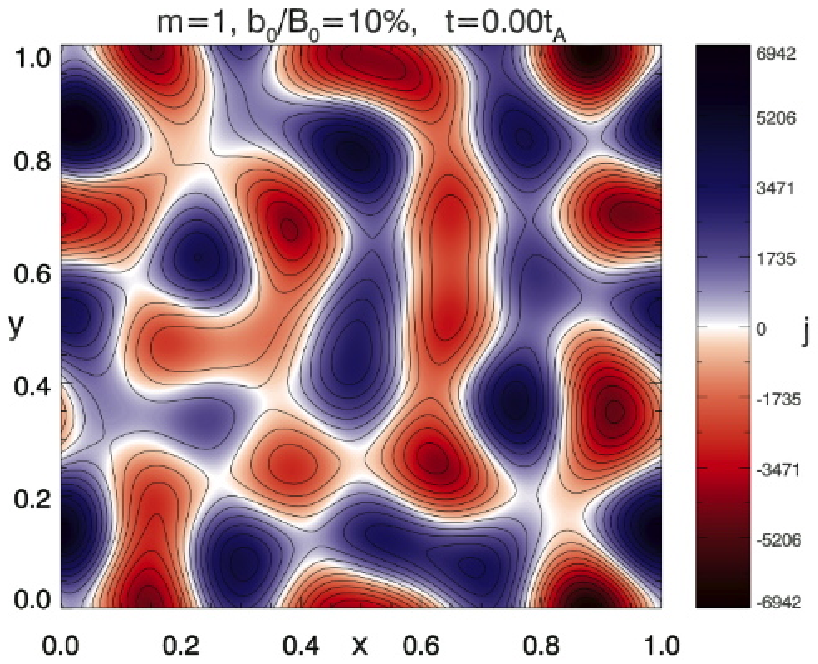}
\includegraphics[width=.33\textwidth]{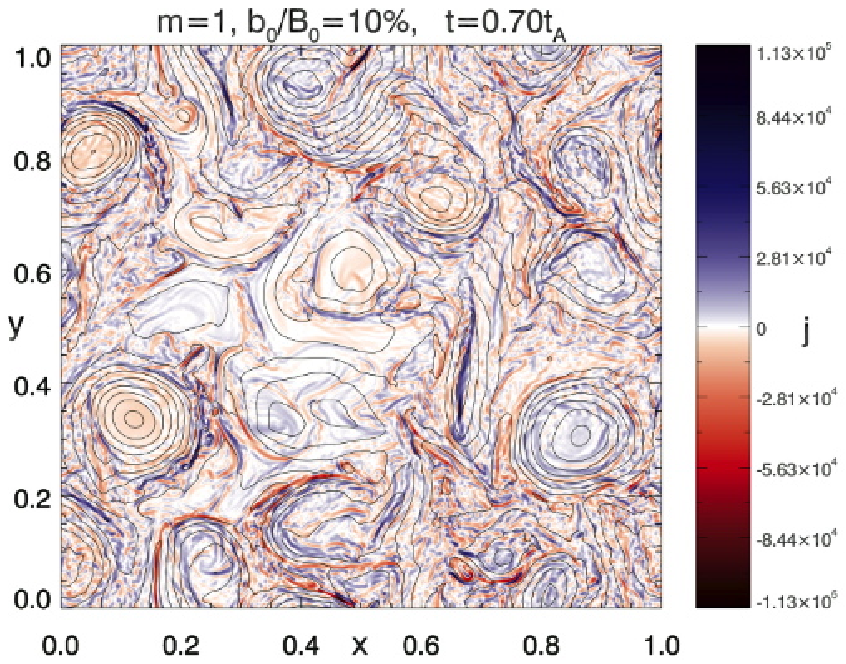}
\includegraphics[width=.33\textwidth]{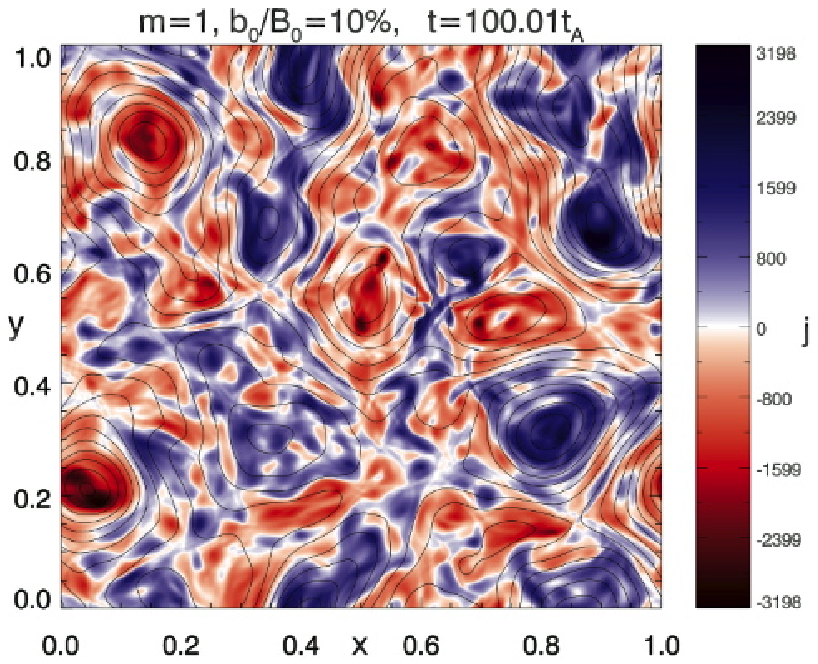} \\[1em]
\includegraphics[width=.33\textwidth]{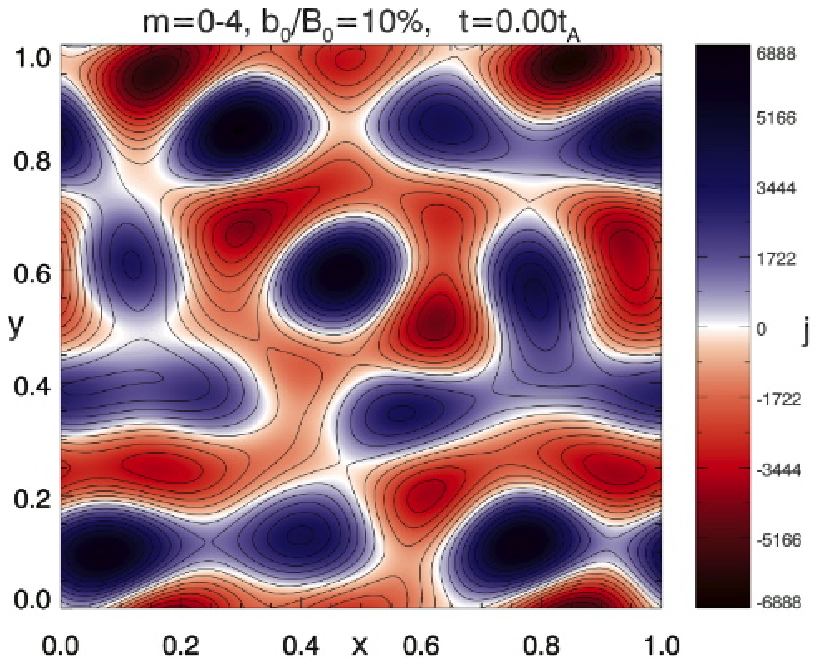}
\includegraphics[width=.33\textwidth]{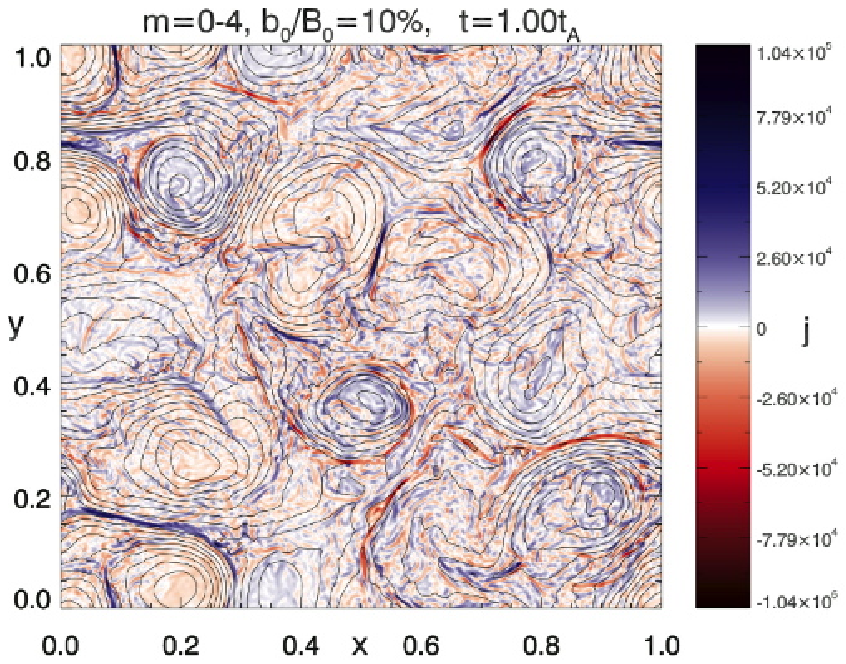}
\includegraphics[width=.33\textwidth]{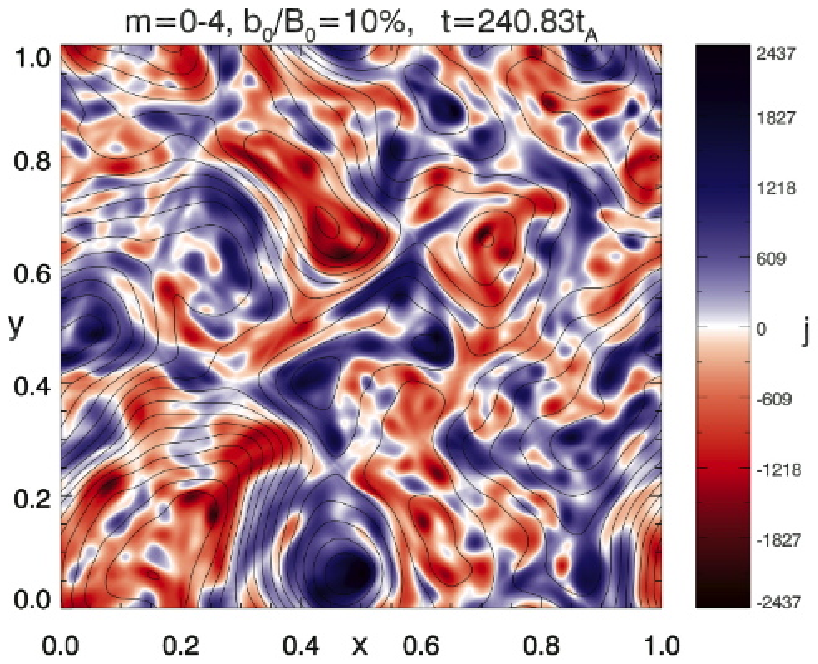}
\caption{\textsf{Runs~C} (m=1) and \textsf{D} (m=0--4): 
Magnetic field lines of the orthogonal magnetic 
field component $\mathbf{b}$ and current density $j$ in the mid-plane
$z=5$ at selected times for runs~C (\emph{top row}) and D (\emph{bottom row}) 
with $b_0/B_0 = 10\%$.
The left panels show the initial condition at $t=0$, the central panels
show the fields at the time of maximum dissipation, while the right panels
show the fields at a later time when they have relaxed (asymptotic regime).
\label{fig:fig7}}
\end{figure*}

As mentioned in Section~\ref{sec:sred}, for X-ray bright solar coronal
loops photospheric motions have a low frequency,
giving rise to a coronal magnetic field dominated by the parallel $m=0$ mode,
and their low frequency can then be approximated with zero.
In general photospheric motions characterized by a surface convective
timescale $\tau_{sc}$ will not have a single harmonic at the frequency
$1/\tau_{sc}$, rather the amplitude of its Fourier transform will peak
at the frequency  $1/\tau_{sc}$ but include many other harmonics.

For longer loops (with longer Alfv\'en crossing times 
$\tau_A=L_z/v_A$ comparable or longer than granulation timescales),
and for loops on other active stars with magnetized coronae and
outer convective envelopes, photospheric motions can have frequencies 
closer to resonances higher than $\nu_0 = 0$.
Furthermore also when photospheric motions have a dominant
low frequency, higher frequency modes will be present, although
with smaller amplitudes \citep{2008ApJ...685..606N}
that contribute considerably less heating \citep{1997ApJ...490..442M}.
In all these cases photospheric motions will give rise to
parallel modes higher than zero ($m \ge 1$) for the coronal magnetic field.
These will be the dominant modes when the photospheric frequency is resonant, 
and give a small contribution to the magnetic field
when photospheric motions frequencies are close to zero.
Additionally higher parallel modes can also be generated by 
nonlinear dynamics also when starting with a zero parallel mode
\citep{2007ApJ...662..701B}, and  by disturbances stemming from 
chromospheric dynamics \citep{2007Sci...318.1574D}.

\begin{figure*}
\centering
\includegraphics[width=.33\textwidth]{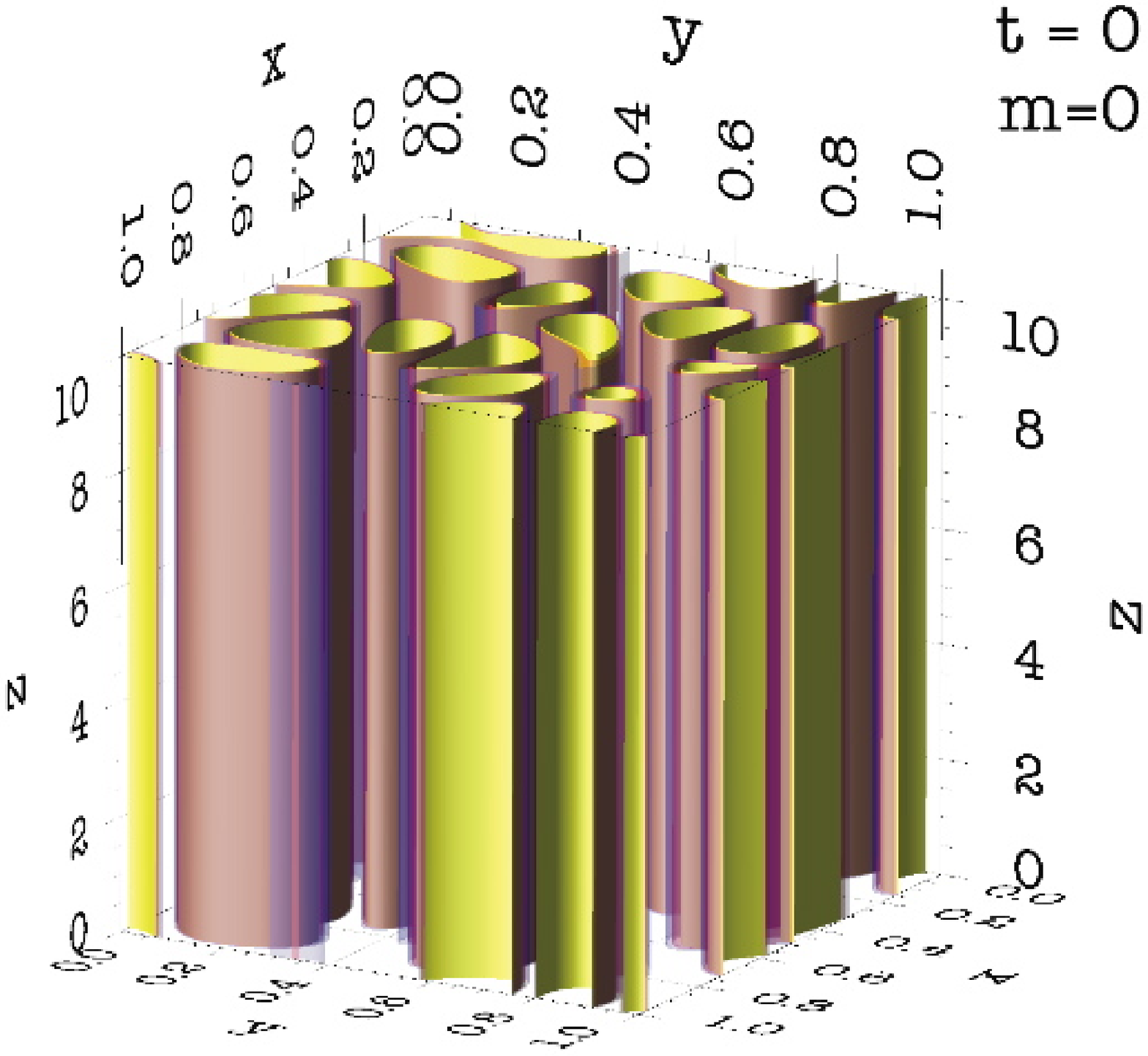}
\includegraphics[width=.33\textwidth]{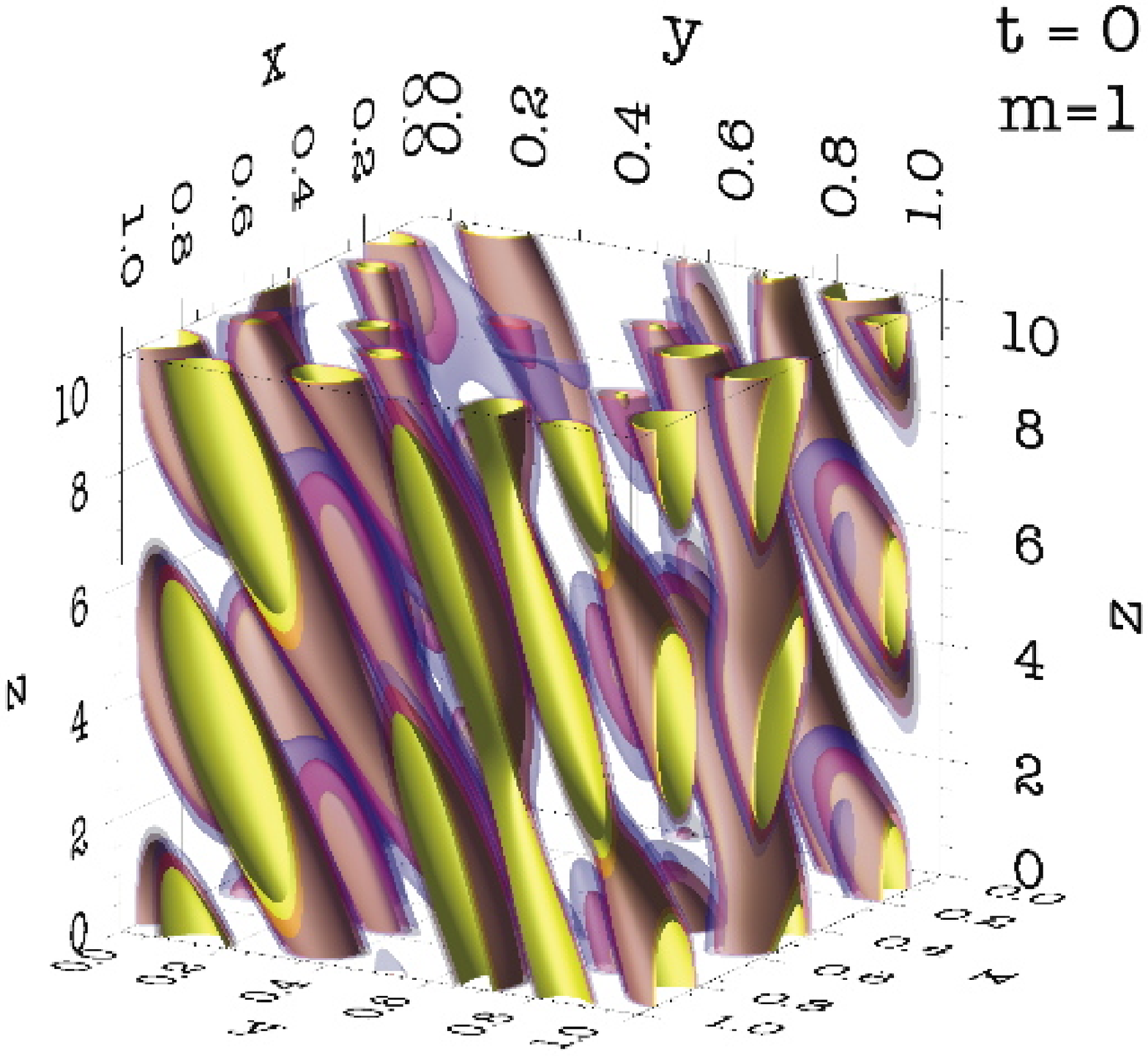}
\includegraphics[width=.33\textwidth]{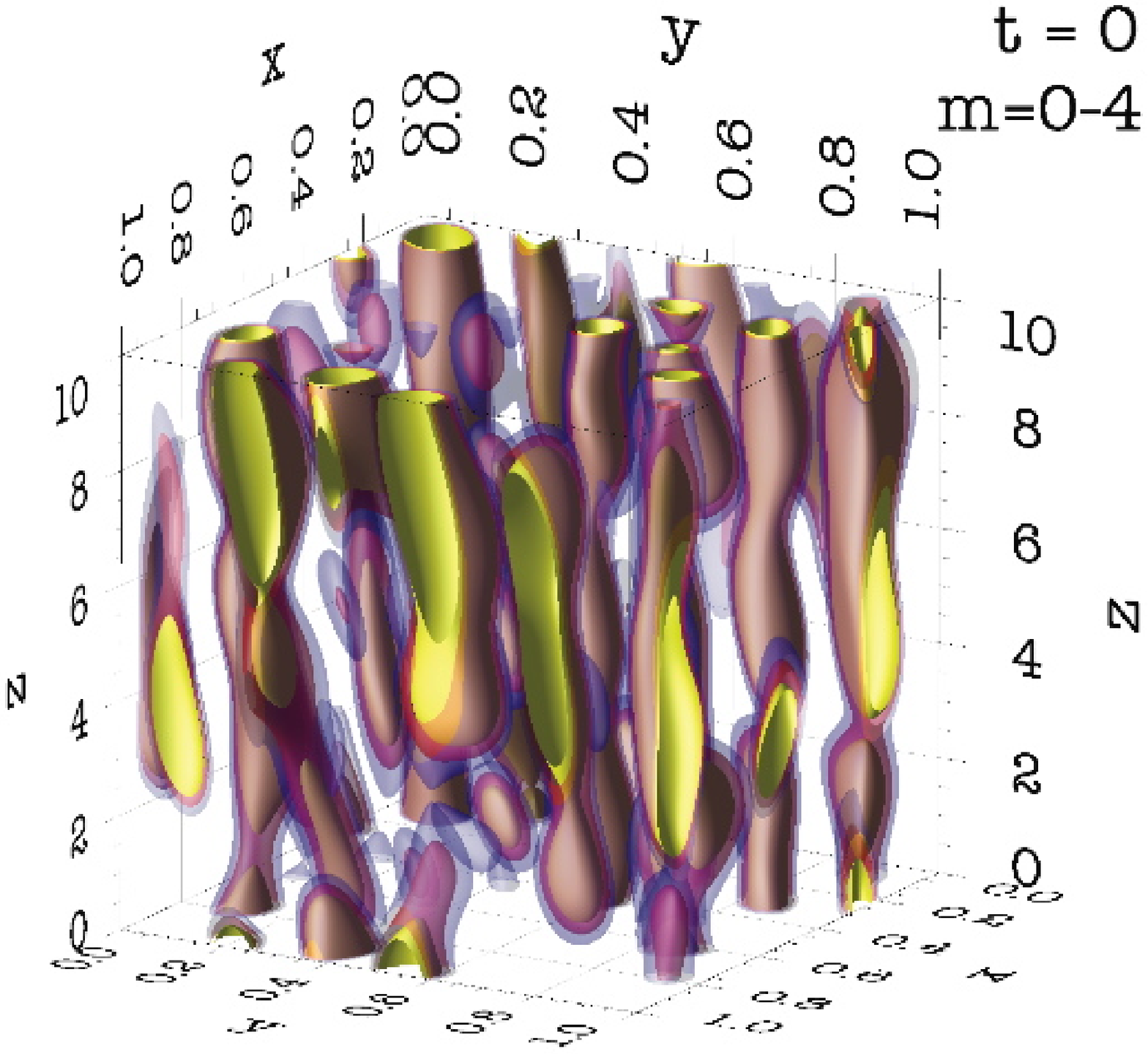} \\[1em]
\includegraphics[width=.33\textwidth]{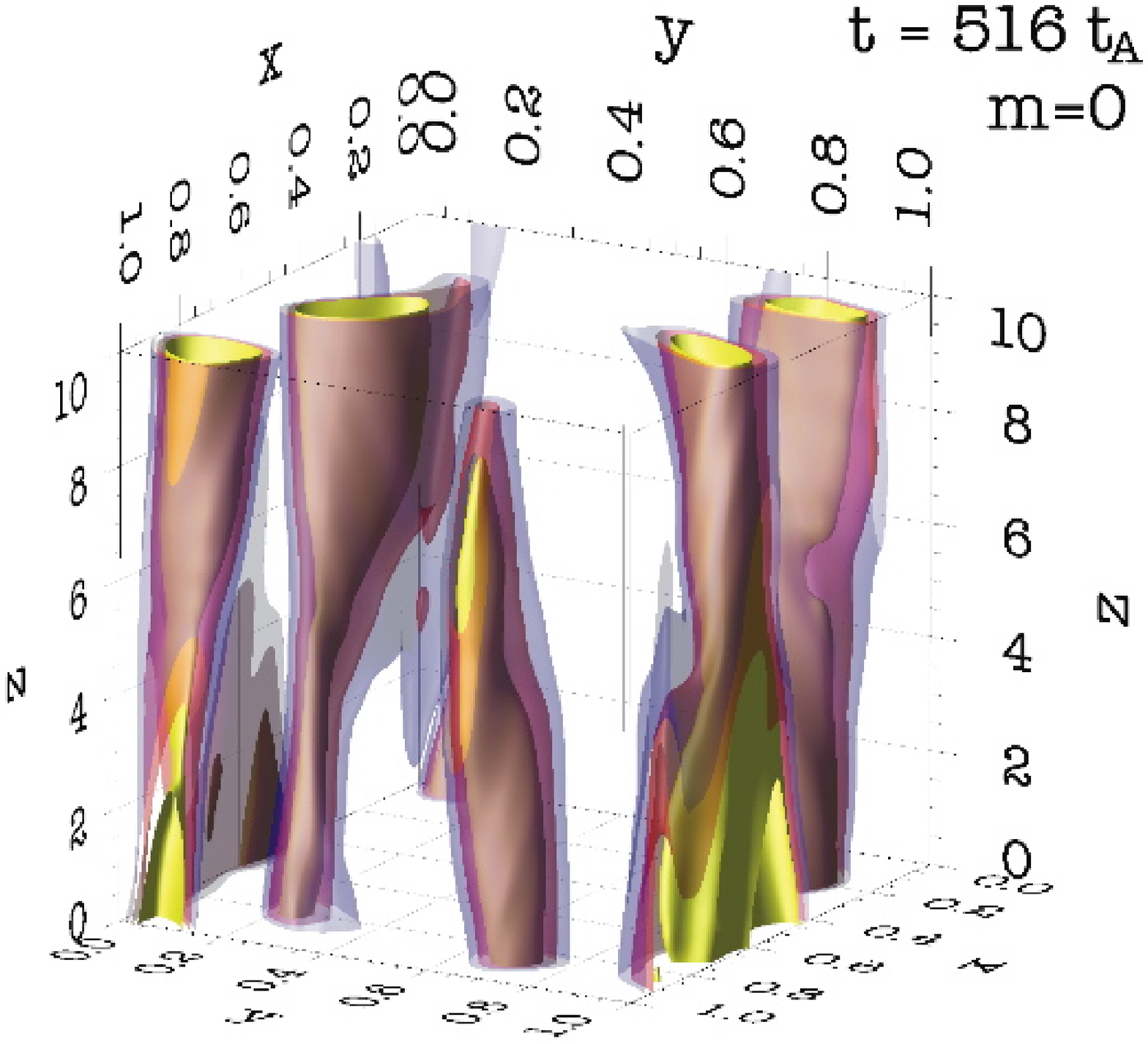}
\includegraphics[width=.33\textwidth]{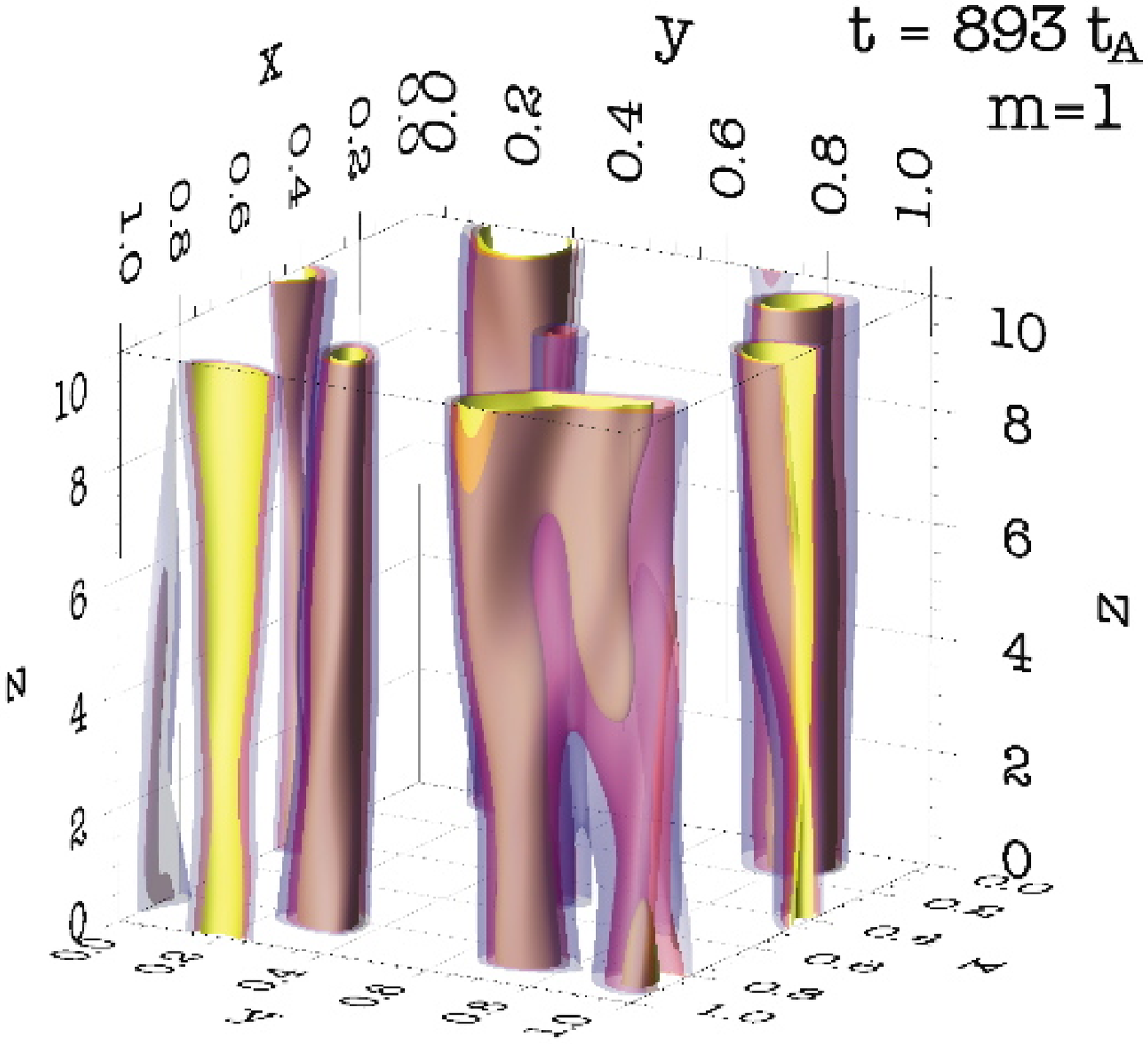}
\includegraphics[width=.33\textwidth]{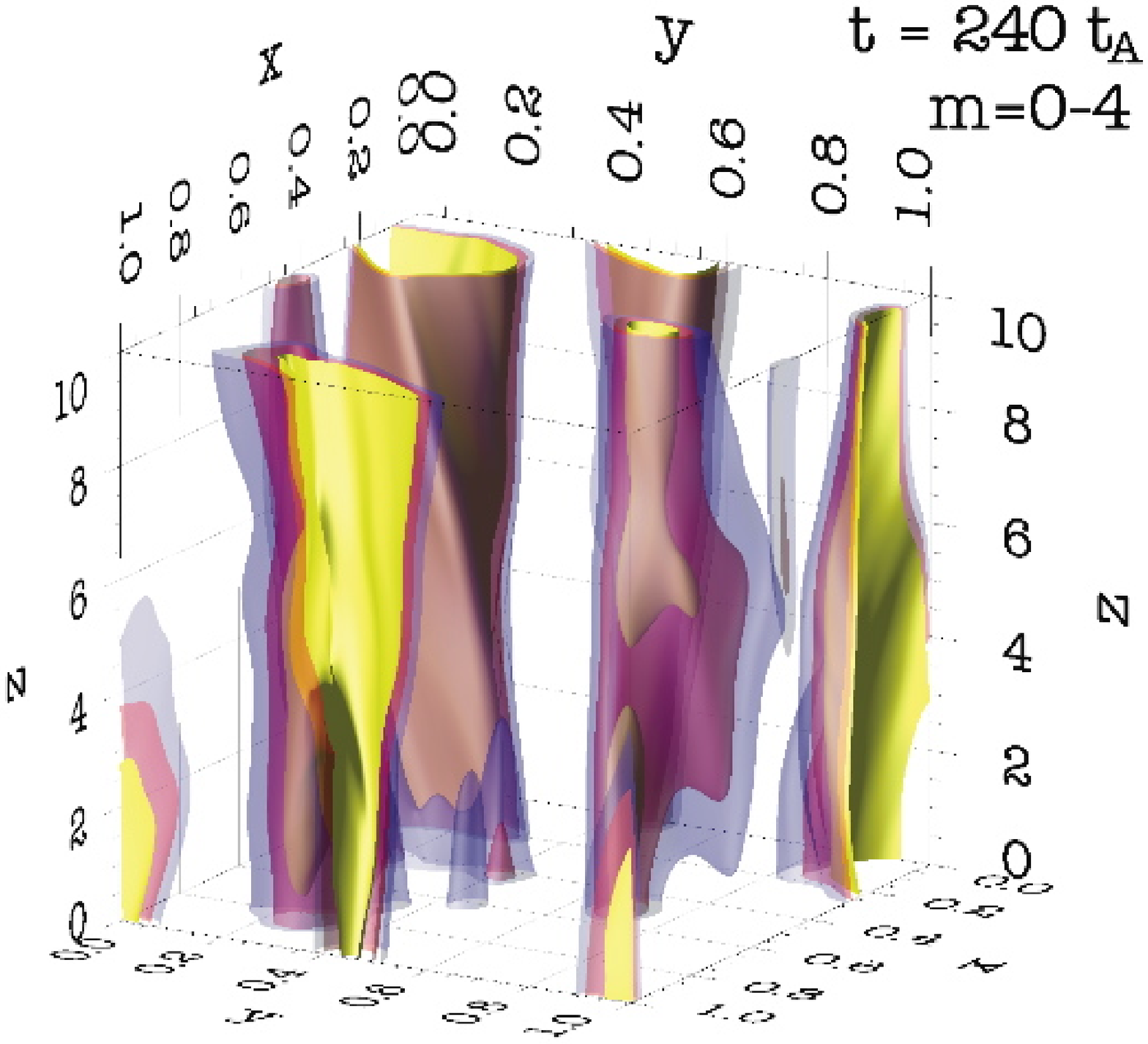}
\caption{Isosurfaces of the magnetic potential $\psi$ (in yellow and transparent 
red and blue colors)
for the three-dimensional simulations with line-tied boundary conditions
runs~B, C and D, with respectively single parallel modes $m=0$, $m=1$ and 
all modes $m \in [0,4]$, and initial conditions with $b_0/B_0 = 10\%$.
The elongated structure of $\psi$ along $z$ shows that 
in  the \emph{asymptotic regime} the fields 
relax into an equilibrium with a strong parallel m=0 mode
not only for run~B, that initially has only the mode m=0,
but for all the initial conditions considered, including
runs~C that initially has only the parallel mode m=1 and run~D
that is started with many parallel modes. Snapshots
times, left to right, are respectively 
$t = 516\, \tau_A$,  $893\, \tau_A$ and $240\, \tau_A$.
The magnetic potential $\psi$ is defined in Section~\ref{sec:pm}.
The computational box has been rescaled for an improved visualization,
but the axial length is ten times longer that the perpendicular cross section length.
\label{fig:figpsi}}
\end{figure*}

It is therefore of interest to consider initial conditions with modes
higher than $m=0$, and in the numerical simulations analyzed in 
this section the initial magnetic field
has \emph{only the parallel mode} $m=1$ (corresponding to the
resonant frequency $\nu_2 = \nu_A$), while large-scale 
perpendicular modes are set as in previous simulations
with wavenumbers between 3 and 4
(Section~\ref{sec:initial}, Equation~[\ref{eq:pot4}]).
Thus unlike run~B with $m=0$, the out-of-equilibrium initial magnetic field 
now varies also along $z$, with both terms in the equilibrium equation~(\ref{eq:eqb}) 
not vanishing ($\partial_z j \ne 0$, $\mathbf{b} \cdot \nabla j \ne 0$).
The magnetic field lines of the orthogonal component $\mathbf{b}$
and current density $j$ are shown in Figure~\ref{fig:fig7} (top row)
at time $t=0$ in the mid-plane $z=5$,
while isosurfaces of the magnetic potential at $t=0$ are shown in 
Figure~\ref{fig:figpsi} (central column).  In both figures the case with
$b_0/B_0 = 10\%$ is considered. This is one of a series of simulations collectively
labeled as runs~C, with same parameters for the initial condition 
except the magnetic field intensity $b_0$ (the multiplicative factor in 
Equation~[\ref{eq:pot4}]) that spans the range $0.1\% \le b_0/B_0\le10\%$.

The time evolution of total energy for runs~C 
with different values of $b_0$ is shown in 
Figure~\ref{fig:figc1} (left panel). The run with $b_0/B_0 = 10\%$
has a similar behavior to run~B with same magnetic
field intensity (cf.\ Figure~\ref{fig:fig1}). Its energy decays
approximately as $E \propto t^{-1}$, with current sheets
forming in physical space (Figure~\ref{fig:fig7}, top row) and
dissipating $\sim 92\%$ of the initial energy,
a slightly higher value respect to the corresponding run~B.
Subsequently the system relaxes to an asymptotic state with 
$b/B_0 \sim 2.87\%$.

The analysis of the equilibria set forth in Section~\ref{sec:sred}
shows that \emph{the only dynamically accessible equilibria are
those with variation length-scale greater than approximately the loop length}
$z_\ell \gtrsim L_z$. These have structures very elongated
in the axial direction, and therefore dominated by the parallel mode
$m=0$. When initial conditions do not include the 
$m=0$ mode, their higher modes will necessarily have to transfer part of their energy
to the mode $m=0$ via nonlinear dynamics in order to relax to equilibrium.
Furthermore most of the energy of the modes with $m \ge 1$ that is not
converted into the parallel zero mode must be dissipated for the relaxed
state to be close to a reduced MHD equilibrium with $z_\ell  \gtrsim L_z$,
with a predominant parallel zero mode.

In fact the isosurfaces of the magnetic potential $\psi$ in Figure~\ref{fig:figpsi}
show that both runs~B and C with $b_0/B_0 = 10\%$
(respectively in the left and central columns)
relax to a lower energy state with structures very elongated
along $z$ (the computational box has been rescaled for an improved visualization,
but the axial length is \emph{ten times longer} that the perpendicular cross section length),
even though their initial conditions are radically different, consisting
exclusively of mode $m=0$ for run~B and $m=1$ for run~C.

Consequently similar dynamics will occur for all runs~C independently
from the value of $b_0/B_0$, as shown in Figure~\ref{fig:figc1},
because all of them do not have a mode $m=0$ in their initial
magnetic field. Thus they all decay while part of the energy initially in the $m=1$ mode 
is either transferred to the $m=0$ mode (and partially also to higher order modes)
or dissipated (including dissipation of the higher order modes generated during the 
nonlinear dynamics).
On the contrary when the initial condition is made only of mode $m=0$,
the system is very close to an equilibrium for $b_0/B_0 \lesssim 3\%$ 
(runs~B, Figure~\ref{fig:fig1}) since this condition implies $z_\ell \gtrsim L_z$
and the magnetic field is quasi-invariant along $z$, so that no substantial
dynamics occur when $b_0/B_0 \lesssim 3\%$.

Figure~\ref{fig:figc1} shows that for runs~C energy starts
decaying at later times for smaller values of $b_0/B_0$.
The longer timescales for dissipation to occur and energy to start
decaying is consistent with the decrease of the strength of
nonlinear interactions for lower values of $b_0/B_0$. For
instance the eddy turnover time increases as $t_\ell \sim 2\ell/b_0$,
since no velocity is initially present and this soon becomes
of the order of $b_0/2$ (because for resonant frequencies velocity is in equipartition
with the magnetic field). Subsequently velocity strongly decreases once a zero
mode is created and the system relaxes. 

Clearly the fraction of magnetic energy dissipated during the decay
depends on several factors. The most important of these is
how much energy in transferred to the parallel zero mode, since
all higher modes will be largely dissipated in order to reach an equilibrium
with $z_\ell \gtrsim L_z$. A detailed analysis of the energy fluxes
between these modes goes beyond the scope of the present paper, and
might be carried out in future work. Nevertheless it is clear 
from Figure~\ref{fig:figc1} that runs~C with $4\% \le b_0/B_0 \le 10\%$
generate a zero mode quickly, while for $b_0/B_0 \le 1\%$
a large fraction of the initial energy is dissipated with only
a small fraction transferred to the zero mode.
In all cases in the asymptotic stage, when the mode $m=0$
is the strongest mode, the relaxed magnetic field intensity $b$ is below
the stability threshold~(\ref{eq:eqrp}) with $b/B_0 \lesssim 3\%$.

In spite of all the aforementioned differences, the longer \emph{decay} timescales
for runs~C with lower values of $b_0/B_0$ render \emph{similar}
the behavior of the system \emph{forced} by photospheric
motions for both cases with a velocity that is constant in time (zero frequency) 
and a velocity with higher resonant frequencies.
In fact considering a straightened loop with initially
only the guide field $B_0$, if a constant velocity ($\nu_0=0$)
is applied at the photospheric boundaries, the magnetic field
will grow linearly in time initially (Equation~[\ref{eq:bp}]),
because until the orthogonal component of the magnetic field 
does not reach the critical value $b \sim \ell B_0/L_z$ (Equation~[\ref{eq:eqrp}])
the system is very close to equilibrium and nonlinear terms can be neglected.
The linear growth of the magnetic field is derived indeed from 
the \emph{linearized} reduced MHD equations~(\ref{eq:els}).

In similar fashion if the photospheric velocity frequency is a higher resonance
$\nu_n = n \nu_A/2$, with $n \ge 1$, again nonlinear terms can be 
neglected initially because for low values of the magnetic field intensity $b$
the decay timescales are much longer than the linear growth
of the magnetic field, with the amplitude doubling every Alfv\'enic
crossing time $\tau_A$. Equations~(\ref{eq:linres1})-(\ref{eq:linres2}) are obtained
also for the resonant frequencies from the linearized reduced MHD equations, 
analogously to the constant forcing case (Equations~[\ref{eq:els}]-[\ref{eq:bp}]).
A statistical steady state will finally be obtained when the energy flux injected
in the system at the boundary by photospheric motions
is balanced by a similar energy flux from the large toward
the small scales (to form current sheets), in similar fashion to
the constant velocity case \citep{2007ApJ...657L..47R}.

\subsection{Runs D: modes m=0--4} \label{sec:run04}

In this section we analyze numerical simulations (runs~D) in which the initial
magnetic field $\mathbf{b}$ includes all parallel modes $m \in [0,4]$,
while large-scale perpendicular modes are set as in previous simulations
with wavenumbers between 3 and 4 (Section~\ref{sec:initial}, Equation~[\ref{eq:pot4}]).
The parallel zero mode is the strongest, with higher parallel modes having less energy.
For all runs the fraction of magnetic energy in the \emph{parallel mode}~$m$, indicated
with $\mathscr{E}_m$, is set in the initial magnetic field (Equation~[\ref{eq:pot4}])
so that $\mathscr{E}_m/\mathscr{E}_0 = (m+1)^{-2.6}$, corresponding to a progressively
smaller energy for higher modes. Explicitly 
$\mathscr{E}_m/\mathscr{E}_0$ = 16.5\%, 5.7\%, 2.7\%, 1.5\% for $m \in [1,4]$.
This specific choice of values is arbitrary, but
the presence of multiple parallel modes of decreasing weight is chosen
to represent the coronal field induced by photospheric motions
in which multiple frequencies are present. Although
to date there are no measurements of the spectrum of photospheric velocities,
the presence of higher frequency modes with decreasingly smaller amplitudes
is expected. 
Since smaller granules are expected to have faster convective timescales
this is partially confirmed by the recent detection of mini-granular
structures with their size distributed as a power law with an approximate 
Kolmogorov index $\sim -5/3$ and dominant on scales smaller than $600$\,km
\citep{2012ApJ...756L..27A}.

As shown in Figure~\ref{fig:figc1} (right panel)
the dynamics are analogous to those of run~B
with only the parallel mode $m=0$ (cf.\ Figure~\ref{fig:fig1}).
The dissipated energy decreases for lower values of $b_0/B_0$,
but unlike runs~B a small but discernible energy dissipation occurs
also for very small ratios of $b_0/B_0$.
Each run~D dissipates a larger fraction of magnetic energy
respect to the corresponding run~B with same $b_0/B_0$
before reaching the asymptotic regime (cf.\ insets
in Figures~\ref{fig:figc1} and \ref{fig:fig1}), because initially
a fraction of the energy in runs~D is in parallel modes higher than zero,
and they will be dissipated during the decay so that the system
can relax to equilibrium (dominated by the mode $m=0$ with variation
length-scale $z_\ell \gtrsim L_z$).

Dynamics in physical space are also similar to those of run~B,
as shown in Figure~\ref{fig:fig7} (bottom row) for $b_0/B_0 = 10\%$,
with current sheets forming and dissipating energy thus leading to 
a relaxed field with larger magnetic islands through an inverse cascade.
The evolution of the three-dimensional structure of the magnetic potential $\psi$,
shown in Figure~\ref{fig:figpsi} (right column), is similar to those of runs~B and C. 
The relaxed magnetic potential at $t = 240\, \tau_A$ has a structure elongated along~$z$,
with a strong $m=0$ parallel mode similarly to runs~B and C,
even if its structure at time $t=0$ is considerably more complex due to the presence 
of multiple parallel scales ($m \in [0,4]$).

\section{Conclusions and Discussion} \label{sec:cad}

Equilibria and dynamics of the magnetically confined regions of solar and stellar atmospheres
have been investigated with a reduced MHD cartesian model 
to advance our understanding of the mechanism 
that powers the X-ray activity of the Sun, late-type main
sequence stars, and more in general of stars with
a magnetized corona and an outer convective envelope.

Since equilibria play a pivotal role in understanding the dynamics
of this system, their structure has been analyzed in detail in Section~\ref{sec:eq}.
The mapping between the solutions of the 2D Euler equation $\mathbf{u}_{2D}(\mathbf{x}, t)$
and reduced MHD equilibria $\mathbf{b}_{eq}(\mathbf{x}_\perp, z)$ (Equation~[\ref{eq:etr}]:
$t \rightarrow z$, $\mathbf{u}_{2D} \rightarrow \mathbf{b}_{eq}/B_0$)
allows to formulate a heuristic quantitative analysis of the structure of the reduced MHD equilibria. 
The \emph{inverse cascade} developed by the solutions of the Euler equation \emph{in time},
corresponds to an \emph{asymmetric structure along} $z$ of the reduced MHD equilibria,
as pictorially summarized in Figure~\ref{fig:fig_ic}.

In similar fashion to 2D velocity vortices of scale $\ell$ that double their size in about one
eddy turnover time $t_\ell \sim \ell/u_\ell$ (Equation~[\ref{eq:tk41}]),  a reduced
MHD equilibrium with orthogonal magnetic field of intensity $b_{bd}$
at the bottom boundary $z=0$, and made of magnetic
islands of scale $\ell$, will have progressively larger magnetic islands at larger $z$, 
doubling their transverse scale over the axial spatial distance
\begin{equation} \label{eq:pvs2}
z_{\ell} \sim \frac{B_0}{b_{bd}}\, \ell.
\end{equation}
This represents the \emph{parallel variation length-scale}
of the equilibrium solution, and \emph{measures quantitatively
its asymmetry}. An equilibrium is strongly asymmetric along $z$
if the variation scale is smaller than the loop length $z_\ell < L_z$,
but if the variation scale is greater than approximately the loop length $z_\ell \gtrsim L_z$
then the equilibrium is quasi-invariant along $z$ (at the loop scale).

On the other hand in reduced MHD any spatial variation of the
physical fields along $z$ is rapidly propagated away at the
Alfv\'en speed $B_0$ (the fastest speed in the system) 
by the linear terms $\propto B_0\partial_z$ in Equations~(\ref{eq:eq1})-(\ref{eq:eq2}).
Therefore the \emph{physical solutions cannot develop strongly asymmetric
structures} along~$z$, as confirmed by boundary forced and decaying numerical
simulations with line-tying boundary conditions and a strong guide field
\citep[e.g.,][]{1996JGR...10113445G, 1999ApJ...527L..63D, 2008ApJ...677.1348R,
2011A&A...536A..67W, 2012A&A...544L..20D}. This implies that physical solutions are close
to an equilibrium or not depending on the intensity of the orthogonal component
of the magnetic field $b$, that regulates the value of the parallel equilibrium
variation scale $z_\ell$ (\ref{eq:pvs2}) for a given guide field of intensity $B_0$.
Consequently \emph{the reduced MHD equilibria that can be accessed
dynamically are those quasi-invariant along $z$ with an
orthogonal magnetic field component $\mathbf{b}$ for which the parallel length-scale
is greater than approximately the loop length} $z_\ell \gtrsim L_z$. 

The simulations reported by \cite{2013ApJ...773L...2R}, analyzed in detail in 
Section~\ref{sec:run0} (runs~A--B),  are in agreement with this picture. 
They considered initial magnetic fields $\mathbf{b_0}$ invariant along $z$
(i.e., only the parallel mode $m=0$ is present initially and $\partial j_0 = 0$),
but with different intensities and non-vanishing Lorentz force for the perpendicular magnetic field, i.e., 
$\mathbf{b}_0 \cdot \nabla j_0 \ne 0$, and identified the \emph{magnetic field intensity threshold}
$b \sim \ell B_0/ L_z$, \emph{corresponding to the critical variation scale} $z_\ell \sim L_z$. 
Initial conditions with $ b_0 > \ell B_0/ L_z$ ($z_\ell < L_z$) have dynamics increasingly similar
to 2D MHD turbulence decay for larger values of $b_0$, 
with the orthogonal magnetic field forming current sheets
and dissipating energy that decays as a power-law in time with $E \propto t^{-\alpha}$
($\alpha \sim 1$ for $b_0/B_0=10\%$, Figure~\ref{fig:fig1}). 
As shown in Equation~(\ref{eq:ss}) in the 2D case the evolution of fields
with different initial intensities is self-similar  in time with
$\mathbf{b}'(t) = \mathbf{b}(t\cdot b'_0/b_0)\, b'_0/b_0$,
implying that they all decay with same power-law index and relax to asymptotic 
fields with intensities proportional to their initial value
$b'_\infty / b'_0 = b_\infty/b_0$. 
But in stark contrast with the 2D case the line-tied 3D simulations decay with progressively
shallower power-laws (Figure~\ref{fig:fig1}) for weaker initial magnetic fields 
with smaller ratios $b_0/B_0$, relaxing to asymptotic equilibria 
for which the ratio $b_\infty/b_0$ is not independent from $b_0$.
Instead $b_\infty \sim \ell B_0/ L_z$, corresponding to a variation scale approximately equal 
to the loop length $z_\ell \sim L_z$ (as explained in Section~\ref{sec:run0} the orthogonal scale 
$\ell$ increases for stronger magnetic field because an inverse cascade of magnetic energy 
occurs).
Furthermore initial magnetic fields with intensity below the threshold $ b_0 \lesssim \ell B_0/ L_z$ 
($z_\ell \gtrsim L_z$) show little dynamics with no significant decay nor current sheets 
formation and dissipation.

Therefore these simulations
confirm numerically that the dynamically accessible equilibria are those
quasi-invariant along~$z$ (i.e., with a dominant $m=0$ mode)
with magnetic field intensity smaller than the threshold $b \lesssim \ell B_0/ L_z$, and
corresponding parallel variation length-scale larger than approximately the loop 
length $z_\ell \gtrsim L_z$.
The nature of this equilibria is radically different from the classic reduced MHD equilibria considered
in plasma and solar physics in the framework of \emph{linear instabilities} (kink, tearing, etc.), 
that typically are strictly invariant along~$z$ ($\partial_z = 0$) and in the reduced MHD framework
have a vanishing orthogonal Lorenz force component with $\mathbf{b} \cdot \nabla j = 0$. 
This condition is satisfied by very symmetric fields, e.g., a sheared field, or circular field lines 
\citep[examples can be found in][and references therein]
{1983ApJ...264..635P, 1993PhFlB...5.2858L}.
But our initial magnetic fields (Section~\ref{sec:initial}) have non-vanishing
orthogonal Lorentz forces ($\mathbf{b} \cdot \nabla j \ne 0$), a property that stems from 
the complexity and disorder of photospheric motions.
As shown in Figure~\ref{fig:fig_p1} and \ref{fig:fig_p3} both terms in the equilibrium 
Equation (\ref{eq:eqb}) \emph{do not vanish as the system relaxes to equilibrium}, 
with their rms $\sigma_z$ and $\sigma_\perp$
(Equation~[\ref{eq:rmsz}]) getting asymptotically equal, while the rms of their sum $\sigma_{eq}$ 
(Equation~[\ref{eq:rmseq}]) vanishes asymptotically as a power-law.
Consequently \emph{the system does not relax to a classic linearly unstable equilibrium}
with $\partial_z j= 0$ and $\mathbf{b} \cdot \nabla j = 0$, as confirmed by a visual
inspection of the orthogonal magnetic field $\mathbf{b}$ in Figure~\ref{fig:fig5} (right column).

The simulations (runs~B) have very different dynamics whether
their initial parallel variation scales are larger or smaller than
the critical length-scale $z_\ell \sim L_z$.
For $z_\ell < L_z$, with $b > \ell B_0/L_z$,  the initial magnetic field is very far from the corresponding
equilibrium, as this is too asymmetric along~$z$ and is therefore dynamically
inaccessible. The only way for the out-of-equilibrium field to reach an equilibrium 
is therefore to decay to a lower energy configuration with smaller $b$ until the critical length scale 
$z_\ell \sim L_z$ is reached, and this necessarily implies the formation of current sheets
and dissipation through nonlinear dynamics  
\citep[a magnetically dominated nonlinear MHD turbulent cascade analyzed in depth in][]
{2011PhRvE..83f5401R}. 
On the contrary initial magnetic fields with $z_\ell \gtrsim L_z$,  for which $b \lesssim \ell B_0/L_z$,
are very close to the corresponding equilibrium because they are both 
elongated along $z$. Thus the field simply
readjusts to the close equilibrium with no significant nonlinear dynamics, current sheet formation
nor dissipation, as shown in Figures~\ref{fig:fig1}--\ref{fig:fig5} and particularly in
Figures~\ref{fig:fig_p1} and \ref{fig:fig_p3}. Strictly speaking also in this case a very
small energy dissipation occurs, but it is negligible,  does not involve the formation
of significantly stronger currents, and additionally nonlinearities are diminished in close
proximity to equilibrium. 

The quasi-static evolution of the magnetic field is then restricted only to  field
intensities smaller than approximately $b \lesssim \ell B_0/L_z$, while stronger
fields are necessarily out-of-equilibrium and develop turbulent dynamics with subsequent
current sheet formation and energy dissipation. 

Consequently \emph{two distinct stages} can be identified in the dynamics of 
an initially uniform and strong axial magnetic field 
$B_0 \mathbf{\hat{e}}_z$ shuffled at its footpoints by
a constant or low frequency photospheric velocity $\mathbf{u}_{ph}$
(see Section~\ref{sec:sred} for a discussion on forcing frequencies).
To mimic the solenoidal component of the photospheric horizontal
velocity (the irrotational component cannot twist the field lines), the 
incompressible velocity at the boundary
$\mathbf{u}_{ph}$ is made up of distorted vortices
\citep[see][for a specific example]{2008ApJ...677.1348R}
with $\mathbf{u}_{ph} \cdot \nabla \omega_{ph} \ne 0$, given
the general complexity and disorder of photospheric motions.
\emph{At first} photospheric motions generate an orthogonal 
coronal magnetic field component that grows linearly in time
and is a mapping of the photospheric velocity, i.e., 
$\mathbf{b} = \mathbf{u}_{ph}\, t/\tau_A$ (Equation~[\ref{eq:bp}]).
Until its intensity remains below the threshold $b \sim \ell B_0/L_z$ the 
field is essentially in equilibrium and \emph{nonlinearities do not develop},
leading to its linear growth in time. In fact neglecting nonlinear terms in
the reduced MHD equations, the linear growth follows from the
remaining linear terms (Equation~[\ref{eq:els}]) and boundary conditions.
When the magnetic field intensity crosses the threshold 
the variation length-scale of the corresponding equilibrium
becomes smaller than the loop length $z_\ell \lesssim L_z$.
The structure of the equilibrium becomes  then too asymmetric along~$z$,
while the dynamically induced magnetic field (Equation~[\ref{eq:bp}])
is quasi-invariant along~$z$. The magnetic field is then too distant from its 
corresponding equilibrium that cannot be accessed dynamically unless the field
intensity decreases.
The Lorentz force components that were in equilibrium during the quasi-static
stage now cannot reach a balance with each other, the magnetic field
is therefore in non-equilibrium and nonlinear dynamics develop.

\cite{1988ApJ...330..474P} had conjectured a two-stage process
to account for the inferred  Poynting flux in active regions, estimated by
\cite{1977ARA&A..15..363W} at about
$S_z \sim 10^7$~erg~cm$^{-2}$~s$^{-1}$. In fact if
current sheet formation and energy dissipation would
be effective at an earlier stage, with too weak magnetic fields,
then the Poynting flux would be too small to sustain
the X-ray activity of active regions.
Reverting now to conventional units, the time and space averaged
Poynting flux is given by 
$\langle S_z \rangle \sim S/\ell^2 \sim \rho v_{A,\parallel}\, u_{ph}\, v_{A,\perp}$
(see Section~\ref{sec:energy}, density $\rho$ and Alfv\'en velocities are introduced
from dimensional calculations),
where $v_{A,\parallel} = B_0/\sqrt{4\pi \rho}$  and $ v_{A,\perp}= b/\sqrt{4\pi \rho}$
are the Alfv\'en velocities associated respectively to the guide field $B_0$ and 
the orthogonal magnetic field component $\mathbf{b}$.
Introducing the threshold magnetic field intensity $b \sim \ell B_0/L_z$
(or the associated Alfv\'en velocity) we obtain for the Poynting flux:
\begin{equation}
  \langle S_z \rangle \sim \rho\, v^2_{A,\parallel} u_{ph} \frac{\ell}{L_z}
              = \frac{B_0^2 u_{ph} \ell  }{4\pi L_z}.
\end{equation}
This coincides with the strong guide field regime of the scaling relation obtained by
\cite{2008ApJ...677.1348R} (Equation~[68] with $\alpha \gg 1$)
for boundary forced simulations, that yields for typical solar active region loops
an energy flux $\sim 1.6 \times 10^6$~erg~cm$^{-2}$~s$^{-1}$, in the lower
range of the constraint inferred by \cite{1977ARA&A..15..363W}.
But recent fully compressible MHD simulations with similar setup, 
that include the integration of an energy equation with thermal 
conduction and energy losses provided by optically 
thin radiation, and in addition have \emph{density stratification} (with strong gradients from 
the chromosphere to the corona), exhibit Poynting fluxes of the order of
$\sim 10^7$~erg~cm$^{-2}$~s$^{-1}$, and most importantly
an X-ray emission that matches the physical 
properties of the observed radiation \citep{dahlburgugarte}.

Additionally numerical simulations with initial magnetic fields
not invariant along $z$ have been carried out (runs~C and D).
Higher parallel modes can be present for both the Sun
and other active stars of interest.
They can be generated by the nonlinear dynamics
even when photospheric motions have a low frequency, or they can
be directly excited by photospheric motions in long loops, that
on some stars have observationally inferred lengths of the order
of several stellar radii 
\citep{2005ApJS..160..469F, 2008ApJ...688..437G, 2010Natur.463..207P}.

Runs~C include only the parallel mode $m=1$, while for runs~D all modes
$m \in [0,4]$ are present. As in previously discussed runs~B the initial
magnetic field $\mathbf{b}_0$ is not in equilibrium, now
with both terms in the equilibrium Equation~(\ref{eq:eqb}) non-vanishing
($\mathbf{b}_0 \cdot \nabla j_0 \ne 0$ and $\partial_z j_0 \ne 0$).
Remarkably these initial conditions decay to equilibria with 
structures similar to those of runs~B (whose initial magnetic fields included
only the mode $m=0$), as shown in Figure~\ref{fig:figpsi}. 
\emph{Independently from the structure
of the initial magnetic field}, and from the specific modes that it includes,
\emph{the final equilibrium is always quasi-invariant along~$z$ with parallel
variation length-scale larger than approximately the loop length} $z_\ell \gtrsim L_z$.
Its structure is elongated along the axial direction, a strong $m=0$ parallel mode is present, 
and the magnetic field intensity  is smaller than the threshold value $b \lesssim \ell B_0/L_z$.
This further confirms the analysis of the equilibria performed in Section~\ref{sec:sred}, i.e.,
that the reduced MHD equilibria dynamically accessible in a line-tied configuration
are elongated along the axial direction with a dominant  $m=0$  parallel mode.
It also implies that nonlinear dynamics can transfer energy from higher parallel modes 
to the mode $m=0$ even when this is not initially present (runs~C), a process that can be
of interest also in periodic turbulence \citep{2011PhRvE..84e6330A, 2012PhRvE..85c6406S},
and has been conjectured to play a role in the dynamics that lead to the acceleration 
of the solar wind \citep{2001PhPl....8.2377D}.

In contrast to runs~B, with initial conditions invariant along~$z$, that decay only
for magnetic field intensities above the threshold $b \sim \ell B_0/L_z$, in runs~C
a decay is always observed independently from the intensity of the initial magnetic
field (Figure~\ref{fig:figc1}). Since the accessible equilibria are quasi-invariant along~$z$ 
and the initial magnetic field of runs~C does not contain the $m=0$ mode but only the 
$m=1$ mode, then it always decays. The asymptotic stage is 
reached when the mode $m=0$ has been generated and excess energy in higher 
modes dissipated. The intensity of the relaxed magnetic field depends on the
ability of nonlinear dynamics to transfer energy among higher parallel modes
and from these to the $m=0$ mode, but the longer decay timescales for lower values of $b_0/B_0$
are consistent with a decrease of the strength of nonlinear interactions 
(e.g., the eddy turnover time decreases ad $t_\ell \sim 2\ell/b_0$,
see Section~\ref{sec:run1}). The longer nonlinear timescales
render the effect of a high-frequency resonant photospheric forcing
similar in many aspects to that of a constant photospheric velocity
(see Section~\ref{sec:run1} for a more complete discussion).
In fact if a photospheric velocity with a higher resonant frequency
$\nu_n = n \nu_A/2$ with $n \ge 1$ is applied at the boundary, 
nonlinear terms can be neglected initially because for low 
values of the magnetic field intensity $b$
the decay timescales are much longer than the linear growth
of the magnetic field.   A statistical steady state
will finally be obtained when the energy flux injected
in the system at the boundary by photospheric motions
is balanced by a similar energy flux from the large toward
the small scales (to form current sheets), in similar fashion to
the constant velocity case \citep{2007ApJ...657L..47R}.

When initial conditions include higher parallel modes $m \in [0,4]$ and
the $m=0$ mode has the largest amplitude (runs~D, Section~\ref{sec:run04})
the dynamics are very similar to runs~B with only the $m=0$ mode 
in the initial conditions (cf. Figures~\ref{fig:fig1} and \ref{fig:figc1}).
The excess energy in higher parallel modes is either dissipated or 
transferred to the $m=0$ mode and the relaxed fields have structures
similar to runs~B and C (Figure~\ref{fig:figpsi}).

The parallel variation scale $z_{\ell} \sim \ell B_0/b$ introduced
here (Equation~[\ref{eq:pvs2}]) can be interpreted as a \emph{critical length}
or \emph{twist}. In fact given the magnetic field intensity $b$, significant nonlinear 
dynamics will develop only if the loop length is longer than the variation scale $L_z \gtrsim z_\ell$.
On the other hand, fixed the loop length $L_z$ nonlinear dynamics will develop only
if the variation scale is smaller than the loop length $z_\ell \lesssim L_z$, or equivalently
if the field intensity is larger than the  threshold $b \gtrsim \ell B_0/L_z$, that corresponds
to an average twist larger than $\langle \Phi \rangle \sim L_z b/ (\ell B_0) \gtrsim \pi/3$
(this is only an estimate, since the orthogonal field does not have cylindrical symmetry
$b \ne b(r)$, the twist should be computed numerically for sample field lines).

The concept of a critical length or twist has been developed in the study of
several linear instabilities in coronal loops with line-tied boundary conditions, 
including kink and tearing instabilities
\citep{1972SoPh...22..425R, 1979SoPh...64..303H, 1981GApFD..17..297H, 1981PhFl...24.1092E, 
1983SoPh...88..163E, 1989SoPh..119..107V, 1990ApJ...350..428V, 1990ApJ...350..437F, 
1998ApJ...494..840L, 2009PhPl...16d2102H, 2010PhPl...17e5707H}.
It is found that the system is \emph{linearly unstable} for a fixed magnetic field intensity
only if the loop length is larger than a critical value (for kink and other instabilities a critical
twist can be used equivalently).

It is important to remark that although it is useful to regard the parallel variation 
scale $z_\ell$ (Equation~[\ref{eq:pvs2}]) as a critical length, and that an average 
critical twist can be defined,
the dynamics that they help describe are \emph{not linear instabilities}.
In the configurations of interest to this paper the critical length or twist
distinguish two different dynamic regimes in which respectively for small field intensities $b$ 
nonlinear dynamics are suppressed and the evolution can be regarded
as quasi-static, while for stronger intensities nonlinear dynamics develop.
Of course the boundary between these two regimes is not sharp, but gradually
\emph{as $b$ is increased the ``2D'' Lorentz force $\mathbf{b} \cdot \nabla \mathbf{b}$
cannot be balanced by the axial field line tension $B_0 \partial_z \mathbf{b}$, 
leading the system out of equilibrium}.
In particular all the 3D line-tied magnetic fields that we have considered lack the symmetries
of linearly unstable configurations, and for all of them the orthogonal component 
of the magnetic field is not symmetric and its 2D Lorentz force component 
(for which $\mathbf{b} \cdot \nabla j \ne 0$) does not vanish
at all times, from the initial condition to the relaxed
asymptotic stage, as can be seen in 
Figures~\ref{fig:fig5}-\ref{fig:fig_p3}, \ref{fig:fig7} and \ref{fig:figpsi}.

The reason for which the equilibria with $z_\ell \gtrsim L_z$ are not linearly unstable
is that they are close to each other. Therefore adding a perturbation to
the magnetic field simply changes slightly the corresponding equilibrium to which
the field readjusts. For instance the initial condition of run~B with $b_0/B_0=2\%$
is not exactly an equilibrium (since $\partial_z j_0 = 0$ and $\mathbf{b}_0 \cdot \nabla j_0 \ne 0$),
therefore it can be regarded as an equilibrium to which
has been added a small perturbation. But as shown particularly well in Figure~\ref{fig:fig7} 
(bottom panel) no instability of sort is detected, rather the field undergoes a slight
readjustment.
Clearly for a non-vanishing orthogonal magnetic field $\mathbf{b}$ in equilibrium, 
with a progressively smaller 2D Lorentz force component for which
$\mathbf{b} \cdot \nabla j \rightarrow 0$ (with $b \ne 0$),  
the field approaches a symmetric configuration that can be linearly unstable,
since from the equilibrium Equation~(\ref{eq:eqb}) also $\partial_z j \rightarrow 0$
in this limit. But as previously discussed the configurations of interest here
are those with non-vanishing orthogonal 2D Lorentz force component
$\mathbf{b} \cdot \nabla j \ne 0$.

The relaxation of coronal magnetic fields has often
been studied subsequently to a linear instability,
mostly kink modes. Particularly for the cases
that have a strong axial magnetic field the structure
of the lower energy relaxed field appears elongated along~$z$ 
\citep{1990ApJ...361..690M, 1994ApJ...426..742L, 1996A&A...308..935B, 1997SoPh..172..257V,
1998ApJ...494..840L, 2000A&A...360..345B, 2002A&A...387..687G, 2008A&A...485..837B,
2009A&A...506..913H}.
Early boundary forced simulations have been performed by \cite{1998PhPl....5.4028N},
with similar setup as those of \cite{2008ApJ...677.1348R}.
But due to their low resolution they misinterpret as an instability the dynamics that develop
as the threshold $b\sim \ell B_0 /L_z$ is crossed,
when the system gradually transitions from quasi-static evolution to turbulent dynamics. 
From the simulations (runs~B) and the equilibria analysis it is clear that
the forces that are approximatively in balance below the threshold become \emph{gradually} 
unbalanced for larger magnetic field intensities, leading to the \emph{development of nonlinear 
dynamics with no intermediate instability as would occur for instance with a kink mode,
where the nonlinear stage would follow the linear instability}. 
The non-vanishing 2D Lorentz force term $\mathbf{b} \cdot \nabla  \mathbf{b}$ 
(for which $\mathbf{b} \cdot \nabla j \ne 0$), 
that in the 2D case (run~A) sets the system out-of-equilibrium and develops turbulent 
nonlinear dynamics, can be balanced in the 3D line-tied runs~B by the axial field line tension term
$B_0 \partial_z \mathbf{b}$ for field intensities below the threshold $b \lesssim \ell B_0/L_z$.
For larger magnetic field intensities the 2D Lorentz force term is stronger than its axial component,
hence the dynamics develop progressively more akin to the 2D case. 
Ultimately a force balance cannot be reached because the corresponding equilibrium is too asymmetric
along~$z$ and therefore \emph{dynamically inaccessible}, so that for larger intensities
$b$ a larger fraction of \emph{energy must be necessarily dissipated} for the system
to be able to relax and access a new equilibrium.

This two-stage process then provides a fully self-consistent alternative
to coronal heating models based on instabilities.
For instance, since resistive instabilities are slow for macroscopically thick
magnetic shears,
\cite{2005ApJ...622.1191D, 2009ApJ...704.1059D} 
obtain a shear intensity threshold for dissipation supposing that 
nanoflares would occur when photospheric motions shear the 
magnetic field beyond a certain angle, when a secondary ideal 
instability (triggered by the slow primary resistive instability) 
can develop thus accelerating the dynamics.
On the other hand, as discussed in this paper, 
as photospheric motions disorderly twist the field lines, 
once the magnetic field intensity is higher than the threshold
$b \gtrsim \ell B_0/L_z$ magnetically dominated turbulent
dynamics develop, forming current sheets 
that thin down to the dissipative scales on fast
Alfv\'en time-scales \citep{2013ApJ...773L...2R}, 
while triggering the ``ideal'' tearing instability 
\citep[see Introduction;][]{2014ApJ...780L..19P, 2015ApJ...806..131L,
2015ApJ...801..145T}, and leading to dynamics similar to 
so-called plasmoid instability \citep{1978ZhPmR..28..193B,
1986PhFl...29.1520B, 2007PhPl...14j0703L, 2008PhRvL.100w5001L,
2009PhPl...16k2102B}.

Finally, these simulations of decaying magnetic fields
show that, beyond the intensity threshold 
[Equation~(\ref{eq:pvs2})], current sheets form on fast
ideal timescales \emph{because of the nonlinear dynamics that develop}.
This is in stark contrast with the frequent hypothesis of
quasi-static evolution of the coronal magnetic
field subject to footpoint shuffling, that should continuously 
relax to a nearby equilibrium without forming current sheets
\citep[e.g.,][]{1985ApJ...298..421V}.
In the quasi-static scenario the corona could be
heated by the uniformly distributed small-scale current sheets
created by the shredding of the coronal magnetic field
after many successive random walk steps of its
field lines footpoints ~\citep{1986ApJ...311.1001V}.
But this mechanism would lead to current sheet formation
on timescale longer than photospheric convection
(several random walk steps would be required).
While the relaxation simulations presented in this paper and in 
\cite{2013ApJ...773L...2R} show that current sheets form on ideal
Alfv\'en timescales (much faster than convective timescales),
with the footpoints fixed at the photospheric plates where no 
motions are in place (and therefore no footpoint random walk occurs).

\acknowledgments
The author thanks Gene Parker and Marco Velli 
for helpful and insightful discussions, and the anonymous referee
for his remarks.
This research has been supported in part by NASA through a subcontract
with the Jet Propulsion Laboratory, California Institute of Technology, and
NASA LWS grant number NNX15AB88G and NNX15AB89G.
Computational resources supporting this work were provided by the 
NASA High-End Computing (HEC) Program through the 
NASA Advanced Supercomputing (NAS) Division at Ames Research Center.

\end{document}